\documentclass[11pt]{article}
\usepackage{latexsym,amssymb,amsmath,times}
\usepackage{enumitem}
\usepackage{float}

\usepackage{hyperref}

\usepackage[square,sort&compress,comma,numbers]{natbib}
\makeatletter
\@addtoreset{equation}{section} \makeatother

\input amssym.def
\input amssym.tex

\usepackage{tikz}
\usetikzlibrary{decorations.markings}

\newcommand{\half}{{{\textstyle\frac{1}{2}}}}
\newcommand{\quarter}{{{\textstyle\frac{1}{4}}}}
\newcommand{\be}{\begin{equation}}
\newcommand{\ee}{\end{equation} }
\newcommand{\beqa}{\begin{eqnarray} }
\newcommand{\eeqa}{\end{eqnarray} }
\newcommand{\ba}{\begin{array}}
\newcommand{\ea}{\end{array}}
\newcommand{\bpm}{\begin{pmatrix}}
\newcommand{\epm}{\end{pmatrix}}

\newcommand{\Spin}{\mathbf{Spin}}

\newcommand{\Pin}{\mathbf{Pin}}

\newcommand{\deltaS}{\delta_{\varepsilon}}

\newcommand{\rmC}{{\rm C}}

\newcommand{\sF}{\cF}
\newcommand{\sbrF}{\bar{\cF}}

\newcommand\hcL{{\hat{\cal L}}}

\newcommand{\ODD}{\mathbf{O}(D,D)}

\newcommand{\SpinD}{{\Spin(1,D{-1})_{\!\it{L}}}}
\newcommand{\oSpinD}{{{\Spin}(D{-1},1)_{\!\it{R}}}}

\newcommand{\Ott}{\mathbf{O}(10,10)}

\newcommand{\Spint}{{\Spin(1,9)}}
\newcommand{\oSpint}{{{\Spin}(9,1)}}
\newcommand{\Pint}{{\Pin(1,9)}}
\newcommand{\oPint}{{{\Pin}(9,1)}}

\newcommand{\brcF}{\bar{\cF}}
\newcommand{\Cp}{{C_{+}}{}}

\newcommand{\brCp}{{\brC_{+}}{}}

\newcommand\tr{{\rm tr}}
\newcommand\Tr{{\rm Tr}}

\newcommand\cC{{\cal C}}
\newcommand\cD{{\cal D}}

\newcommand\cF{{\cal F}}
\newcommand\cG{{\cal G}}
\newcommand\cH{{\cal H}}

\newcommand\cJ{{\cal J}}

\newcommand\cL{{\cal L}}
\newcommand\cM{{\cal M}}
\newcommand\cN{{\cal N}}

\newcommand\cP{{\cal P}}

\newcommand\cT{{\cal T}}

\newcommand{\sign}{{\small{\mathbf{{c}}}}}
\newcommand{\signp}{{\small{\mathbf{{c^{\prime}}}}}}

\newcommand\tcF{{\tilde{\cal F}}}
\newcommand{\eleven}{{(11)}}

\newcommand\rhop{{\rho^{\prime}}{}}
\newcommand\psip{\psi^{\prime}}
\newcommand\varepsilonp{\varepsilon^{\prime}{}}
\newcommand\brvarepsilon{\bar{\varepsilon}}
\newcommand\brvarepsilonp{\brvarepsilon^{\prime}{}}
\newcommand\brrhop{\brrho^{\prime}{}}
\newcommand\brpsip{\brpsi^{\prime}}

\newcommand\dis{\displaystyle}

\newcommand\twist{\dot}

\def\mA{\twist{A}}
\def\mB{\twist{B}}
\def\mC{\twist{C}}
\def\mD{\twist{D}}
\def\mE{\twist{E}}
\def\mF{\twist{F}}
\def\mG{\twist{G}}

\def\mK{\twist{K}}
\def\mR{\twist{R}}
\def\mL{\twist{L}}
\def\mM{\twist{M}}
\def\mN{\twist{N}}
\def\mT{\twist{T}}
\def\mP{\twist{P}}
\def\mS{\twist{S}}
\def\mX{\twist{X}}
\def\mY{\twist{Y}}
\def\mV{\twist{V}}
\def\mbrV{\twist{\brV}}
\def\mbrP{\twist{\brP}}

\def\mcJ{\twist{\cJ}}
\def\mcH{\twist{\cH}}
\def\mcG{\twist{\cG}}
\def\mcP{\twist{\cP}}
\def\mcL{\twist{\cL}}

\def\mcF{\twist{\cF}}

\def\mbullet{\twist{\bullet}}

\def\mbrcP{\twist{\brcP}}
\def\mrmC{\twist{\rmC}}

\def\md{\twist{d}}

\def\mpartial{\twist{\partial}}
\def\mna{\twist{\nabla}}

\def\mGamma{\twist{\Gamma}}
\def\mPhi{\twist{\Phi}}
\def\mbrPhi{\twist{\brPhi}}

\def\mcD{\twist{\cD}}

\def\msF{\twist{\sF}}
\def\msbrF{{\twist{\bar{\sF\,}}\!}}

\newcommand\mcPp{{\twist{\cal{P}}{}^{\prime}}}
\newcommand\mcPq{{\twist{\bar{\cal{P}}}{}^{\prime}}}

\newcommand\cPp{{\cal{P}^{\prime}}}
\newcommand\cPq{{\bar{\cal{P}}^{\prime}}}

\def\breta{\bar{\eta}}
\def\bralpha{\bar{\alpha}}
\def\brbeta{\bar{\beta}}
\def\brgamma{\bar{\gamma}}

\def\brrho{\bar{\rho}}
\def\brpsi{\bar{\psi}}

\def\brp{{\bar{p}}}
\def\brq{{\bar{q}}}
\def\brr{{\bar{r}}}
\def\brs{{\bar{s}}}

\def\brPhi{{{\bar{\Phi}}}}
\def\brDelta{{{\bar{\Delta}}}}

\def\brC{\bar{C}}

\def\brF{\bar{F}}

\def\brP{\bar{P}}

\def\brV{\bar{V}}

\def\brcF{\bar{\cF}}

\def\brcP{\bar{\cP}}

\newcommand{\DO}{\mathbf{\nabla}}

\newcommand{\na}{{\nabla}}

\begin{document}
\begin{titlepage}
\title{%\vskip -60pt
%\vskip 20pt
\vskip 2cm Supersymmetric  gauged Double Field Theory:\\Systematic derivation by  virtue  of \textit{Twist}}

\author{\sc  Wonyoung Cho${}^{\dagger}$,  J. J.  Fern\'andez-Melgarejo${}^{\sharp}$, Imtak Jeon${}^{\flat}$ ~and~ Jeong-Hyuck Park${}^{\dagger}$}
\date{}
\maketitle \vspace{-1.0cm}
\begin{center}
%\texttt{ycho@sogang.ac.kr,  josejuan@physics.harvard.edu,\\ imtakjeon@kias.re.kr,  park@sogang.ac.kr}\\
~\\
${}^{\dagger}$Department of Physics, Sogang University, Mapo-gu,  Seoul 121-742, Korea\\
${}^{\sharp}$Department of Physics, Harvard University, Cambridge, MA 02138, USA \\
${}^{\flat}$Korea Institute for Advanced Study,  Dongdaemun-gu, Seoul 130-722, Korea\\
%~\\
~{}\\
%\texttt{imtak@,\quad kanghoon@sogang.ac.kr,\quad park@sogang.ac.kr,\quad yjsuh@sogang.ac.kr}
%\texttt{}\\
%{{\texttt{~{{{imtak@sogang.ac.kr\,,~kanghoon@sogang.ac.kr\,,~park@sogang.ac.kr}}}}}}
~~~\\~\\
\end{center}
\begin{abstract}
\vskip0.2cm
\noindent  In a completely  systematic and geometric  way, we derive   maximal and half-maximal  supersymmetric gauged  double field theories in lower than ten dimensions.  To this end, we   apply a simple twisting ansatz to  the $D{=10}$ ungauged maximal  and  half-maximal  supersymmetric double field theories    constructed previously within the  so-called  semi-covariant formalism. The twisting ansatz may not satisfy the   section condition. Nonetheless,  all the features    of the semi-covariant  formalism, including  its complete covariantizability, are still valid after the twist under  alternative consistency conditions. The twist allows  gaugings  as  supersymmetry preserving   deformations of the $D=10$ untwisted  theories  after Scherk-Schwarz-type  dimensional reductions. The maximal supersymmetric  twist requires an extra condition  to  ensure both  the    Ramond-Ramond gauge symmetry  and the $32$ supersymmetries  unbroken.

\end{abstract}

%%%
%%{\small
%%\begin{flushleft}
%%~~\\
%%~~~~~~~~\textit{PACS}: 04.60.Cf, 02.40.-k\\~\\
%%~~~~~~~~\textit{Keywords}:  $\cM$-theory,  Duality, U-geometry.
%%\end{flushleft}}
%%%

\thispagestyle{empty}

%%%%
%%%
%%02.40.-k 	Geometry, differential geometry, and topology
%%11.25.-w 	Strings and branes
%%04.60.Cf 	Gravitational aspects of string theory
%%04.65.+e 	Supergravity 
%%%
\end{titlepage}
\newpage
\tableofcontents %%
%%\begin{document} --> JHEP
%\twocolumn

\section{Introduction}
A characteristic of Double Field Theory 
(DFT)~\cite{Siegel:1993th,Hull:2009mi,Hohm:2010jy,Hohm:2010pp} 
is the \textit{section condition}, a second order differential constraint  imposed on arbitrary fields and their products, 
such that   the $\ODD$ invariant Laplacian should be trivial,
\be
\partial_{A}\partial^{A}\sim 0\,.
\label{seccon0}
\ee
While DFT employs  doubled spacetime coordinates~\cite{Duff:1989tf,Tseytlin:1990nb,Tseytlin:1990va} manifesting  the $\ODD$  structure of T-duality, the section condition ensures that DFT lives not on the doubled $(D{+D})$-dimensional  space but on a $D$-dimensional null hyperspace, \textit{i.e.~}\textit{section}.

The geometric insight behind  the section condition was proposed in \cite{Park:2013mpa}  to claim that the coordinate space in DFT  is \textit{doubled yet  gauged}: a gauge orbit rather than a point  in  the doubled coordinate space corresponds to a physical point.  Within this picture, the exponentiation of the generalized Lie derivative which is the infinitesimal DFT-diffeomorphism generator was  shown  in \cite{Park:2013mpa}  to agree with the then-known simple ansatz of the tensorial finite diffeomorphism {\`{a} la Hohm and Zwiebach~\cite{Hohm:2012gk}, \textit{c.f.~}\cite{Lee:2013hma,Hohm:2013bwa,Berman:2014jba,
Hull:2014mxa,Naseer:2015tia}. This `coordinate gauge symmetry' was also soon successfully    realized   on a string worldsheet as a usual gauge symmetry~\cite{Lee:2013hma}
(\textit{c.f.~}\cite{Hull:2006qs,Hull:2006va,Berman:2013eva})
, where the spacetime coordinates are dynamical fields.  The constructed string  action couples to an arbitrarily curved  generalized metric and  is still completely covariant with respect to the coordinate gauge symmetry, DFT-diffeomorphisms, world-sheet diffeomorphisms, world-sheet Weyl symmetry and $\ODD$ T-duality. While  it reduces to the conventional string action upon the Riemannian parametrization of the generalized metric,  it can also go beyond the Riemannian regime.  In this way,   DFT is a stringy  gravitational theory which is defined self-consistently adopting the doubled-yet-gauged  coordinate system.

On the other hand, somewhat contrary to the  geometric significance of the section condition, it has been also observed   that,  in order to correctly   reproduce  a variety of  the   known gauged supergravities in lower than ten-dimensions  it is  necessary to consider ``relaxing" the section condition. 
In the supergravity literature,  a powerful way  of the  gauging  has  been  the embedding tensor method~\cite{deWit:2005ub} which allows for a systematic classification of all possible supersymmetric deformations as for gaugings. However, while some of the gaugings  can be obtained by a Scherk-Schwarz dimensional
reduction of the eleven- or ten-dimensional supergravities~\cite{Scherk:1979zr,Kaloper:1999yr},  a class of gaugings has been  known to have no such a higher dimensional origin.  This mystery got a new spin when  Geissb{\"u}hler~\cite{Geissbuhler:2011mx} (\textit{c.f.~}\cite{Aldazabal:2011nj})  realized   the necessity of introducing   section-condition-breaking terms for DFT to reproduce  the complete classification of  the  deformations of ${N=4}$, ${D=4}$ supergravity~\cite{Schon:2006kz}. The section condition was broken by terms which depend on both the ordinary and the dual coordinates of the  internal manifold, \textit{c.f.~}\cite{Dibitetto:2012rk}.  
This was an indication that DFT may go beyond the ordinary supergravity or \textit{Generalized Geometry}~\cite{Coimbra:2011nw,Coimbra:2012yy}. 
 Possible  modifications  of the section condition were  soon investigated  by Grana and Marques~\cite{Grana:2012rr} who looked for  a set of  consistency conditions for the closure of the  generalized Lie derivative twisted by  the Scherk-Schwarz ansatz.  Since then there have been   a few proposals  made toward  the underlying geometric principle,  notably  the flux formulation~\cite{Geissbuhler:2013uka} and  the  torsionful deformation of the semi-covariant formalism by Berman and Lee~\cite{Berman:2013cli}.

         In the flux formulation of DFT~\cite{Berman:2013uda,Geissbuhler:2013uka,Aldazabal:2013sca}, the basic building blocks constituting  the Lagrangian  are `fluxes'  which are  diffeomorphic  scalars, as in \textit{e.g.~}\cite{Siegel:1993th,Hohm:2010xe,Hohm:2011nu}. Yet, they are not local Lorentz   covariant. The  $\SpinD\times\oSpinD$ local Lorentz symmetry is only forced at the whole action level.  On the other hand, in the semi-covariant formulation of DFT~\cite{Jeon:2010rw,Jeon:2011cn}, once proper `projections' are imposed, the semi-covariant derivatives and the semi-covariant curvatures all become  completely covariant with respect to both diffeomorphisms and the local Lorentz symmetry, besides the $\ODD$ T-duality. Within this setup,   the maximal as well as half-maximal $D=10$ supersymmetric double field theories (SDFT)  have been constructed to the full order in fermions~\cite{Jeon:2011sq,Jeon:2012hp} where \textit{each term in the Lagrangians is completely covariant}, see also   the earlier formulation within Generalized Geometry~\cite{Coimbra:2011nw,Coimbra:2012yy}}.   Berman and Lee then modified the semi-covariant formalism to be apt for the twisted generalized Lie derivative by introducing torsionful semi-covariant derivative connections~\cite{Berman:2013cli}.  However, it is fair to say that while these proposals all opened up novel aspects of the section condition and hence DFT itself,  many ingredients were introduced \textit{ad hoc} by hand. Deeper systematic understanding has been  desirable.

It is the purpose of the present paper to  propose such a  geometric scheme  to twist  the maximal and the half-maximal supersymmetric double field theories of  Refs.\cite{Jeon:2011sq,Jeon:2012hp}  and  systematically   derive the gauged supersymmetric double field theories. Essentially, as our main results, we show that \textit{the semi-covariant formalism itself can be  twisted by the Scherk-Schwarz ansatz,  without any arbitrariness.  This  enables us to address  readily  the supersymmetric completions.}     The  twisted and hence gauged  maximal as well as half-maximal supersymmetric  double field theories are then completely fixed by requiring the  supersymmetry to be  unbroken. Each term in the constructed Lagrangian is  completely covariant with respect to the twisted diffeomorphisms, the $\Spint\times\oSpint$ local Lorentz symmetries, and  a subgroup of $\Ott$  which preserves the structure constant. This complete covariance also ensures    the internal coordinate independence.  \\

The organization of the paper is as follows. 
\begin{itemize}
\item In section~\ref{secReview},  we  revisit    with care the semi-covariant  formulation of the ungauged or untwisted double field theory~\cite{Jeon:2010rw,Jeon:2011cn} and its  supersymmetric extensions~\cite{Jeon:2012kd,Jeon:2011sq,Jeon:2012hp}. While reviewing them  in a self-contained manner, we spell, for later use  of twist,   all the  relevant  exact    formulas which hold without assuming  any section condition. Such formulas have not been fully spelled elsewhere before.  

\item In section~\ref{secTwist}, we twist the double field theory with a simple Scherk-Schwarz  ansatz. Following closely  Grana and Marques~\cite{Grana:2012rr}, we analyze a set of consistency conditions for the closure of the twisted generalized Lie derivatives, which we call \textit{twistability conditions}.  We show  that all the nice properties  of the semi-covariant formalism, including its complete covariantizability, are still valid after the twist under the twistability conditions. In particular, we verify that the consistent definition of the twisted Ramond-Ramond cohomology    requires one additional condition which is, after the diagonal gauge fixing of the twofold local Lorentz symmetries,  consistent with the previous work by Geissb{\"u}hler \textit{et al.}~\cite{Geissbuhler:2013uka}.

\item Section~\ref{secSDFTtwist} contains our main  results. Readers may  want to  have a glance of our final results therein, before reading the preparatory sections, \ref{secReview} and \ref{secTwist}.  We present the maximal and  the half-maximal supersymmetric gauged double field theories as the twists of the  $\cN=2$ and  the $\cN=1$, $D=10$  supersymmetric double field theories~\cite{Jeon:2011sq,Jeon:2012hp}.  In particular, we show the twisted maximal  supersymmetric invariance  calls for the same extra condition which the twisted  R-R gauge symmetry  demands as well.

\item In section~\ref{secDiscussion} we conclude with comments.
\end{itemize}
Although our supersymmetry analyses are explicit only up to  the leading order,  we argue in  section~\ref{secTwist} that the full order supersymmetric completions are guaranteed to work, as the higher order fermionic terms are immune to the ``relaxation'' of the section condition.\\

\textit{{\textbf{Conventions.}}}   Equations   which hold  due to the original   section condition~(\ref{seccon0}) and the alternative twistability   conditions  are   denoted  differently with the two distinct   symbols, `$\,\sim\,$' and `$\,\equiv\,$' respectively, besides the strict equality, `$\,=\,$'.  For the sake of simplicity we   shall often adopt a matrix notation  to suppress  contracted  indices, \textit{e.g.~}  $(P\partial_{A}P)_{B}{}^{C}=P_{B}{}^{E}\partial_{A}P_{E}{}^{C}$. Our index conventions  follow \cite{Jeon:2012hp} and are  summarized in Table~\ref{TABindices}.  In Table~\ref{TABderivative},  we also list  various derivatives  which are explained  and used   throughout the paper.

{\scriptsize{\begin{table}[h]
\begin{center}
\begin{tabular}{c|c|c}
Name~&~Schematic formula~&~Debut equation\\
\hline
Semi-covariant derivative~&~$\na_{A}=\partial_{A}+\Gamma_{A}$~&~(\ref{asemicov})\\
Master semi-covariant derivative~&~$\cD_{A}=\partial_{A}+\Gamma_{A}+\Phi_{A}+\brPhi_{A}$~&~(\ref{mastersemicovD})\\
R-R cohomology  differential operators~&~$\cD_{\pm}$~&~(\ref{Dpm})\\
U-derivative~&~$\mD_{\mA}=\mpartial_{\mA}+\Omega_{\mA}$~&~(\ref{Uderiv})\\
U-twisted master semi-covariant derivative~&~$\mcD_{\mA}=\mD_{\mA}+\mGamma_{\mA}+\mPhi_{\mA}+\mbrPhi_{\mA}$~&~(\ref{UtwistMaster})\\
\end{tabular}
\caption{Various derivatives  employed  in the present paper.} 
\label{TABderivative} 
\end{center}
\end{table}}}

{\scriptsize{\begin{table}[h]
\begin{center}
\begin{tabular}{c|c|c}
%\hline
Index~&~Representation~&Raising \& Lowering Indices\\
\hline
$A,B,\cdots$& Untwisted  $\Ott$ vector &$\cJ_{AB}=\left(\tiny{\mathbf{\ba{cc}0&1\\1&0\ea}}\right)$\\
$\mA,\mB,\cdots$& Twisted  $\Ott$ vector &$\mcJ_{\mA\mB}=\left(\tiny{\mathbf{\ba{cc}0&1\\1&0\ea}}\right)$\\
$p,q,\cdots$&$\Spint$ vector~&$\eta_{pq}=\mbox{diag}(-++\cdots+)$ \\
$\brp,\brq,\cdots$&$\oSpint$  vector~&$\breta_{\brp\brq}=\mbox{diag}(+--\cdots-)$ \\
$\alpha,\beta,\cdots$&$\Spint$  spinor~&$\Cp_{\alpha\beta}$\,,~~$(\gamma^{p})^{T}=C_{+}\gamma^{p}C_{+}^{-1}$\\
$\bralpha,\brbeta,\cdots$&$\oSpint$  spinor~&$\brCp_{\bralpha\brbeta}$\,,~~$(\brgamma^{\brp})^{T}=\brC_{+}\brgamma^{\brp}\brC_{+}^{-1}$\\
%\hline
\end{tabular}
\caption{Index  for  each symmetry   representation and the corresponding ``metric" which  raises or lowers  its position.  
Only  the capital $\Ott$ indices are to be twisted. The `$\,+\,$' subscripts  of the charge conjugation matrices indicate that they are chosen to be symmetric.   The doubling of the local Lorentz symmetries, $\Spint\rightarrow\Spint\,\times\,\oSpint$, is crucial to achieve the unification of IIA and IIB supergravities within the unique  $\cN=2$, $D=10$ untwisted SDFT~\cite{Jeon:2012hp}. } 
\label{TABindices}
\end{center}
\end{table}}}

%\newpage

%%%%%%%%%%%%%%%%%%%%%%%%%%%%%%%%%%%%%%%%%%%%%%%%%%%%%%%%%%%
%%%%%%%%%%%%%%%%%%%%%%%%%%%%%%%%%%%%%%%%%%%%%%%%%%%%%%%%%%%
\section{The   semi-covariant formulation of  ungauged  DFT\label{secReview}}
In this preparatory section, we  revisit the   semi-covariant  formulation of ungauged or untwisted double field theory~\cite{Jeon:2010rw,Jeon:2011cn} and its  supersymmetric extensions~\cite{Jeon:2012kd,Jeon:2011sq,Jeon:2012hp}. Our goal is threefold:  to review them in a self-contained manner, to locate the exact places  where the original section condition is  assumed, and to  collect, for later use  of twist,     precise   formulas which hold without assuming  any section condition. Such formulas have not been fully spelled in the literature before.  Every formula which holds up to the original section condition will be denoted by the symbol, `$\,\sim\,$', rather than by the strict equality, `$\,=\,$'.

In particular, we pay attention to  two strictly-different yet section-condition-equivalent  semi-covariant four-index curvatures,  namely $\cG_{ABCD}$ and $S_{ABCD}$, and analyze their differences exactly without assuming the section condition.

%%%%%%%%%%%%%%%%%%%%%%%%%%%%%%%%%%%%%%%%%%

\subsection{Coordinate gauge symmetry, section condition and diffeomorphism}
\begin{itemize}

\item \textit{{\textbf{Doubled-yet-gauged spacetime.}}} The spacetime is  formally doubled, being  $(D{+D})$-dimensional.  However, \underline{the   doubled spacetime  coordinates  are gauged}:  the coordinate space is equipped with   an   {equivalence relation}, 
\be
x^{A}~\sim~x^{A}+\phi\partial^{A}\varphi\,,
\label{aCGS}
\ee
which is called `coordinate gauge symmetry'~\cite{Park:2013mpa,Lee:2013hma}. In (\ref{aCGS}),  $\phi$ and $\varphi$ are arbitrary  functions in DFT.  

Each equivalence class, or gauge orbit defined by the equivalence  relation~(\ref{aCGS}),  represents a single physical point, and  diffeomorphism symmetry means  an invariance under arbitrary reparametrizations of the gauge orbits.

\item \textit{{\textbf{Section condition: realization of the coordinate gauge symmetry.}}} The equivalence relation  (\ref{aCGS}) is realized  in DFT  by enforcing  that,   arbitrary  functions and   their arbitrary  derivative descendants,  denoted here collectively  by $\Phi$,    are  invariant under the coordinate gauge symmetry  \textit{shift}~\cite{Park:2013mpa,Lee:2013hma},  
\be
\ba{ll}
\Phi(x+\Delta)\sim \Phi(x)\,,\quad&\quad\Delta^{A}=\phi\partial^{A}\varphi\,.
\ea
\label{aTensorCGS}
\ee
This invariance is  equivalent, \textit{i.e.~}sufficient~\cite{Park:2013mpa} and necessary~\cite{Lee:2013hma} to   the {section condition},
\be
\partial_{A}\partial^{A}\sim 0\,.
\label{aseccon}
\ee
Acting on arbitrary functions, $\Phi$, $\Phi^{\prime}$,  and their products, the section condition leads to  the weak constraint,  $\partial_{A}\partial^{A}\Phi\sim 0$ as well as the strong constraint, $\partial_{A}\Phi\partial^{A}\Phi^{\prime}\sim 0$.

\item \textit{{\textbf{Diffeomorphism.}}}  Diffeomorphism symmetry in  DFT    is generated by a generalized Lie derivative~\cite{Siegel:1993th,Gualtieri:2003dx,Grana:2008yw}, 
\be
\hcL_{X}T_{A_{1}\cdots A_{n}}:=X^{B}\partial_{B}T_{A_{1}\cdots A_{n}}+\omega\partial_{B}X^{B}T_{A_{1}\cdots A_{n}}+\sum_{i=1}^{n}(\partial_{A_{i}}X_{B}-\partial_{B}X_{A_{i}})T_{A_{1}\cdots A_{i-1}}{}^{B}{}_{A_{i+1}\cdots  A_{n}}\,,
\label{tcL}
\ee
where   $\omega$ denotes  the weight of the field, $T_{A_{1}\cdots A_{n}}$.  
In particular, the generalized Lie derivative of the $\ODD$ invariant metric is    trivial, 
\be
\hcL_{X}\cJ_{AB}=0\,.
\ee 
The commutator of the generalized Lie derivatives is closed  by C-bracket~\cite{Siegel:1993th,Hull:2009zb} up to the section condition,
\be
\ba{c}
\left[\hcL_{X},\hcL_{Y}\right]\sim\hcL_{[X,Y]_{\rmC}}\,,\\
{}[X,Y]^{A}_{\rmC}:= X^{B}\partial_{B}Y^{A}-Y^{B}\partial_{B}X^{A}+\half Y^{B}\partial^{A}X_{B}-\half X^{B}\partial^{A}Y_{B}\,,
\ea
\label{agB2}
\ee
since the following strict equality holds without resorting to  the section condition~\cite{Jeon:2010rw},
\be
\ba{ll}
\left([\hcL_{X},\hcL_{Y}]-\hcL_{[X,Y]_{\rmC}}\right)T_{A_{1}\cdots A_{n}}=&\half(X^{N}\partial^{M}Y_{N}
-Y^{N}\partial^{M}X_{N})\partial_{M}
T_{A_{1}\cdots A_{n}}\\
{}&+\half\omega(X^{N}\partial_{M}\partial^{M}Y_{N}
-Y^{N}\partial_{M}\partial^{M}X_{N})
T_{A_{1}\cdots A_{n}}\\
{}&\!\!\!+\sum_{i=1}^{n}(\partial_{M}Y_{A_{i}}\partial^{M}X_{B}
-\partial_{M}X_{A_{i}}\partial^{M}Y_{B})
T_{A_{1}\cdots A_{i-1}}{}^{B}{}_{A_{i+1}\cdots  A_{n}}\,,
\ea
\label{Cbracketclosure}
\ee
of which the right hand side  clearly vanishes   upon the section condition. We shall come back to this expression when we perform the twist.

\end{itemize}

%%%%%%%%%%%%%%%%%%%%%%%%%%%%%%%%%%%
\subsection{Dilaton, vielbeins and projectors}

\begin{itemize}

\item \textit{{\textbf{Dilaton and  a pair of  vielbeins.}}}  The geometric objects  in DFT come from the closed string  NS-NS sector and consist of  a dilaton, $d$, and a pair of vielbeins,   $V_{Ap}$, $\brV_{A\brp}$.   
While  the vielbeins  are   weightless,   the dilaton gives rise to  the $\ODD$  invariant \textit{integral measure}  with weight one~\cite{Hull:2009zb},  after exponentiation, 
\be
e^{-2d}\,.
\label{ameasure}
\ee
The vielbeins  satisfy the following  four   \textit{defining properties}~\cite{Jeon:2011cn,Jeon:2011vx} (see also \cite{Siegel:1993th,Grana:2008yw}):
\be
\ba{llll}
V_{Ap}V^{A}{}_{q}=\eta_{pq}\,,\quad&\quad
\brV_{A\brp}\brV^{A}{}_{\brq}=\breta_{\brp\brq}\,,\quad&\quad
V_{Ap}\brV^{A}{}_{\brq}=0\,,\quad&\quad V_{Ap}V_{B}{}^{p}+\brV_{A\brp}\brV_{B}{}^{\brp}=\cJ_{AB}\,.
\ea
\label{defV}
\ee
That is to say, they are normalized, orthogonal and complete.  The vielbeins  are $\ODD$ vectors as their  indices  indicate. In fact,  they are   the only  $\ODD$ non-singlet  field variables even   in the supersymmetric extensions  of DFT~\cite{Jeon:2011sq,Jeon:2012hp}.  As a solution to  (\ref{defV}),  they can be  parametrized     in terms of ordinary    zehnbeins  and $B$-field,   in various ways up to $\ODD$ rotations and field redefinitions, \textit{e.g.~}\cite{Jeon:2012kd,Andriot:2013xca,Andriot:2014qla}. 

Due  to the defining properties of (\ref{defV}), arbitrary variations of the vielbeins   meet 
\be
\ba{ll}
\delta V_{Ap}=\brP_{A}{}^{B}\delta V_{Bp}+V_{A}{}^{q}\delta V_{B[p}V^{B}{}_{q]}\,,~~~~&~~~~
\delta\brV_{A\brp}=P_{A}{}^{B}\delta\brV_{B\brp}+\brV_{A}{}^{\brq}\delta\brV_{B[\brp}\brV^{B}{}_{\brq]}\,.
\ea
\label{useful2}
\ee

\item \textit{{\textbf{Projectors.}}} The  vielbeins  generate a pair of symmetric, orthogonal and complete two-index projectors,\footnote{The difference of the two projectors, , $P_{AB}-\brP_{AB}=\cH_{AB}$,  corresponds to the ``generalized metric" in \cite{Hohm:2010pp}, which can be also independently defined as a symmetric $\ODD$ element, \textit{i.e.~}$\cH_{AB}=\cH_{BA}$, $\cH_{A}{}^{B}\cH_{B}{}^{C}=\delta_{A}^{~C}$. However, in the `full order' supersymmetric extensions of DFT~\cite{Jeon:2011sq,Jeon:2012hp} where \textit{e.g.~}the 1.5 formalism works, it appears that the projectors are more fundamental than the ``generalized metric".} 
\be
\ba{ll}
P_{AB}=P_{BA}=V_{A}{}^{p}V_{Bp}\,,\quad&\quad
\brP_{AB}=\brP_{BA}=\brV_{A}{}^{\brp}\brV_{B\brp}\,,
\ea
\label{def2P}
\ee
satisfying 
\be
\ba{lll}
P_{A}{}^{B}P_{B}{}^{C}=P_{A}{}^{C}\,,\quad&\quad
\brP_{A}{}^{B}\brP_{B}{}^{C}=\brP_{A}{}^{C}\,,\quad&\quad
P_{A}{}^{B}\brP_{B}{}^{C}=0\,,\\
P_{A}{}^{B}+\brP_{A}{}^{B}=\delta_{A}{}^{B}\,,\quad&\quad 
\tr(P)=P_{A}{}^{A}=D\,,
\quad&\quad\tr(\brP)=\brP_{A}{}^{A}=D\,.
\ea
\label{projection}
\ee
Further, the two-index projectors  generate a pair of six-index  projectors,
\be
\ba{l}
\cP_{ABC}{}^{DEF}:=P_{A}{}^{D}P_{[B}{}^{[E}P_{C]}{}^{F]}+\textstyle{\frac{2}{D-1}}P_{A[B}P_{C]}{}^{[E}P^{F]D}\,,\\
\bar{\cP}_{ABC}{}^{DEF}:=\brP_{A}{}^{D}\brP_{[B}{}^{[E}\brP_{C]}{}^{F]}+\textstyle{\frac{2}{D-1}}\brP_{A[B}\brP_{C]}{}^{[E}\brP^{F]D}\,,
\ea
\label{P6}
\ee
which  satisfy the `projection' property,
\be
\ba{ll}
\cP_{ABC}{}^{DEF}\cP_{DEF}{}^{GHI}=\cP_{ABC}{}^{GHI}\,,\quad&\quad
\brcP_{ABC}{}^{DEF}\brcP_{DEF}{}^{GHI}=\brcP_{ABC}{}^{GHI}\,,
\ea
\ee
 symmetric and traceless  properties,
\be
\ba{lll}
\cP_{ABCDEF}=\cP_{DEFABC}\,,\quad&\quad\cP_{ABCDEF}=\cP_{A[BC]D[EF]}\,,
\quad&\quad P^{AB}\cP_{ABCDEF}=0\,,\\
\brcP_{ABCDEF}=\brcP_{DEFABC}\,,\quad&\quad\brcP_{ABCDEF}=\brcP_{A[BC]D[EF]}\,,
\quad&\quad \brP^{AB}\brcP_{ABCDEF}=0\,,
\ea
\label{symP6}
\ee
as well as further properties like 
\be
\ba{ll}
\cP_{[AB]C}{}^{DEF}=\cP_{CAB}{}^{[EF]D}\,,\quad&\quad
\brcP_{[AB]C}{}^{DEF}=\brcP_{CAB}{}^{[EF]D}\,,\\
\multicolumn{2}{c}{
\textstyle{\frac{2}{3}}\cP_{[AB]C}{}^{DEF}+
\textstyle{\frac{1}{3}}\cP_{CAB}{}^{DEF}=
\cP_{[ABC]}{}^{DEF}=\cP_{[ABC]}{}^{[DEF]}\,,}\\
\multicolumn{2}{c}{
\textstyle{\frac{2}{3}}\brcP_{[AB]C}{}^{DEF}+
\textstyle{\frac{1}{3}}\brcP_{CAB}{}^{DEF}=
\brcP_{[ABC]}{}^{DEF}=\brcP_{[ABC]}{}^{[DEF]}\,.}
\ea
\label{asymP6}
\ee

In addition to the six-index  projection operators~(\ref{P6}), we also set for later use,
\be
\ba{l}
\cPp_{CAB}{}^{FDE}:=\brP_{C}{}^{F}P_{[A}{}^{[D}P_{B]}{}^{E]}+\textstyle{\frac{2}{D-1}}P_{C[A}P_{B]}{}^{[D}\brP^{E]F}\,,\\
\cPq_{CAB}{}^{FDE}:=P_{C}{}^{F}\brP_{[A}{}^{[D}\brP_{B]}{}^{E]}+\textstyle{\frac{2}{D-1}}\brP_{C[A}\brP_{B]}{}^{[D}P^{E]F}\,.
\ea
\label{P6pq}
\ee

\end{itemize}

%%%%%%%%%%%%%%%%%%%%%%%%%%%%%%%%%%%
\subsection{Semi-covariant derivatives, curvatures and their complete covariantizations\label{SECsemi}}

\begin{itemize}

%%%%%%%%%%%%%%%%%%%%%

\item  \textit{{\textbf{Semi-covariant derivative and the torsionless connection.}}}  The semi-covariant derivative is defined by~\cite{Jeon:2010rw,Jeon:2011cn} 
\be
\na_{C}T_{A_{1}A_{2}\cdots A_{n}}
:=\partial_{C}T_{A_{1}A_{2}\cdots A_{n}}-\omega_{{\scriptscriptstyle{T\,}}}\Gamma^{B}{}_{BC}T_{A_{1}A_{2}\cdots A_{n}}+
\sum_{i=1}^{n}\,\Gamma_{CA_{i}}{}^{B}T_{A_{1}\cdots A_{i-1}BA_{i+1}\cdots A_{n}}\,.
\label{asemicov}
\ee
It satisfies  the Leibniz rule and is   compatible   with the $\ODD$ invariant constant metric, 
\be
\na_{A}\cJ_{BC}=0\,.
\label{nacJ}
\ee
We choose the connection to be the torsionless one from Ref.\cite{Jeon:2011cn}:\footnote{The connection~(\ref{Gammao})  of  \cite{Jeon:2011cn} was reviewed further from a slightly different angle in \cite{Hohm:2011si}.} 
\be
\ba{ll}
\Gamma_{CAB}=&2\left(P\partial_{C}P\brP\right)_{[AB]}
+2\left({{\brP}_{[A}{}^{D}{\brP}_{B]}{}^{E}}-{P_{[A}{}^{D}P_{B]}{}^{E}}\right)\partial_{D}P_{EC}\\
{}&-\textstyle{\frac{4}{D-1}}\left(\brP_{C[A}\brP_{B]}{}^{D}+P_{C[A}P_{B]}{}^{D}\right)\!\left(\partial_{D}d+(P\partial^{E}P\brP)_{[ED]}\right)\,,
\ea
\label{Gammao}
\ee
which is a unique solution to the  following five constraints~\cite{Jeon:2011cn}:
\begin{eqnarray}
&\na_{A}P_{BC}=0\,,\quad\quad\na_{A}\brP_{BC}=0\,,\label{acompP}\\
&\na_{A}d=-\half e^{2d}\na_{A}(e^{-2d})=\partial_{A}d+\half\Gamma^{B}{}_{BA}=0\,,\label{acompd}\\
&\Gamma_{ABC}+\Gamma_{ACB}=0\,,\label{aGBC}\\
&\Gamma_{ABC}+\Gamma_{BCA}+\Gamma_{CAB}=0\,,\label{aGABC}\\
&\cP_{ABC}{}^{DEF}\Gamma_{DEF}=0\,,\quad\quad\bar{\cP}_{ABC}{}^{DEF}\Gamma_{DEF}=0\,.\label{akernel}
\end{eqnarray}
The first two relations, (\ref{acompP}), (\ref{acompd}), are the  compatibility conditions with  the dilaton and the projectors, \textit{i.e.~}the whole NS-NS sector. 
The third constraint~(\ref{aGBC}) is the  compatibility condition with the $\ODD$ invariant constant metric,  (\ref{nacJ}).  The next cyclic   property, (\ref{aGABC}),  makes  the semi-covariant derivative   compatible  with  the generalized Lie derivative as well as with  the C-bracket, 
\be
\ba{ll}
\hcL_{X}(\partial)=\hcL_{X}(\na)\,,\quad&\quad
[X,Y]_{\rmC}(\partial)=[X,Y]_{\rmC}(\na)\,.
\ea
\label{untwtorsionless}
\ee  
The last  formulae~(\ref{akernel})  are  projection conditions   which   ensure the uniqueness.

While the torsionless connection satisfies all the five constraints, (\ref{acompP}\,--\,\ref{akernel}) and thus uniquely determined,  a generic torsionful connection meets only the first three conditions, (\ref{acompP}),  (\ref{acompd}),  (\ref{aGBC}), and decomposes into the torsionless connection and  torsions~\cite{Coimbra:2011nw,Jeon:2011sq},
\be
\Gamma_{CAB}+\Delta_{Cpq}V_{A}{}^{p}V_{B}{}^{q}+\brDelta_{C\brp\brq}\brV_{A}{}^{\brp}\brV_{B}{}^{\brq}\,.
\label{torsionfulG}
\ee
In order to maintain (\ref{acompd}), the torsions must satisfy
\be
\ba{llll}
\Delta_{Apq}=\Delta_{A[pq]}\,,\quad&\quad \Delta_{Apq}V^{Ap}=0\,,\quad&\quad
\brDelta_{A\brp\brq}=\brDelta_{A[\brp\brq]}\,,\quad&\quad \brDelta_{A\brp\brq}\brV^{A\brp}=0\,.
\ea
\label{torsion}
\ee
In the full order supersymmetric extensions of DFT~\cite{Jeon:2011sq,Jeon:2012hp},   they are given by quadratic fermions.

It is worth while to note 
\be
P_{I}{}^{A}\brP_{J}{}^{B}\Gamma_{CAB}=(P\partial_{C}P\brP)_{IJ}\,,
\ee
such that
\be
\Gamma^{C}{}_{p\brq\,}\partial_{C}=V^{A}{}_{p}\brV^{B}{}_{\brq}\Gamma^{C}{}_{AB}\partial_{C}\,\sim\,0\,.
\label{scGp}
\ee

\item\textit{{\textbf{Spin connections and semi-covariant master derivative.}}}
The master semi-covariant derivative~\cite{Jeon:2011vx},
\be
\cD_{A}:=\na_{A}+\Phi_{A}+\brPhi_{A}
=\partial_{A}+\Gamma_{A}+\Phi_{A}+\brPhi_{A}\,,
\label{mastersemicovD}
\ee
generalizes the  semi-covariant derivative, $\na_{A}$~(\ref{asemicov}), to include the spin connections, $\Phi_{A}$ and  $\brPhi_{A}$,   for the two local Lorentz groups, $\SpinD$ and $\oSpinD$ respectively.

By definition, it is compatible with the vielbeins,
\be
\ba{c}
\cD_{A}V_{Bp}=\partial_{A}V_{Bp}+\Gamma_{AB}{}^{C}V_{Cp}+\Phi_{Ap}{}^{q}V_{Bq}=0\,,\\
\cD_{A}\brV_{A\brp}=\partial_{A}\brV_{B\brp}+\Gamma_{AB}{}^{C}\brV_{C\brp}+\brPhi_{A\brp}{}^{\brq}\brV_{B\brq}=0\,,
\ea
\ee
and, from (\ref{acompd}),  also with  the dilaton,
\be
\cD_{A}d=\na_{A}d=0\,.
\ee
The connections are then related to each other by
\be
\ba{ll}
\Phi_{Apq}=\Phi_{A[pq]}=V^{B}{}_{p}\na_{A}V_{Bq}\,,\quad&\quad
\brPhi_{A\brp\brq}=\brPhi_{A[\brp\brq]}=\brV^{B}{}_{\brp}\na_{A}\brV_{B\brq}\,,
\ea
\label{PhibrPhi}
\ee
and
\be
\ba{ll}
\Gamma_{ABC}&=V_{B}{}^{p}(\partial_{A}V_{Cp}+\Phi_{Ap}{}^{q}V_{Cq})+
\brV_{B}{}^{\brp}(\partial_{A}\brV_{C\brp}+\brPhi_{A\brp}{}^{\brq}\brV_{C\brq})\\
{}&=V_{B}{}^{p}\partial_{A}V_{Cp}+\brV_{B}{}^{\brp}\partial_{A}\brV_{C\brp}+\Phi_{ABC}+\brPhi_{ABC}\,.
\ea
\ee
Consequently,  their generic infinitesimal variations satisfy
\be
\ba{ll}
\delta\Phi_{Apq}=\cD_{A}(V^{B}{}_{p}\delta V_{Bq})+V^{B}{}_{p}V^{C}{}_{q}\delta\Gamma_{ABC}\,,~~~&~~~
\delta\brPhi_{A\brp\brq}=\cD_{A}(\brV^{B}{}_{\brp}\delta \brV_{B\brq})+\brV^{B}{}_{\brp}\brV^{C}{}_{\brq}\delta\Gamma_{ABC}\,.
\ea
\label{useful3}
\ee

The master semi-covariant derivative is also compatible with all the constant metrics and the gamma matrices in Table~\ref{TABindices},
\be
\ba{lllll}
\cD_{A}\cJ_{BC}=0\,,\quad&~~~\cD_{A}\eta_{pq}=0\,,\quad&~~~
\cD_{A}\breta_{\brp\brq}=0\,,\quad&~~~\cD_{A}(\gamma^{p})^{\alpha}{}_{\beta}=0\,,\quad&~~~
\cD_{A}(\brgamma^{\brp})^{\bralpha}{}_{\brbeta}=0\,.
\ea
\ee
The well known relation between  the spinorial and the vectorial representations of the spin connections follows
\be
\ba{ll}
\Phi_{A}{}^{\alpha}{}_{\beta}=\quarter\Phi_{Apq}(\gamma^{pq})^{\alpha}{}_{\beta}\,,\quad&\quad 
\brPhi_{A}{}^{\bralpha}{}_{\brbeta}=\quarter\brPhi_{A\brp\brq}(\brgamma^{\brp\brq})^{\bralpha}{}_{\brbeta}\,.
\ea
\label{Phisv}
\ee

\item\textit{{\textbf{Semi-covariant four-index curvatures.}}} The usual  ``field strengths'' of the three connections, 
\be
\ba{l}
R_{CDAB}=\partial_{A}\Gamma_{BCD}-\partial_{B}\Gamma_{ACD}+\Gamma_{AC}{}^{E}\Gamma_{BED}-\Gamma_{BC}{}^{E}\Gamma_{AED}\,,\\
F_{ABpq}=\partial_{A}\Phi_{Bpq}-\partial_{B}\Phi_{Apq}+\Phi_{Apr}\Phi_{B}{}^{r}{}_{q}-\Phi_{Bpr}\Phi_{A}{}^{r}{}_{q}\,,\\
\brF_{AB\brp\brq}=\partial_{A}\brPhi_{B\brp\brq}-\partial_{B}\brPhi_{A\brp\brq}+\brPhi_{A\brp\brr}\brPhi_{B}{}^{\brr}{}_{\brq}-\brPhi_{B\brp\brr}\brPhi_{A}{}^{\brr}{}_{\brq}\,,
\ea
\label{threeF}
\ee 
are, from   $\,[\cD_{A},\cD_{B}]V_{Cp}=0\,$  and   $\,[\cD_{A},\cD_{B}]\brV_{C\brp}=0$,~    related to each other by 
\be
R_{ABCD}=F_{CDpq}V_{A}{}^{p}V_{B}{}^{q}+\brF_{CD\brp\brq}\brV_{A}{}^{\brp}\brV_{B}{}^{\brq}=F_{CDAB}+\brF_{CDAB}\,.
\label{RFF}
\ee
This implies
\be
\ba{ll}
R_{ABCD}=R_{[AB][CD]}\,,\quad&\quad R_{p\brq CD}= V^{A}{}_{p}\brV^{B}{}_{\brq}R_{ABCD}=0\,.
\ea
\label{Rprop}
\ee
Following \cite{Jeon:2011kp}, replacing the ordinary or the naked derivatives in (\ref{threeF}) by the semi-covariant derivatives we define 
\be
\ba{l}
\sF_{ABpq}:=\na_{A}\Phi_{Bpq}-\na_{B}\Phi_{Apq}+\Phi_{Ap}{}^{r}\Phi_{Brq}-\Phi_{Bp}{}^{r}\Phi_{Arq}\,,\\
\sbrF_{AB\brp\brq}
:=\na_{A}\brPhi_{B \brp\brq}-\na_{B}\brPhi_{A \brp\brq}+\brPhi_{A \brp}{}^{\brr}\brPhi_{B \brr\brq}-\brPhi_{B \brp}{}^{\brr}\brPhi_{A \brr\brq}\,,
\ea
\label{FABpq}
\ee
which are, with  the torsion-free  condition~(\ref{aGABC}), related to (\ref{threeF}) by
\be
\ba{ll}
\sF_{AB pq}=F_{ABpq}-\Gamma^{C}{}_{AB}\Phi_{Cpq}\,,\quad&\quad
\sbrF_{AB \brp\brq}
=\brF_{AB \brp\brq}-\Gamma^{C}{}_{AB}\brPhi_{C\brp\brq}\,,
\ea
\ee
and appear in the commutators of the master semi-covariant derivatives,
\be
\ba{ll}
{}[\cD_{A},\cD_{B}]T_{p}=\cF_{ABpq}T^{q}-\Gamma^{C}{}_{AB}\partial_{C}T_{p}\,,\quad&\quad
[\cD_{A},\cD_{B}]T_{\brp}=\brcF_{AB\brp\brq}T^{\brq}-\Gamma^{C}{}_{AB}\partial_{C}T_{\brp}\,.
\ea
\label{comcF}
\ee
Further, they can be rewritten in terms of the master semi-covariant derivatives, then to carry some  opposite signs in comparison to (\ref{FABpq}), 
\be
\ba{l}
\sF_{AB pq}=\cD_{A}\Phi_{Bpq}-\cD_{B}\Phi_{Apq}-\Phi_{Ap}{}^{r}\Phi_{Brq}+\Phi_{Bp}{}^{r}\Phi_{Arq}\,,\\
\sbrF_{AB \brp\brq}
=\cD_{A}\brPhi_{B \brp\brq}-\cD_{B}\brPhi_{A \brp\brq}-\brPhi_{A \brp}{}^{\brr}\brPhi_{B \brr\brq}+\brPhi_{B \brp}{}^{\brr}\brPhi_{A \brr\brq}\,.
\ea
\ee
Hence,  
contracted with the vielbeins -- which are compatible with $\cD_{A}$ but not with $\na_{A}$ --  we may write
\be
\ba{l}
\sF_{ABCD}=\sF_{ABpq}V_{C}{}^{p}V_{D}{}^{q}=\na_{A}\Phi_{BCD}-\na_{B}\Phi_{ACD}-\Phi_{AC}{}^{E}\Phi_{BED}+\Phi_{BC}{}^{E}\Phi_{AED}\,,\\
\sbrF_{ABCD}=\sbrF_{AB\brp\brq}\brV_{C}{}^{\brp}\brV_{D}{}^{\brq}=\na_{A}\brPhi_{BCD}-\na_{B}\brPhi_{ACD}-\brPhi_{AC}{}^{E}\brPhi_{BED}+\brPhi_{BC}{}^{E}\brPhi_{AED}\,.
\ea
\label{FABCD}
\ee

Now we are ready to define two kinds of {semi-covariant four-index curvatures}:
\begin{itemize}
\item \textit{Semi-covariant four-index curvature of the spin connections}, \textit{c.f.~}\cite{Geissbuhler:2013uka}, 
\be
\cG_{ABCD}:=\half\left[(\sF+\sbrF)_{ABCD}
+(\sF+\sbrF)_{CDAB}+
(\Phi+\brPhi)^{E}{}_{AB}(\Phi+\brPhi)_{ECD}\right]\,.
\label{defcG}
\ee
\item \textit{Semi-covariant  Riemann curvature of the diffeomorphic connection}~\cite{Jeon:2010rw,Jeon:2011cn},
\be
S_{ABCD}:=\half\left(R_{ABCD}+R_{CDAB}-\Gamma^{E}{}_{AB}\Gamma_{ECD}\right)\,.
\label{asemicovS}
\ee
\end{itemize}
These  two four-index curvatures  are closely   related to each other, 
\be
\ba{ll}
\cG_{ABCD}&=S_{ABCD}+\half(\Gamma-\Phi-\brPhi)_{EAB}(\Gamma-\Phi-\brPhi)^{E}{}_{CD}\\
{}&=S_{ABCD}+
\half(V_{A}{}^{p}\partial_{E}V_{B p}
+\brV_{A}{}^{\brp}\partial_{E}\brV_{B\brp})(V_{C}{}^{q}\partial^{E}V_{Dq}
+\brV_{C}{}^{\brq}\partial^{E}\brV_{D\brq})\,,
\ea
\label{relationcGS}
\ee
such that  upon the section condition we have
\be
\cG_{ABCD}\sim S_{ABCD}\,.
\label{cGSsim}
\ee
As a bonus, this  implies that,  up to the section condition $\cG_{ABCD}$ is local Lorentz invariant as $S_{ABCD}$ is so. Note that while $F_{ABpq}$ and $\brF_{AB\brp\brq}$ are local Lorentz covariant, $\cF_{ABpq}$ and  $\brcF_{AB\brp\brq}$ are  not.

A notable difference between $\cG_{ABCD}$ and $S_{ABCD}$ is that  while the latter  can be expressed   in terms of the dilaton and the projectors, the former  cannot be defined thoroughly    by them: it requires the vielbeins.  In the following section, we shall see that it is $\cG_{ABCD}$ rather than  $S_{ABCD}$ that survives to serve as the semi-covariant curvature  after the twist.

It is  worth while to note that, in the expressions of $\Phi_{Apq}$, $\brPhi_{A\brp\brq}$ (\ref{PhibrPhi}), $\sF_{ABpq}$, $\sbrF_{AB\brp\brq}$ (\ref{FABpq}) and $\cG_{ABCD}$ (\ref{defcG}), the ordinary naked derivative and the $\Gamma$-connection are completely `confined' into the semi-covariant derivative. On the other hand, it is not the case with   $R_{ABCD}$, $F_{ABpq}$, $\brF_{AB\brp\brq}$ and $S_{ABCD}$.

A crucial defining property of the semi-covariant  Riemann curvature is that, under arbitrary transformation of the   connection,  it   transforms as 
\be
\delta S_{ABCD}=\na_{[A}\delta\Gamma_{B]CD}+\na_{[C}\delta\Gamma_{D]AB}
-\textstyle{\frac{3}{2}}\Gamma_{[ABE]}\delta\Gamma^{E}{}_{CD}
-\textstyle{\frac{3}{2}}\Gamma_{[CDE]}\delta\Gamma^{E}{}_{AB}\,.
\label{Svar}
\ee
Surely for the torsion-free connection~(\ref{Gammao}), the last two terms are  absent and  only the first two total derivative terms remain,
\be
\delta S_{ABCD}=\na_{[A}\delta\Gamma_{B]CD}+\na_{[C}\delta\Gamma_{D]AB}\,.
\label{Svar0}
\ee
Yet,  in the full order supersymmetric extensions of DFT~\cite{Jeon:2011sq,Jeon:2012hp}, the connection includes  
bi-fermionic  torsions and  the above general relation~(\ref{Svar}) enables  the `1.5 formalism' to work.

Without necessity of the section condition, $S_{ABCD}$  satisfies~\cite{Jeon:2010rw},
\be
S_{ABCD}=S_{[AB][CD]}=S_{CDAB}\,,
\ee
and, especially for the torsionless connection, a Bianchi identity,\footnote{See Eq.(2.46) of \cite{Jeon:2010rw} for a simple proof of the Bianchi identity.} 
\be
S_{A[BCD]}=0\,.
\label{BianchiS}
\ee
Further,  for    the torsionless connection~(\ref{Gammao}),  one can show by a brute-force method,\footnote{To obtain (\ref{vanishingS}),  we have used the computer algebra, \textit{Cadabra}~\cite{Peeters:2006kp,Peeters:2007wn}. 
 } 
\be
\ba{l}
(P^{AB}P^{CD}+\brP^{AB}\brP^{CD})S_{ACBD}
=4\partial_{A}\partial^{A}d-4\partial_{A}d\partial^{A}d
+\half\partial_{A}P_{CD}\partial^{A}P^{CD}\,\sim\,0\,,\\
P_{I}{}^{A}P_{J}{}^{B}\brP_{K}{}^{C}\brP_{L}{}^{D}S_{ABCD}=
\half(P\partial_{A}P\brP)_{IL}(P\partial^{A}P\brP)_{JK}
-\half(P\partial_{A}P\brP)_{IK}(P\partial^{A}P\brP)_{JL}\,\sim\,0\,,
\\
P_{I}{}^{A}\brP_{J}{}^{B}P_{K}{}^{C}\brP_{L}{}^{D}S_{ABCD}=-\half(P\partial_{A}P\brP)_{IJ}(P\partial^{A}P\brP)_{KL}\,\sim\,0\,,\\
P_{I}{}^{A}\brP_{J}{}^{B}(P-\brP)^{CD}S_{ACBD}=-\half P_{I}{}^{A}\brP_{J}{}^{B}\partial_{C}\partial^{C}P_{AB}
+(P\partial_{C}P\brP)_{IJ}\partial^{C}d\,\sim\,0\,,
\ea
\label{vanishingS}
\ee
of which the right hand sides all vanish upon the  section condition, $\partial_{A}\partial^{A}\sim0$.

It follows, from (\ref{cGSsim}), that identical relations hold for $\cG_{ABCD}$, either by the strict equality or up to the section condition, for example,
\be
\ba{ll}
\cG_{ABCD}=\cG_{[AB][CD]}=\cG_{CDAB}\,,\quad&\quad
\cG_{A[BCD]}\sim 0\,.
\ea
\label{BianchicG}
\ee

%%%%%%%%%%%%%%%%%%%%%%%%%%%%%%%%%%%%%%%%%%%%%%%%

\item \textit{{\textbf{Complete  covariantizations.}}}
The ordinary derivative of a covariant  tensor is no longer covariant under diffeomorphisms. The   difference between its actual diffeomorphic  transformation  and the  generalized Lie derivative  reads precisely, 
\be
\ba{l}
{}(\delta_{X}-\hcL_{X})\partial_{C}T_{A_{1}\cdots A_{n}}
=\left[\partial_{C}\,,\,\hcL_{X}\right]T_{A_{1}\cdots A_{n}}\\
=\partial^{B}X_{C}\partial_{B}T_{A_{1}\cdots A_{n}}+
\omega_{{\scriptscriptstyle{T\,}}}\partial_{C}\partial_{B}X^{B}
T_{A_{1}\cdots A_{n}}+
\dis{\sum_{i=1}^{n}\,2\partial_{C}\partial_{[A_{i}}X_{B]}
T_{A_{1}\cdots A_{i-1}}{}^{B}{}_{A_{i+1}\cdots A_{n}}\,.}
\ea
\label{anomalpartial}
\ee
Especially for the connection we have  
\be
\ba{ll}
(\delta_{X}{-\hcL_{X}})\Gamma_{CAB}=& 2\big[(\cP+\brcP)_{CAB}{}^{FDE}-\delta_{C}^{~F}\delta_{A}^{~D}
\delta_{B}^{~E}\big]\partial_{F}\partial_{[D}X_{E]}\\
{}&+2(\cPp-\cPq)_{CAB}{}^{FDE}\partial^{G}P_{FD}\,\partial_{G}X_{E}
+2P_{[A}{}^{D}\brP_{B]}{}^{E}\partial^{G}P_{DE}\,\partial_{G}X_{C}\\
{}&+\textstyle{\frac{2}{D-1}}(P_{C[A}P_{B]}{}^{E}+
\brP_{C[A}\brP_{B]}{}^{E})(\partial_{G}\partial^{G}X_{E}-2\partial^{G}d\,\partial_{G}X_{E})\,.
\ea
\label{dhcLG}
\ee
It follows that
\be
(\delta_{X}{-\hcL_{X}})\Gamma^{A}{}_{AB}=
-2\partial^{C}d\partial_{C}X_{B}+\partial_{B}
\partial_{C}X^{C}=-2(\delta_{X}{-\hcL_{X}})\partial_{B}d\,.
\ee
Further, using 
\be
\ba{ll}
{}\left[\na_{C}\,,\,\hcL_{X}\right]T_{A_{1}\cdots A_{n}}=&\partial^{B}X_{C}\partial_{B}T_{A_{1}\cdots A_{n}}
+\omega_{{\scriptscriptstyle{T\,}}}(\partial_{C}\partial_{B}X^{B}+\hcL_{X}\Gamma^{B}{}_{BC})
T_{A_{1}\cdots A_{n}}\\
{}&+
\dis{\sum_{i=1}^{n}\,\left(2\partial_{C}\partial_{[A_{i}}X_{B]}
-\hcL_{X}\Gamma_{CA_{i}B}\right)
T_{A_{1}\cdots A_{i-1}}{}^{B}{}_{A_{i+1}\cdots A_{n}}\,,}
\ea
\ee
we may obtain an exact expression of the diffeomorphic   anomaly  of the semi-covariant derivative,
\be
\ba{ll}
(\delta_{X}-\hcL_{X})(\na_{C}T_{A_{1}\cdots A_{n}})=&\partial^{B}X_{C}\partial_{B}T_{A_{1}\cdots A_{n}}
+\omega_{{\scriptscriptstyle{T\,}}}\left[\partial_{C}\partial_{B}X^{B}-(\delta_{X}-\hcL_{X})\Gamma^{B}{}_{BC}\right]
T_{A_{1}\cdots A_{n}}\\
{}&+
\dis{\sum_{i=1}^{n}\,\left[2\partial_{C}\partial_{[A_{i}}X_{B]}+(\delta_{X}-\hcL_{X})\Gamma_{CA_{i}B}\right]
T_{A_{1}\cdots A_{i-1}}{}^{B}{}_{A_{i+1}\cdots A_{n}}\,,}
\ea
\label{dhcLnab}
\ee
into which  (\ref{dhcLG}) can be readily substituted.

Lastly for the  semi-covariant Riemannian  curvature of the torsionless connection, from 
\be
\hcL_{X}S_{ABCD}=\na_{[A}\hcL_{X}\Gamma_{B]CD}
-2\na_{[A}\partial_{B]}\partial_{[C}X_{D]}
-\partial_{E}X_{[A}\partial^{E}\Gamma_{B]CD}
~+~\Big[\,(A,B)\,\leftrightarrow\,(C,D)\,\Big]\,,
\ee
we get an exact  formula,
\be
\!\!(\delta_{X}-\hcL_{X})S_{ABCD}=\na_{[A}(\delta_{X}-\hcL_{X})\Gamma_{B]CD}
+2\na_{[A}\partial_{B]}\partial_{[C}X_{D]}
+\partial_{E}X_{[A}\partial^{E}\Gamma_{B]CD}
~+\Big[(A,B)\leftrightarrow(C,D)\Big]\,.
\label{dhcLS}
\ee

Now, we consider imposing  the section condition, $\partial_{A}\partial^{A}\sim0$. Clearly, from    (\ref{dhcLG}), (\ref{dhcLnab}), (\ref{dhcLS}) and (\ref{relationcGS}), we note
\begin{eqnarray}
&&(\delta_{X}{-\hcL_{X}})\Gamma_{CAB}\sim 2\big[(\cP+\brcP)_{CAB}{}^{FDE}-\delta_{C}^{~F}\delta_{A}^{~D}
\delta_{B}^{~E}\big]\partial_{F}\partial_{[D}X_{E]}\,,\\
&&(\delta_{X}{-\hcL_{X}})\na_{C}T_{A_{1}\cdots A_{n}}\sim
\dis{\sum_{i=1}^{n}2(\cP{+\brcP})_{CA_{i}}{}^{BDEF}
\partial_{D}\partial_{E}X_{F}\,T_{A_{1}\cdots A_{i-1} BA_{i+1}\cdots A_{n}}\,,}
\label{diffeoanormalous}
\end{eqnarray}
and for the four-index curvatures, 
\be
\ba{ll}
(\delta_{X}-\hcL_{X})\cG_{ABCD}
&\sim(\delta_{X}-\hcL_{X})S_{ABCD}\\
&\sim 2\na_{[A}\Big(\!(\cP{+\brcP})_{B][CD]}{}^{EFG}\partial_{E}\partial_{F}X_{G}\Big)
+\Big[(A,B)\leftrightarrow(C,D)\Big]\,.
\ea
\label{anomalcurv}
\ee
Thus,  upon the section condition, it is the six-index projection operators that dictate    the  anomalies    in the  diffeomorphic transformations    of   the semi-covariant derivative and the semi-covariant  curvatures. This also explains or motivates  the naming,  `\textit{semi-covariant}': we say a tensor is semi-covariant if its diffeomorphic anomaly, if any,  is governed  by the six-index projectors.

The  anomalous terms can be easily projected out through appropriate contractions with the two-index projectors. In this manner, the \textit{completely covariant derivatives}  are given by
\be
\ba{ll}
P_{C}{}^{D}{\brP}_{A_{1}}{}^{B_{1}}\cdots{\brP}_{A_{n}}{}^{B_{n}}
\DO_{D}T_{B_{1}\cdots B_{n}}\,,~&~
{\brP}_{C}{}^{D}P_{A_{1}}{}^{B_{1}}\cdots P_{A_{n}}{}^{B_{n}}
\DO_{D}T_{B_{1}\cdots B_{n}}\,,\\
P^{AB}{\brP}_{C_{1}}{}^{D_{1}}\cdots{\brP}_{C_{n}}{}^{D_{n}}\DO_{A}T_{BD_{1}\cdots D_{n}}\,,~&~
\brP^{AB}{P}_{C_{1}}{}^{D_{1}}\cdots{P}_{C_{n}}{}^{D_{n}}\DO_{A}T_{BD_{1}\cdots D_{n}}\quad~~(\mbox{divergences})\,,\\
P^{AB}{\brP}_{C_{1}}{}^{D_{1}}\cdots{\brP}_{C_{n}}{}^{D_{n}}
\DO_{A}\DO_{B}T_{D_{1}\cdots D_{n}}\,,~&~
{\brP}^{AB}P_{C_{1}}{}^{D_{1}}\cdots P_{C_{n}}{}^{D_{n}}
\DO_{A}\DO_{B}T_{D_{1}\cdots D_{n}}\quad(\mbox{Laplacians})\,.
\ea
\label{covT}
\ee
These can be also freely pull-backed by the vielbeins to take the form:
\be
\ba{llllll}
\cD_{p}T_{\brq_{1}\cdots \brq_{n}}\,,\quad&
\cD_{\brp}T_{q_{1}\cdots q_{n}}\,,\quad&
\cD_{p}T^{p}{}_{\brq_{1}\cdots\brq_{n}}\,,\quad&
\cD_{\brp}T^{\brp}{}_{q_{1}\cdots q_{n}}\,,\quad&
\cD_{p}\cD^{p}T_{\brq_{1}\cdots \brq_{n}}\,,\quad&
\cD_{\brp}\cD^{\brp}T_{q_{1}\cdots q_{n}}\,.
\ea
\label{covT2}
\ee
Similarly we obtain  completely covariant two-index as well as zero-index curvatures from the  semi-covariant four-index curvatures.
\begin{itemize}
\item \textit{Completely covariant  ``Ricci'' curvatures},\footnote{The expression~(\ref{aRicciScalar1}) is  for the   torsionless connection. For  torsionful extension, see \cite{Jeon:2011sq,Jeon:2012hp}  and especially  (A.71) of \cite{Jeon:2012kd}.}
\be
\ba{l}
\cG_{pr\brq}{}^{r}=\half\cF_{\brq r p}{}^{r}=S_{pr\brq}{}^{r}
+\half P^{AB}\partial_{E}V_{Ap}\partial^{E}\brV_{B\brq}\,,\\
\cG_{p\brr\brq}{}^{\brr}
=\half\brcF_{p\brr\brq}{}^{\brr}
=S_{p\brr\brq}{}^{\brr}
+\half\brP^{AB}\partial_{E}V_{Ap}\partial^{E}\brV_{B\brq}\,,
\ea
\label{aRicciScalar1}
\ee
whose sum gives~\cite{Jeon:2012kd}
\be
\ba{lll}
\cG_{p\brq}
=S_{p\brq}+\half \partial_{A}V_{Bp}\partial^{A}\brV^{B}{}_{\brq}
&\sim&S_{p\brq}\,.
\ea
\ee

\item \textit{Completely covariant scalar curvatures},
\be
\ba{lll}
\cG_{pq}{}^{pq}=\sF_{pq}{}^{pq}+\half\Phi_{Epq}\Phi^{Epq}=P^{AC}P^{BD}S_{ABCD}
+\half P^{AB}\partial_{E}V_{Ap}\,\partial^{E}V_{B}{}^{p}&\sim&S_{pq}{}^{pq}\,,\\
\cG_{\brp\brq}{}^{\brp\brq}=\sbrF_{\brp\brq}{}^{\brp\brq}+\half\brPhi_{E\brp\brq}\brPhi^{E\brp\brq}
=\brP^{AC}\brP^{BD}S_{ABCD}+\half \brP^{AB}\partial_{E}\brV_{A\brp}\,\partial^{E}\brV_{B}{}^{\brp}&\sim&S_{\brp\brq}{}^{\brp\brq}\,.
\ea
\label{aRicciScalar2}
\ee
\end{itemize}
In fact,  the two ``Ricci'' curvatures agree to each other upon the section condition:
From the identity~(\ref{vanishingS}),  their difference reads exactly,
\be
\cG_{pr\brq}{}^{r}-\cG_{p\brr\brq}{}^{\brr}=
V^{A}{}_{p}\brV^{B}{}_{\brq}(\partial_{E}P_{AB}\partial^{E}d-\half \partial_{E}\partial^{E}P_{AB})+\half(P-\brP)^{AB}
\partial_{E}V_{Ap}\partial^{E}\brV_{B\brq}\,,
\ee
and hence,
\be
\ba{lllll}
\cG_{pr\brq}{}^{r}&\sim&\cG_{p\brr\brq}{}^{\brr}&\sim&
\half\cG_{p\brq}\,.
\ea
\ee
Similarly the two scalar curvatures are related to each other: From (\ref{vanishingS}), their sum reads exactly, 
\be
\ba{ll}
\cG_{pq}{}^{pq}+\cG_{\brp\brq}{}^{\brp\brq}&=(P^{AC}P^{BD}+\brP^{AC}\brP^{BD})S_{ABCD}
	+\half(P^{CD}\partial_{A}
V_{C p}\partial^{A}V_{D}{}^{p}+\brP^{CD}\partial_{A}
\brV_{C\brp}\partial^{A}\brV_{D}{}^{\brp})\\
{}&=4\partial_{A}\partial^{A}d-4\partial_{A}d\partial^{A}d
+\half(\partial_{A}V_{Bp}\partial^{A}V^{Bp}+\partial_{A}
\brV_{B\brp}\partial^{A}\brV^{B\brp})\,,
\ea
\label{trivialid}
\ee
and hence, upon the section condition,
\be
\cG_{pq}{}^{pq}+\cG_{\brp\brq}{}^{\brp\brq}~\sim~0\,.
\label{cGpluscG}
\ee
The other   formulas in (\ref{vanishingS}) also imply  
 a pair of `trivial'   four-index  covariant quantities, 
\be
\ba{ll}
\cG_{pq\brr\brs}=\cG_{ABCD}V^{A}{}_{p}V^{B}{}_{q}\brV^{C}{}_{\brr}\brV^{D}{}_{\brs}\sim0\,,\quad&\quad
\cG_{p\brq r\brs}=\cG_{ABCD}V^{A}{}_{p}\brV^{B}{}_{\brq}V^{C}{}_{r}\brV^{D}{}_{\brs}\sim0\,.
\ea
\label{cGo22}
\ee

\end{itemize}

%%%%%%%%%%%%%%%%%%%%%%%%%%%%%%%%%%%
\subsection{Fermions, Ramond-Ramond cohomology  and completely covariant Dirac operators\label{SECRR}}

\begin{itemize}

\item  \textit{{\textbf{Fermions and Ramond-Ramond cohomology .}}}  In addition to the  NS-NS sector composed of the dilaton and the pair of vielbeins, the $\cN=2$ $D=10$ supersymmetric extension of DFT~\cite{Jeon:2012hp} calls for a Ramond-Ramond potential, a pair of dilatinos and a pair of gravitinos: The fundamental fields of the supersymmetric  theory are precisely,
\be
\ba{lllllll}
d\,,\quad&V_{Ap}\,,\quad&\brV_{A\brp}\,,\quad&\cC^{\alpha}{}_{\bralpha}\,,\quad&\rho^{\alpha}\,,\quad\rhop^{\bralpha}\,,\quad&\psi^{\alpha}_{\brp}\,,\quad&\psip_{p}{}^{\bralpha}\,.
\ea
\ee
The whole R-R sector is represented by a single potential, $\cC^{\alpha}{}_{\bralpha}$. As its indices indicate (\textit{c.f.~}Table~\ref{TABindices}), it assumes  the  bi-fundamental  spinorial   representation of $\Spint\times\oSpint$~\cite{Coimbra:2011nw,Coimbra:2012yy,Jeon:2012kd}.

All the fermions, \textit{i.e.~}dilatinos, gravitinos and supersymmetry parameters,  are    not twenty, but ten-dimensional Majorana-Weyl spinors. The chirality of the theory reads  with two arbitrary sign factors, $\sign$, $\signp$    ($\sign^{2}=\signp{}^{2}=1$),
\be
\ba{llll}
{\gamma^{\eleven}\psi_{\brp}}=\sign\,\psi_{\brp}\,,\quad &\quad 
\gamma^{\eleven}\rho=-\sign\,\rho\,,\quad&\quad 
\brgamma^{\eleven}\psi^{\prime}_{p}=\signp\psi^{\prime}_{p}\,,\quad&\quad 
\brgamma^{\eleven}\rhop=-\signp\rhop\,,\\
\gamma^{\eleven}\varepsilon=\sign\,\varepsilon\,,\quad&\quad \brgamma^{\eleven}\varepsilonp=\signp\varepsilonp\,,\quad&\quad{\gamma^{\eleven}\cC\brgamma^{\eleven}=\sign\signp\,\cC\,.}&{}
\ea
\label{chirality}
\ee
\textit{A priori}, there are four different sign choices. But, they are all  {equivalent} up to the field redefinitions through  $\Pint\times\oPint$ rotations. That is to say,  $\cN=2$ $D=10$ SDFT is  \textit{chiral} with respect to  \textit{both} $\Spint$ {and} $\oSpint$, and  the theory is unique.  
Without loss of generality, henceforth we  set 
\be
\ba{ll}
\sign=+1\,,\quad&\quad\signp=+1\,.
\ea
\ee 
Although  the theory is unique,  the Riemannian  solutions are twofold and can be identified as type IIA or IIB supergravity backgrounds~\cite{Jeon:2012hp}.   The theory  admits also non-Riemannian backgrounds~\cite{Lee:2013hma} (\textit{c.f.~}math literature~\cite{13044294}).

The R-R field strength, $\cF^{\alpha}{}_{\bralpha}$, and its charge conjugation   are defined by~\cite{Jeon:2012kd}
\be
\ba{ll}
\cF:=\cD_{+}\cC\,,\quad&\quad
\brcF:=\brC^{-1}_{+}(\cF)^{T}C_{+}\,.
\ea
\label{RRFLUX}
\ee
Here  $\cD_{+}$ corresponds to one of the two completely covariant   differential operators, $\cD_{\pm}$, which are defined by the torsionless connection~(\ref{Gammao}) to  act  on an arbitrary 
$\Spint\times\oSpint$ bi-fundamental field, $\cT^{\alpha}{}_{\brbeta}$\,: 
\be
\cD_{\pm}\cT:=\gamma^{p}\cD_{p}\cT\pm\gamma^{\eleven}\cD_{\brp}\cT\brgamma^{\brp}\,,
\label{Dpm}
\ee
where we put  $\cD_{p}=V^{A}{}_{p}\cD_{A}$, $\cD_{\brp}=\brV^{A}{}_{\brp}\cD_{A}$ and with (\ref{Phisv}),  $\cD_{A}\cT=\partial_{A}\cT+\Phi_{A}\cT-\cT\brPhi$.

The crucial property of the differential operators, $\cD_{\pm}$,  is that upon the section condition they are nilpotent~\cite{Jeon:2012kd}.  Straightforward computation can  show
\be
\ba{ll}
(\cD_{\pm})^{2}\cT=&-\quarter(\cG_{pq}{}^{pq}+\cG_{\brp\brq}{}^{\brp\brq})\cT+\partial_{A}\partial^{A}\cT-2\partial_{A}d\partial^{A}\cT\\
{}&+\half V_{Bp}\partial_{A}V^{B}{}_{q}\gamma^{pq}\partial^{A}\cT-\half\brV_{B\brp}\partial_{A}\brV^{B}{}_{\brq}\partial^{A}\cT\brgamma^{\brp\brq}\\
{}&+\quarter(V_{Bp}\partial_{A}\partial^{A}V^{B}{}_{q}-2V_{Bp}\partial_{A}V^{B}{}_{q}\partial^{A}d)\gamma^{pq}\cT\\
{}& -\quarter(\brV_{B\brp}\partial_{A}\partial^{A}\brV^{B}{}_{\brq}-2V_{B\brp}\partial_{A}\brV^{B}{}_{\brq}\partial^{A}d)\cT\brgamma^{\brp\brq}\\
      {}&+\textstyle{\frac{1}{8}}\cG_{pqrs}\gamma^{pqrs}\cT-\quarter\cG_{pq\brr\brs}
      \gamma^{pq}\cT\brgamma^{\brr\brs}+\textstyle{\frac{1}{8}}\cG_{\brp\brq\brr\brs}\cT\brgamma^{\brp\brq\brr\brs}\\
      {}&\pm\quarter\gamma^{\eleven}(
      \cG_{p\brq\brr\brs}\gamma^{p}\cT\brgamma^{\brq\brr\brs}
      -\cG_{pqr\brs}\gamma^{pqr}\cT\brgamma^{\brs})\\
     {}& \pm\quarter\gamma^{\eleven}\left[2(\cG^{\brr}{}_{p\brr\brq}-
      \cG^{r}{}_{pr\brq})\gamma^{p}\cT\brgamma^{\brq}
      -4V_{Bp}\partial_{A}\brV^{B}{}_{\brq}\gamma^{p}\partial^{A}\cT\brgamma^{\brq}\right]\,.
\ea
\label{Dsquare}
\ee
Thus, up to the section condition,  with   (\ref{relationcGS}), (\ref{BianchicG}), (\ref{vanishingS}), (\ref{trivialid}),  each term on the right hand side above   vanishes and    the nilpotency of the differential operators follows
\be
(\cD_{\pm})^{2}\cT\sim0\,.
\label{nilpotent}
\ee
This defines the  R-R cohomology consistently  coupled to the NS-NS sector in an $\ODD$ covariant manner~\cite{Jeon:2012kd}. In particular, the R-R gauge transformations are given by the same nilpotent differential operator, 
\be
\ba{lll}
\delta\cC=\cD_{+}\Lambda&\quad\longrightarrow\quad&
\delta\cF=(\cD_{+})^{2}\Lambda\sim 0\,.
\ea
\ee

In a similar fashion to (\ref{diffeoanormalous}), upon the section condition,   
the spin  connections  transform anomalously  under diffeomorphisms,
\be
\ba{ll}
(\delta_{X}-\hcL_{X})\Phi_{Apq}\sim2\cP_{Apq}{}^{DEF}\partial_{D}\partial_{[E}X_{F]}\,,\quad&\quad
(\delta_{X}-\hcL_{X})\brPhi_{A\brp\brq}\sim2\brcP_{A\brp\brq}{}^{DEF}\partial_{D}\partial_{[E}X_{F]}\,.
\ea
\label{spinanomal}
\ee
Thus, like (\ref{covT}), these anomalous terms can be easily projected out, such that 
 the following  modules of the spin connections are completely covariant under diffeomorphisms, 
\be
\ba{llllll}
\brP_{A}{}^{B}\Phi_{Bpq}\,,~&~P_{A}{}^{B}\brPhi_{B\brp\brq}\,,~&~\Phi_{A[pq}V^{A}{}_{r]}\,,~&~
\brPhi_{A[\brp\brq}\brV^{A}{}_{\brr]}\,,~&~\Phi_{Apq}V^{Ap}\,,~&~
\brPhi_{A\brp\brq}\brV^{A\brp}\,.
\ea
\label{covPhi}
\ee 
The completely covariant Dirac operators are then, with respect to both diffeomorphisms and local Lorentz symmetries, as follows~\cite{Jeon:2011vx,Jeon:2011sq}. 
\be
\ba{lllll}
\gamma^{p}\cD_{p}\rho=\gamma^{A}\cD_{A}\rho\,,~&~
\gamma^{p}\cD_{p}\psi_{\brp}\,,~
&~\cD_{\brp}\rho\,,~&~\cD_{\brp}\psi^{\brp}=\cD_{A}\psi^{A}\,,
~&~\brpsi^{A}\gamma_{p}(\cD_{A}\psi_{\brq}-\half\cD_{\brq}\psi_{A})\,,\\
\brgamma^{\brp}\cD_{\brp}\rhop=\brgamma^{A}\cD_{A}\rhop\,,~&~
\brgamma^{\brp}\cD_{\brp}\psi^{\prime}_{p}\,,~&~
\cD_{p}\rhop\,,~&~\cD_{p}\psip{}^{p}=\cD_{A}\psip{}^{A}\,,~&~
\brpsi^{\prime A}\brgamma_{\brp}(\cD_{A}\psi^{\prime}_{q}-\half\cD_{q}\psi^{\prime}_{A})\,.
\ea
\label{covDirac}
\ee
One can also show that $\cD_{\pm}\cT$ (\ref{Dpm}) are completely covariant too.

\item \textit{{\textbf{Completely covariant curvatures from completely covariant derivatives}}}

From  (\ref{scGp}), (\ref{comcF}) and  the relation,
\be
\cG_{p\brq AB}=\cG_{ABp\brq}=\half(\cF+\brcF)_{p\brq CD}\,,
\ee   
the completely covariant ``Ricci''  curvatures~(\ref{aRicciScalar1}), are related to the commutators of the completely covariant differential operators~(\ref{covT2}),
\be
\ba{lll}
\cG_{pr\brq}{}^{r}T^{p}=\half\cF_{\brq r p}{}^{r}T^{p}=\half[\cD_{p},\cD_{\brq}]T^{p}+\half
\Gamma^{C}{}_{p\brq}\partial_{C}T^{p}&\sim&\half[\cD_{p},\cD_{\brq}]T^{p}\,,\\
\cG_{p\brr\brq}{}^{\brr}T^{\brq}=\half\brcF_{p\brr\brq}{}^{\brr}T^{\brq}=-\half[\cD_{p},\cD_{\brr}]T^{\brr}-\half\Gamma^{C}{}_{p\brq}\partial_{C}T^{\brq}&\sim&-\half[\cD_{p},\cD_{\brq}]T^{\brq}\,.
\ea
\label{vectorREP}
\ee

In a similar fashion  to (\ref{comcF}), we may obtain the expressions for the  commutators of the master semi-covariant differential operators which act  on  spinors,  $\varepsilon^{\alpha}$ and   $\varepsilonp^{\bralpha}$, in   $\SpinD$ and $\oSpinD$ representations respectively, 
\be
\ba{l}  
{}[\cD_{A},\cD_{B}]\varepsilon=\quarter\cF_{ABpq}\gamma^{pq}\varepsilon-\Gamma^{C}{}_{AB}\partial_{C}\varepsilon\,,\\
{}[\cD_{A},\cD_{B}]\varepsilonp=\quarter\brcF_{AB\brp\brq}\brgamma^{\brp\brq}\varepsilonp-\Gamma^{C}{}_{AB}\partial_{C}\varepsilon^{\prime}\,.
\ea
\label{comcD}
\ee
These immediately  imply  
\be
\ba{l}
{}[\gamma^{p}\cD_{p},\cD_{\brq}]\varepsilon=\cG_{pr\brq}{}^{r}\gamma^{p}\varepsilon-\half
\cG_{\brq prs}\gamma^{prs}\varepsilon-\Gamma^{C}{}_{p\brq}\gamma^{p}\partial_{C}\varepsilon\,,\\
{}[\cD_{p},\brgamma^{\brq}\cD_{\brq}]\varepsilon^{\prime}=-\cG_{p\brr\brq}{}^{\brr}\gamma^{\brq}\varepsilonp+\half
\cG_{p\brq\brr\brs}\brgamma^{\brq\brr\brs}\varepsilonp-\Gamma^{C}{}_{p\brq}\brgamma^{\brq}\partial_{C}\varepsilonp\,,
\ea
\label{comforCUR}
\ee
and further give
\be
\ba{l}
{}\gamma^{pq}[\cD_{p},\cD_{q}]\varepsilon=(\quarter\cF_{pqrs}\gamma^{pqrs}-\cF_{prq}{}^{r}\gamma^{pq}-\half \cF_{pq}{}^{pq}) \varepsilon-\Gamma^{A}{}_{pq}\gamma^{pq}\partial_{A}\varepsilon\,,\\
{}\brgamma^{\brp\brq}[\cD_{\brp},\cD_{\brp}]\varepsilon^{\prime}=(\quarter\brcF_{\brp\brq\brr\brs}\brgamma^{\brp\brq\brr\brs}-\brcF_{\brp\brr\brq}{}^{\brr}\gamma^{\brp\brq}-\half \brcF_{\brp\brq}{}^{\brp\brq}) \varepsilon^{\prime}
-\Gamma^{A}{}_{\brp\brq}\brgamma^{\brp\brq}\partial_{A}\varepsilon^{\prime}\,.
\ea
\label{comcFvar}
\ee
Then,  combined with the following relations, 
\be
\ba{rl}
{}\cD_{A}\cD^{A}\varepsilon&=\partial_{A}\partial^{A}\varepsilon-2\partial_{A}d\partial^{A}\varepsilon-\textstyle{\frac{1}{8}}\Phi_{Apq}\Phi^{Apq}\varepsilon+\quarter(\cD_{A}\Phi^{A}{}_{pq})\gamma^{pq}\varepsilon\\
{}&\quad+\half\Phi_{Apq}\gamma^{pq}\partial^{A}\varepsilon+\textstyle{\frac{1}{16}}\Phi_{Apq}\Phi^{A}{}_{rs}\gamma^{pqrs}\varepsilon\,,\\
{}\cD_{A}\cD^{A}\varepsilonp&=\partial_{A}\partial^{A}\varepsilonp-2\partial_{A}d\partial^{A}\varepsilonp-\textstyle{\frac{1}{8}}\brPhi_{A\brp\brq}\brPhi^{A\brp\brq}\varepsilonp+
\quarter(\cD_{A}\brPhi^{A}{}_{\brp\brq})\brgamma^{\brp\brq}\varepsilonp\\
{}&\quad+\half\brPhi_{A\brp\brq}\brgamma^{\brp\brq}\partial^{A}\varepsilonp+\textstyle{\frac{1}{16}}\brPhi_{A\brp\brq}\brPhi^{A}{}_{\brr\brs}\brgamma^{\brp\brq\brr\brs}\varepsilonp\,,\\
\cD_{A}\Phi^{A}{}_{pq}&=-2R_{[p}{}^{A}{}_{q]A}+2\Phi_{A[p}{}^{r}\Gamma^{A}{}_{q]r}+\partial_{A}(V^{B}{}_{p}\partial^{A}V_{Bq})-2\partial_{A}dV^{B}{}_{p}\partial^{A}V_{Bq}\\
{}&=2\cF_{[p}{}^{r}{}_{q]r}+\partial_{A}(V^{B}{}_{p}\partial^{A}V_{Bq})-2\partial_{A}dV^{B}{}_{p}\partial^{A}V_{Bq}\,,\\
\cD_{A}\brPhi^{A}{}_{\brp\brq}&=-2R_{[\brp}{}^{A}{}_{\brq]A}+2\brPhi_{A[\brp}{}^{\brr}\Gamma^{A}{}_{\brq]\brr}+\partial_{A}(\brV^{B}{}_{\brp}\partial^{A}\brV_{B\brq})-2\partial_{A}d\brV^{B}{}_{\brp}\partial^{A}\brV_{B\brq}\\
{}&=2\brcF_{[\brp}{}^{\brr}{}_{\brq]\brr}+\partial_{A}(\brV^{B}{}_{\brp}\partial^{A}\brV_{B\brq})-2\partial_{A}d\brV^{B}{}_{\brp}\partial^{A}\brV_{B\brq}\,,%%\\
%%\cF_{[p}{}^{r}{}_{q]r}=&-R_{[p}{}^{r}{}_{q]r}+\Phi_{A[p}{}^{r}\Gamma^{A}{}_{q]r}\,,\\
%%\brcF_{[\brp}{}^{\brr}{}_{\brq]\brr}=&-R_{[\brp}{}^{\brr}{}_{\brq]\brr}+\brPhi_{A[\brp}{}^{\brr}\Gamma^{A}{}_{\brq]\brr}\,,
%%%
\ea
\ee
we can derive the following identities, 
\be
\ba{ll}
\left[(\gamma^{p}\cD_{p})^{2}
+\cD_{\brp}\cD^{\brp}\right]\varepsilon=&-\quarter\cG_{pq}{}^{pq}\varepsilon+\textstyle{\frac{1}{8}} \cG_{pqrs}\gamma^{pqrs}\varepsilon
+\partial_{A}\partial^{A}\varepsilon-2\partial_{A}d\partial^{A}\varepsilon\\
{}&\!\!+\quarter\left[\partial_{A}(V^{B}{}_{p}\partial^{A}V_{Bq})-2\partial_{A}dV^{B}{}_{p}\partial^{A}V_{Bq}\right]\gamma^{pq}\varepsilon
+\half V^{B}{}_{p}\partial_{A}V_{Bq}\gamma^{pq}\partial^{A}\varepsilon\,,\\
\left[(\brgamma^{\brp}\cD_{\brp})^{2}
+\cD_{p}\cD^{p}\right]\varepsilonp=&-\quarter\cG_{\brp\brq}{}^{\brp\brq}\varepsilonp+\textstyle{\frac{1}{8}} \cG_{\brp\brq\brr\brs}\brgamma^{\brp\brq\brr\brs}\varepsilonp
+\partial_{A}\partial^{A}\varepsilonp-2\partial_{A}d\partial^{A}\varepsilonp\\
{}&\!\!+\quarter\left[\partial_{A}(\brV^{B}{}_{\brp}\partial^{A}\brV_{B\brq})-2\partial_{A}d\brV^{B}{}_{\brp}\partial^{A}\brV_{B\brq}\right]\brgamma^{\brp\brq}\varepsilonp
+\half\brV^{B}{}_{\brp}\partial_{A}\brV_{B\brq}\brgamma^{\brp\brq}\partial^{A}\varepsilonp\,.
\ea
\label{Dirac2}
\ee
Therefore, upon the section condition  the completely covariant   ```Ricci"' and scalar curvatures~(\ref{aRicciScalar1}), (\ref{aRicciScalar2}) are   related to the  completely covariant Dirac operators~(\ref{covDirac}), \textit{c.f.~Generalized Geometry}~\cite{Coimbra:2011nw},
\be
\ba{llllll}
{}[\gamma^{p}\cD_{p},\cD_{\brq}]\varepsilon&\sim&\cG_{pr\brq}{}^{r}\gamma^{p}\varepsilon\,,\quad&\quad
{}[\cD_{p},\brgamma^{\brq}\cD_{\brq}]\varepsilon^{\prime}&\sim&=-\cG_{p\brr\brq}{}^{\brr}\gamma^{\brq}\varepsilonp\,,
\ea
\label{GG1}
\ee
and
\be
\ba{llllll}
(\gamma^{p}\cD_{p})^{2}\varepsilon
+\cD_{\brp}\cD^{\brp}\varepsilon&\sim&-\quarter\cG_{pq}{}^{pq}\varepsilon\,,\quad&\quad
(\brgamma^{\brp}\cD_{\brp})^{2}\varepsilonp
+\cD_{p}\cD^{p}\varepsilonp&\sim&-\quarter\cG_{\brp\brq}{}^{\brp\brq}\varepsilonp\,.
\ea
\label{GG2}
\ee
While the completely covariant ``Ricci" curvatures can be identified from both  vectorial  and spinorial commutators, (\ref{vectorREP}) and  (\ref{GG1}),  it appears that  the completely covariant  scalar curvatures can be only identified in the spinorial representation~(\ref{GG2}).

\end{itemize}

%%%%%%%%%%%%%%%%%%%%%%%%%%%%%%%%%%%
\subsection{DFT action and  supersymmetric extensions \label{SECSUSYextensions}}

\begin{itemize}

\item  \textit{{\textbf{Pure DFT action.}}} The bosonic action of  the untwisted   DFT for the NS-NS sector, or the \textit{pure DFT},     is  given by the fully covariant scalar curvature, 
\be
\dis{\int_{\Sigma_{D}}}e^{-2d}\,(P^{AC}P^{BD}-\brP^{AC}\brP^{BD})S_{ABCD}\,,
\label{aactionR}
\ee
where the integral is taken over a section, $\Sigma_{D}$. The dilaton and the projector equations of motion correspond to the vanishing of the scalar curvature \textit{i.e.~}the Lagrangian itself and the ``Ricci"  curvature, $S_{p\brq}$, respectively.

It is  precisely this expression of (\ref{aactionR}) that ensures the 	`1.5 formalism' in the full order   supersymmetric  extensions of DFT with   torsionful connections~\cite{Jeon:2011sq,Jeon:2012hp}.  In fact, 
without imposing the section condition, the scalar curvature in the Lagrangian~(\ref{aactionR})  precisely agrees with the original DFT Lagrangian~(\cite{Hohm:2010pp}) written  in terms of the generalized metric, $\cH=P-\brP$,
\be
\ba{ll}
S_{pq}{}^{pq}-S_{\brp\brq}{}^{\brp\brq}=&
\textstyle{\frac{1}{8}}\cH^{AB}\partial_{A}\cH_{CD}\partial_{B}\cH^{CD}+\half\cH^{AB}\partial^{C}\cH_{AD}\partial^{D}\cH_{BC}-\partial_{A}\partial_{B}\cH^{AB}\\
{}&-4\cH^{AB}\partial_{A}d\partial_{B}d+
4\cH^{AB}\partial_{A}\partial_{B}d+
4\partial_{A}\cH^{AB}\partial_{B}d\,.
\ea
\label{SgH}
\ee
However, due to the  relations~(\ref{cGSsim}), (\ref{vanishingS}) which hold for the torsionless connection~(\ref{Gammao}),  there exist  alternative  section-condition-equivalent expressions  for the action, \textit{e.g.~}$(P-\brP)^{AB}S_{AEB}{}^{E}$~\cite{Jeon:2011cn}, or  
replacing $S_{ABCD}$ by  the spinorial curvature,
\be
\ba{lll}
\dis{\int_{\Sigma_{D}}}e^{-2d}\,(\cG_{pq}{}^{pq}-\cG_{\brp\brq}{}^{\brp\brq})&\sim&2\dis{\int_{\Sigma_{D}}}e^{-2d}\,\cG_{pq}{}^{pq}\,.
\ea
\label{aactioncG}
\ee 
These  agree with (\ref{aactionR}) upon the section condition, yet strictly differ by section-condition-vanishing purely bosonic terms:
\be
\ba{ll}
\cG_{pq}{}^{pq}-\cG_{\brp\brq}{}^{\brp\brq}&=S_{pq}{}^{pq}-S_{\brp\brq}{}^{\brp\brq}
	+\half P^{AB}\partial_{E}
V_{A p}\partial^{E}V_{B}{}^{p}-\half\brP^{AB}\partial_{E}
\brV_{A\brp}\partial^{E}\brV_{B}{}^{\brp}\\
{}&=S_{pq}{}^{pq}-S_{\brp\brq}{}^{\brp\brq}
+\half\left(\partial_{E}
V_{A p}\partial^{E}V^{A p}-\partial_{E}
\brV_{A\brp}
\partial^{E}\brV^{A\brp}\right)\,,
\ea
\label{GminusG}
\ee
and
\be
2\cG_{pq}{}^{pq}=S_{pq}{}^{pq}-S_{\brp\brq}{}^{\brp\brq}+4\partial_{A}\partial^{A}d
-4\partial_{A}d\partial^{A}d+\partial_{E}
V_{A p}\partial^{E}V^{A p}\,.
\label{2G}
\ee
The second equality of (\ref{GminusG})  follows from (\ref{defV}), (\ref{def2P}) and an identity,
\be
\brP^{AB}\partial_{E}
V_{A p}\partial^{E}V_{B}{}^{p}=P^{AB}\partial_{E}
\brV_{A\brp}\partial^{E}\brV_{B}{}^{\brp}\,.
\ee

\item  \textit{{\textbf{The full order supersymmetric  extensions.}}}  Based on the semi-covariant formalism revisited above, the $\cN=2$ (maximal)  $D=10$ supersymmetric double field theory has been constructed   to the full order in fermions~\cite{Jeon:2012hp}, 
\be
\cL^{\scriptscriptstyle\cN=2}_{\scriptscriptstyle D=10}(\cJ_{AB},\partial_{A},d,V_{Ap},\brV_{A\brp},\cC,\rho,\psi_{\brp},\rhop,\psip_{p})\,.
\label{N2sdftL}
\ee
By truncating the R-R potential and  the primed fermions, the $\cN=1$ (half-maximal)  $D=10$ supersymmetric double field theory~\cite{Jeon:2011sq} is also readily obtainable, 
\be
\cL^{\scriptscriptstyle\cN=1}_{\scriptscriptstyle D=10}(\cJ_{AB},\partial_{A},d,V_{Ap},\brV_{A\brp},\rho,\psi_{\brp})\,.
\label{N1sdftL}
\ee

Generically, the    supersymmetric double field theory Lagrangians decompose  into three parts,
\be
\cL_{\scriptscriptstyle \rm{SDFT}}=\cL_{[2,0]}+\cL_{[1,2]}+\cL_{[0,4]}\,,
\ee
where the subscript indices     denote the powers of the derivatives and the fermions separately, such that for  $\cL_{[a,b]}$, $a+b/2=2$, as the derivatives and the fermions have the mass dimensions one and half respectively, while the Lagrangian has the mass dimension two.  Further, the supersymmetry parameter, $\varepsilon$, has the mass dimension minus half, such that under supersymmetry transformations, each part of the Lagrangian transforms schematically as
\be
\ba{lll}
\delta_{\varepsilon}(\cL_{[2,0]})=\Delta_{\varepsilon[2,1]}\,,\quad&\quad
\delta_{\varepsilon}(\cL_{[1,2]})=\Delta^{\prime}_{\varepsilon [2,1]}+\Delta_{\varepsilon[1,3]}\,,\quad&\quad
\delta_{\varepsilon}(\cL_{[0,4]})=\Delta^{\prime}_{\varepsilon[1,3]}+
\Delta_{\varepsilon[0,5]}\,,
\ea
\ee
where $a+b/2=5/2$ for each $\Delta^{(\prime)}_{\varepsilon[a,b]}$.

In particular, the supersymmetry   of the $\cN=1$, $D=10$ SDFT~\cite{Jeon:2011sq}  amounts to the following algebraic identities,  
\be
\ba{rll}
\Delta_{\varepsilon[2,1]}+\Delta_{\varepsilon[2,1]}^{\prime}&=&
\partial_{A}K_{[1,1]}^{A}+
\left[~\bullet\partial_{A}\partial^{A}\bullet
~+~\partial_{A}\bullet\partial^{A}\bullet~\right]\,,\\
\Delta_{\varepsilon[1,3]}+\Delta_{\varepsilon[1,3]}^{\prime}&=&
\partial_{A}K_{[0,3]}^{A}\,,\\
\Delta_{\varepsilon[0,5]}&=&0\,,
\ea
\ee
such that the Lagrangian is invariant up to total derivatives and the  section condition,
\be
\delta_{\varepsilon}\cL^{\scriptscriptstyle\cN=1}_{\scriptscriptstyle D=10}=
\partial_{A}\left(K_{[1,1]}^{A}+K^{A}_{[0,3]}\right)+
\left[~\bullet\partial_{A}\partial^{A}\bullet
~+~\partial_{A}\bullet\partial^{A}\bullet~\right]_{[2,1]}\,.
\label{var1SDFT}
\ee
On the other hand, the supersymmetry of the $\cN=2$ $D=10$ SDFT~\cite{Jeon:2012hp} means the invariance of the Lagrangian up to total derivatives, the  section condition and the self-duality of the R-R field strength,
\be
\deltaS\cL^{\scriptscriptstyle \cN=2}_{\scriptscriptstyle D=10} =-\textstyle{\frac{1}{8}} e^{-2d}\brV^{A}{}_{\brq}\deltaS V_{Ap}\Tr\left(\gamma^{p}\tcF_{-}\brgamma^{\brq}\overline{\tcF_{-}}\,\right)+\partial_{A}K^{A}
+\left[~\bullet\partial_{A}\partial^{A}\bullet~
+~\partial_{A}\bullet\partial^{A}\bullet~\right]_{[2,1]}\,,
\label{var2SDFT}
\ee
where $\tcF_{-}$ is the self-dual part of the R-R field strength~(\ref{RRFLUX}) defined, to the full order in fermions, by
\be
\tcF_{-}:=\big(1-\gamma^{\eleven}\big)\left(\cF-i\half\rho\brrhop+i\half\gamma^{p}
\psi_{\brq}\brpsi^{\prime}_{p}\brgamma^{\brq}\right)\,,
\label{SD1}
\ee
and $\overline{\tcF_{-}}$ denotes,  like (\ref{RRFLUX}), its charge conjugation,
\be
\overline{\tcF_{-}}=\brC_{+}^{-1}(\tcF_{-})^{T}C_{+}\,.
\ee
A crucial fact about the section-condition-vanishing terms,    $\left[\,\bullet\partial_{A}\partial^{A}\bullet\,
+\,\partial_{A}\bullet\partial^{A}\bullet\,\right]_{[2,1]}$, which is common in  (\ref{var1SDFT}) and  (\ref{var2SDFT}), is that they are strictly linear in the fermions (dilatinos and gravitinos), and hence they are fully obtainable  just from  the leading order supersymmetry transformation rules.

It is also crucial  to note that  the above form of the algebraic identities, (\ref{var1SDFT}) and (\ref{var2SDFT}), still holds, \textit{i.e.}~the supersymmetry is unbroken upon the section  condition,   even if  we deform the Lagrangian by adding arbitrary section-condition-vanishing terms,  which, counting the mass dimensions, should be purely bosonic.  Examples include 
the replacement of $S_{ABCD}$ by $\cG_{ABCD}$ and adding $\cG_{pq}{}^{pq}+\cG_{\brp\brq}{}^{\brp\brq}$~(\ref{trivialid}) to the $\cN=1$ (but not $\cN=2$)   $D=10$ SDFT Lagrangian,  which we shall take below, for  the supersymmetry preserving twist.\\
\end{itemize}

%%%%%%%%%%%%%%%%%%%%%%%%%%%%%%%%%%%%%%%%%%%%%%
%%%%%%%%%%%%%%%%%%%%%%%%%%%%%%%%%%%%%%%%%%%%%%

\section{U-twisted  double field theory\label{secTwist}}
Here we  \textit{twist}  the double field theory formulated within the semi-covariant formalism. Our twist is a DFT generalization of the Scherk-Schwarz twist, based on \cite{Geissbuhler:2011mx,Aldazabal:2011nj,Grana:2012rr,Geissbuhler:2013uka},  and will be from time to time  referred to   as \textit{U-twist}.  In section~\ref{SECUtwist}, we introduce our ansatz of the twist. It involves a scalar and a  local $\ODD$ group element which may not obey the section condition. In section~\ref{SECtwistability},   following closely Grana and Marques~\cite{Grana:2012rr}, from the closure of the U-twisted generalized Lie derivative we derive a set of consistency  conditions which we call \textit{twistability conditions}. They  generalize   the original section condition and slightly differ from \cite{Grana:2012rr}. 
In section~\ref{SECTwistSemi},   we perform the U-twist on the semi-covariant formalism and verify that, with the replacement of $S_{ABCD}$ by $\cG_{ABCD}$, essentially all the nice properties of the semi-covariant formalism, including the complete covariantizability,  survive after the twist,  subject to the twistability conditions.  We also verify  that both the $\cN=2$  supersymmetric invariance  and the nilpotency    of the  differential operators which define the twisted Ramond-Ramond cohomology  commonly  require an extra  condition. Consequently,  the maximal supersymmetric twist of the  $\cN=2$ $D=10$  SDFT requires one more twistability  condition compared to the half-maximal supersymmetric twist of the $\cN=1$ $D=10$ SDFT.

%%%%%%%%%%%%%%%%%%%%%%%%%%%%%%%%%%%%%%%%%%%%%%
%%%%%%%%%%%%%%%%%%%%%%%%%%%%%%%%%%%%%%%%%%%%%%
\subsection{Ansatz for U-twist\label{SECUtwist}}
U-twist  calls for   two local variables, or the \textit{twisting data}: a scalar function,  $\lambda(x)$, and an  $\ODD$ element, $U(x)$. Both of them do \textit{not} necessarily satisfy the section condition~(\ref{aseccon}), but  shall be required to meet  consistency conditions, or  the \textit{twistability conditions}.

The local $\ODD$ element,  $U_{M}{}^{\mN}$,   carries one  undotted (untwisted)   row index and the other  dotted (twisted)  column index, such that, with the  introduction of   an additional  $\ODD$ invariant metric, 
\be
\mcJ_{\mM\mN}=\left(\ba{cc}0&1\\1&0\ea\right)\,,
\ee
it satisfies
\be
U\mcJ U^{t}=\cJ\,.
\ee 
While the dotted (twisted) metric, $\mcJ_{\mM\mN}$,   may coincide  numerically with the undotted (untwisted) metric,  $\cJ_{MN}$ (\textit{c.f.~}Table~\ref{TABindices}), hereafter  we deliberately distinguish them. In particular, the two different kinds of indices will never be contracted.  

\textit{U-twist}  prescribes  substituting  the following expression for  each untwisted (undotted)  field,  $T_{A_{1}\cdots A_{n}}$,  into the    $D=10$  ungauged DFT Lagrangians, \textit{c.f.}~\cite{Grana:2012rr},
\be
T_{A_{1}\cdots A_{n}}=e^{-2\omega\lambda}U_{A_{1}}{}^{\mA_{1}}\cdots U_{A_{n}}{}^{\mA_{n}}\mT_{\mA_{1}\cdots \mA_{n}}\,.
\label{pushback}
\ee
Equivalently,   twisted  fields are defined to carry dotted $\ODD$ indices with  a relevant   weight factor, 
\be
\mT_{\mA_{1}\cdots \mA_{n}}:=e^{2\omega\lambda}(U^{-1})_{\mA_{1}}{}^{B_{1}}\cdots (U^{-1})_{\mA_{n}}{}^{B_{n}}T_{B_{1}\cdots B_{n}}\,.
\label{pullback}
\ee
The derivatives of the untwisted fields then assume a generic  form,
\be
\partial_{C}T_{A_{1}\cdots A_{n}}=e^{-2\omega\lambda}
U_{C}{}^{\mC}U_{A_{1}}{}^{\mA_{1}}\cdots U_{A_{n}}{}^{\mA_{n}}\mD_{\mC}\mT_{\mA_{1}\cdots \mA_{n}}\,,
\label{derivT}
\ee
which naturally leads to the definition of what we call \textit{U-derivative}:   
With the pull-back  of the naked derivative, 
\be
\mpartial_{\mC}=U^{-1}{}_{\mC}{}^{C}\partial_{C}\,,
\label{twistderiv}
\ee
and   a pure gauge ``connection", 
\be
\Omega_{\mC\mA}{}^{\mB}:=\left(U^{-1}\mpartial_{\mC}U\right)_{\mA}{}^{\mB}\,,
\ee
the U-derivative, $\mD_{\mC}$, is defined to act on a  twisted  field by
\be
\mD_{\mC}\mT_{\mA_{1}\cdots \mA_{n}}:=\mpartial_{\mC}\mT_{\mA_{1}\cdots \mA_{n}}
-2\omega\mpartial_{\mC}\lambda\,\mT_{\mA_{1}\cdots \mA_{n}}+\sum_{i=1}^{n}\Omega_{\mC\mA_{i}}{}^{\mB}\mT_{\mA_{1}\cdots \mB\cdots\mA_{n}}\,.
\label{Uderiv}
\ee
In particular, the twist of the $\cN=1$ or the $\cN=2$,  $D=10$ SDFT amounts to inserting  the following expressions  for the dilaton and the  vielbeins into the untwisted Lagrangian, 
\be
\ba{lll}
e^{-2d}=e^{-2\lambda}e^{-2\md}\,,\quad&\quad 
V_{Ap}=U_{A}{}^{\mA}\mV_{\mA p}\,,\quad&\quad
\brV_{A\brp}=U_{A}{}^{\mA}\mbrV_{\mA \brp}\,.
\ea
\label{ONLYTWIST}
\ee
\textit{They  are the only field variables to be twisted}, since other fields (fermions and the R-R potential) are weightless and $\ODD$ singlet.   We shall not put a dot on those effectively untwisted fields for simplicity. The replacement   naturally leads to the twisted, half-maximal or maximal  SDFT Lagrangians, \textit{c.f.~}(\ref{halfmaximal}), (\ref{mtwistSDFT}),
\be
\ba{l}
\!\!\cL^{\scriptscriptstyle\cN=1}_{\scriptscriptstyle D=10}(\cJ_{AB},\partial_{A},d,V_{Ap},\brV_{A\brp},\rho,\psi_{\brp})
=e^{-2\lambda}
\mcL^{\scriptscriptstyle{\rm Half-maximal}}_{\scriptscriptstyle{\rm Twisted~SDFT}}(\mcJ_{\mA\mB},\mD_{\mA},\md,\mV_{\mA p},\mbrV_{\mA\brp},\rho,\psi_{\brp})\,,\\
\!\!\cL^{\scriptscriptstyle\cN=2}_{\scriptscriptstyle D=10}(\cJ_{AB},\partial_{A},d,V_{Ap},\brV_{A\brp},\cC,\rho,\psi_{\brp},\rhop,\psip_{p})
=e^{-2\lambda}
\mcL^{\scriptscriptstyle{\rm Maximal}}_{\scriptscriptstyle{\rm Twisted~SDFT}}(\mcJ_{\mA\mB},\mD_{\mA},\md,\mV_{\mA p},\mbrV_{\mA\brp},\cC,\rho,\psi_{\brp},\rhop,\psip_{p})\,.
\ea
\label{cLtwistcL}
\ee
As seen from the right hand sides of the equalities, --since every   DFT Lagrangian is   $\ODD$ singlet and  possesses   the diffeomorphic weight of unity--~   after the replacement, \textit{i)} the twisting matrix, $U_{A}{}^{\mA}$, effectively drops out in the Lagrangian,  \textit{ii)}  the  $\ODD$ invariant constant metric gets  `dotted',  and  \textit{iii)} the naked derivatives become  the U-derivatives.

The twist translates   the original  section condition as
\be
\mD_{\mA}\mD^{\mA}\sim0\,.
\label{secconD}
\ee
However, we do not  intend to  impose this condition on the twisted theory. Imposing this would lead to a mere equivalent  reformulation of the untwisted double field theory where (\ref{pushback}) corresponded to field redefinition.  In section~\ref{SECtwistability}, we  shall  look for alternative inequivalent conditions, or the \textit{twistability conditions}.  Before doing so, in the next subsection we  pause to collect some useful  properties of the U-derivative, $\mD_{\mC}$. \\

%%%%%%%%%%%%%%%%%%%%%%%%%%%%%%%%%%%%%%%%%%%%%%%%%%%%%%%%%%%
%%%%%%%%%%%%%%%%%%%%%%%%%%%%%%%%%%%%%%%%%%%%%%%%%%%%%%%%%%%
\subsection{Properties of the U-derivative and its connection\label{SECUderivative}}
Since  $U_{A}{}^{\mB}$ is an $\ODD$ element,  we have
\be
\Omega_{\mC\mA\mB}+\Omega_{\mC\mB\mA}=\mpartial_{\mC}\left(\mcJ U^{t}\cJ^{-1}U\mcJ\right)_{\mA\mB}=
\mpartial_{\mC}\mcJ_{\mA\mB}=0\,.
\ee
Hence, the U-derivative ``connection" is skew-symmetric for the last two indices,
\be
\Omega_{\mC\mA\mB}=-\Omega_{\mC\mB\mA}=\Omega_{\mC[\mA\mB]}\,.
\ee
It is  worth while  to note 
\be
\ba{ll}
\Omega_{\mB}{}^{\mB\mA}=\partial^{B}U_{B}{}^{\mA}\,,\quad&\quad
\Omega^{\mB}{}_{\mB\mA}=\partial_{B}(U^{-1})_{\mA}{}^{B}\,,
\ea
\ee
and\footnote{Eq.(\ref{pOmega}) can be derived from the following manipulation,
\[
\mpartial_{\mA}\Omega_{\mB}{}^{\mB\mA}=U^{B}{}_{\mA}\partial_{B}\partial_{C}U^{C\mA}=
\partial_{C}(U^{C}{}_{\mB}\Omega_{\mA\mB}{}^{\mA})-\Omega_{\mC\mB\mA}\Omega^{\mB\mC\mA}
=
-\mpartial_{\mB}\Omega_{\mA}{}^{\mA\mB}-\Omega_{\mA}{}^{\mA\mB}\Omega^{\mC}{}_{\mC\mB}-\Omega_{\mC\mB\mA}\Omega^{\mB\mC\mA}\,.
\]
}
\be
\mpartial_{\mA}\Omega_{\mB}{}^{\mB\mA}=-\half\Omega_{\mA}{}^{\mA\mB}\Omega^{\mC}{}_{\mC\mB}-\half\Omega_{\mC\mB\mA}\Omega^{\mB\mC\mA}\,.
\label{pOmega}
\ee
Pulling back the dotted  derivative index  to  a undotted index, it is useful  to  consider
\be
D_{C}\mT_{\mA_{1}\cdots \mA_{n}}=U_{C}{}^{\mC}\mD_{\mC}\mT_{\mA_{1}\cdots \mA_{n}}=
\partial_{C}\mT_{\mA_{1}\cdots \mA_{n}}-2\omega\partial_{C}\lambda\, \mT_{\mA_{1}\cdots \mA_{n}}
+\sum_{i=1}^{n}\Omega_{C\mA_{i}}{}^{\mB}\mT_{\mA_{1}\cdots \mB\cdots\mA_{n}}\,,
\label{Uderivundot}
\ee
where  naturally we put
\be
\Omega_{C\mA}{}^{\mB}=U_{C}{}^{\mC}\Omega_{\mC\mA}{}^{\mB}=\left(U^{-1}\partial_{C}U\right)_{\mA}{}^{\mB}\,.
\ee
This corresponds to  ``pure gauge"  and thus its ``field strength"  vanishes identically, 
\be
\ba{ll}
0&=\partial_{A}\Omega_{B\mC}{}^{\mD}-\partial_{B}\Omega_{A\mC}{}^{\mD}+\Omega_{A\mC}{}^{\mE}\Omega_{B\mE}{}^{\mD}-\Omega_{B\mC}{}^{\mE}\Omega_{A\mE}{}^{\mD}\\
{}&=D_{A}\Omega_{B\mC}{}^{\mD}-D_{B}\Omega_{A\mC}{}^{\mD}-\Omega_{A\mC}{}^{\mE}\Omega_{B\mE}{}^{\mD}+\Omega_{B\mC}{}^{\mE}\Omega_{A\mE}{}^{\mD}\,.
\ea
\label{vanishingFS}
\ee
By construction,    the U-derivative~(\ref{Uderiv}) can be rewritten as
\be
\ba{ll}
\mD_{\mC}\mT_{\mA_{1}\cdots \mA_{n}}=e^{2\omega\lambda}(U^{-1})_{\mC}{}^{C}(U^{-1})_{\mA_{1}}{}^{A_{1}}\cdots (U^{-1})_{\mA_{n}}{}^{A_{n}}\partial_{C}\left(e^{-2\omega\lambda}U_{A_{1}}{}^{\mB_{1}}\cdots U_{A_{n}}{}^{\mB_{n}}
\mT_{\mB_{1}\cdots \mB_{n}}\right)\,,
\ea
\ee
and is    compatible with the $U$ matrix itself,
\be
\ba{ll}
D_{A}U_{B}{}^{\mC}=\partial_{A}U_{B}{}^{\mC}-U_{B}{}^{\mD}\Omega_{A\mD}{}^{\mC}=0\,,\quad&\quad
\mD_{\mA}U_{B}{}^{\mC}=0\,,\\
D_{A}(U^{-1})_{\mC}{}^{B}=\partial_{A}(U^{-1})_{\mC}{}^{B}+\Omega_{A\mC}{}^{\mD}(U^{-1})_{\mD}{}^{B}=0\,,\quad&\quad\mD_{\mA}(U^{-1})_{\mC}{}^{B}=0\,,
\ea
\label{Ucomp}
\ee
as well as with both the dotted and the undotted $\ODD$ invariant metrics,  
\be
\ba{lll}
D_{C}\mcJ_{\mA\mB}=2\Omega_{C(\mA\mB)}=0\,,\quad&\quad
\mD_{\mC}\mcJ_{\mA\mB}=0\,,\quad&\quad\mD_{\mC}\cJ_{AB}=\mpartial_{\mC}\cJ_{AB}=0\,.
\ea
\ee
Pushing  back the undotted indices of (\ref{vanishingFS}) to the dotted ones, utilizing  the above $U$ matrix compatibility~(\ref{Ucomp}), we also get another useful relation,
\be
\mD_{[\mA}\Omega_{\mB]}{}^{\mC}{}_{\mD}=\Omega_{[\mA}{}^{\mC\mE}\Omega_{\mB]\mE\mD}\,.
\label{usefulOmega}
\ee
We emphasize that,   the $U$ matrix compatibility  is only possible because we distinguish  the dotted and the undotted $\ODD$ indices  deliberately.

Furthermore, from
\be
D_{A}D_{B}\mT_{\mC_{1}\cdots \mC_{n}}=
e^{2\omega\lambda}(U^{-1})_{\mC_{1}}{}^{C_{1}}\cdots (U^{-1})_{\mC_{n}}{}^{C_{n}}\partial_{A}\partial_{B}
\left(e^{-2\omega\lambda}U_{C_{1}}{}^{\mE_{1}}\cdots U_{C_{n}}{}^{\mE_{n}}
\mT_{\mE_{1}\cdots \mE_{n}}\right)\,,
\ee
the U-derivatives are all \textit{commutative},
\be
\ba{lll}
\left[D_{A},D_{B}\right]=0\,,\quad&\quad
\left[D_{A},\mD_{\mB}\right]=0\,,\quad&\quad
\left[\mD_{\mA},\mD_{\mB}\right]=0\,.
\ea
\label{commutativeD}
\ee
This is a crucial result. It means that there is no ordering ambiguity of the U-derivatives, as one might worry while performing the twist,  (\ref{cLtwistcL}). 
Namely,  the `field strength' and the `torsion'  of the U-derivative are all trivial.   It is also worth while to compare  with  the Weitzenb\"ock connection, \textit{e.g.~}\cite{Berman:2013uda}. Although  it appears formally similar to our $\Omega$, there is a  crucial difference: we intentionally  distinguish the dotted indices from the undotted indices, while the Weitzenb\"ock connection and hence the corresponding Weitzenb\"ock derivative do not. Consequently, the Weitzenb\"ock derivatives do not commute, unlike (\ref{commutativeD}),  and the Weizenb\"ock connection is  torsionful. \\

The dilaton, $d$, corresponds to the logarithm of a weightful scalar density. Its U-derivative is then determined    from  
\be
\partial_{A}e^{-2d}=
e^{-2\lambda}D_{A}e^{-2\md}=-2(D_{A}\md)e^{-2\lambda-2\md}\,,
\ee
by
\be
\ba{ll}
D_{A}\md=\partial_{A}\md+\partial_{A}\lambda=\partial_{A}d\,,\quad&\quad
\mD_{\mA}\md=\mpartial_{\mA}\md+\mpartial_{\mA}\lambda=\mpartial_{\mA}d\,.
\ea
\ee
Further, its second  order  derivatives are\footnote{This is also consistent with the following manipulation,
\[
\ba{ll}
\left(-2\partial_{A}\partial_{B}d
+4\partial_{A}d\partial_{B}d\right)e^{-2d}
=\partial_{A}\partial_{B}e^{-2d}&
=\partial_{A}\left(e^{-2\lambda}D_{B}e^{-2\md}\right)=e^{-2\lambda}\left(\partial_{A}-2\partial_{A}\lambda\right)D_{B}e^{-2\md}\\
{}&=e^{-2\lambda}D_{A}D_{B}e^{-2\md}
=e^{-2\lambda}\left(-2D_{A}D_{B}\md+4D_{A}\md D_{B}\md\right) e^{-2\md}\,.
\ea
\]
%%%
%%\[
%%\ba{ll}
%%\left(-2\partial_{A}\partial_{B}d+4\partial_{A}d\partial_{B}d\right)e^{-2d}
%%&=\partial_{A}\partial_{B}e^{-2d}=\partial_{A}\left(e^{-2\lambda}D_{B}e^{-2\md}\right)=
%%e^{-2\lambda}D_{A}D_{B}e^{-2\md}\\
%%{}&
%%=e^{-2\lambda}\left(-2D_{A}D_{B}\md+4D_{A}\md D_{B}\md\right) e^{-2\md}\,.
%%\ea
%%\]
%%%
}
\be
\partial_{A}\partial_{B}d=\partial_{A}(U_{B}{}^{\mB}\mD_{\mB}\md)=D_{A}(U_{B}{}^{\mB}\mD_{\mB}\md)=U_{A}{}^{\mA}U_{B}{}^{\mB}\mD_{\mA}\mD_{\mB}\md=D_{A}D_{B}\md\,,
\ee
and thus, in general,
\be
\partial_{A_{1}}\partial_{A_{2}}\cdots\partial_{A_{n}}d=U_{A_{1}}{}^{\mA_{1}}U_{A_{2}}{}^{\mA_{2}}\cdots U_{A_{n}}{}^{\mA_{n}}\mD_{\mA_{1}}\mD_{\mA_{2}}\cdots\mD_{\mA_{n}}\md=D_{A_{1}}D_{A_{2}}\cdots D_{A_{n}}\md\,.
\ee
~\\

Now, following \cite{Grana:2012rr}, we define two key  quantities out of the twisting data,
\be
f_{\mA}:=\Omega^{\mB}{}_{\mB\mA}-2\mpartial_{\mA}\lambda
=\partial_{C}U^{C}{}_{\mA}-2\mpartial_{\mA}\lambda\,,
\label{f1}
\ee
and the `structure constant',
\be
f_{\mA\mB\mC}:=\Omega_{\mA\mB\mC}+
\Omega_{\mB\mC\mA}+\Omega_{\mC\mA\mB}=f_{[\mA\mB\mC]}\,.
\label{f3}
\ee
Through straightforward computations,  one  can verify
\be
\mpartial^{\mC}\Omega_{\mC\mA\mB}=
\mpartial^{\mC}f_{\mC\mA\mB}+\mpartial_{\mA}f_{\mB}-
\mpartial_{\mB}f_{\mA}
+f^{\mC}\Omega_{\mA\mB\mC}-
f^{\mC}\Omega_{\mB\mA\mC}\,,
\label{partialOmega}
\ee
and
\be
\ba{ll}
f_{\mA\mB\mE}f_{\mC\mD}{}^{\mE}=&\Phi_{\mA\mB\mC\mD}-\Phi_{\mB\mA\mC\mD}+\Phi_{\mA\mC\mB\mD}-
\Phi_{\mB\mC\mA\mD}
+\Phi_{\mB\mA\mD\mC}-\Phi_{\mA\mB\mD\mC}+\Phi_{\mC\mA\mD\mB}-\Phi_{\mC\mB\mD\mA}\\
{}&+\Omega_{E\mA\mB}\Omega^{E}{}_{\mC\mD}-\mpartial_{\mC}\Omega_{\mD\mA\mB}+\mpartial_{\mD}\Omega_{\mC\mA\mB}
-\mpartial_{\mA}\Omega_{\mB\mC\mD}+\mpartial_{\mB}
\Omega_{\mA\mC\mD}\,,
\ea
\ee
where  we set
\be
\Phi_{\mA\mB\mC\mD}=\mpartial_{\mA}U^{E}{}_{\mB}\mpartial_{\mC}U_{E\mD}=\Phi_{\mC\mD\mA\mB}\,.
\ee
In particular, this implies
\be
\Omega_{E[\mA\mB}\Omega^{E}{}_{\mC]\mD}=f_{[\mA\mB}{}^{\mE}f_{\mC]\mD\mE}+\mpartial_{[\mA}f_{\mB\mC]\mD}-\textstyle{\frac{1}{3}}\mpartial_{\mD}f_{\mA\mB\mC}\,.
\label{ffOO3}
\ee
We shall make use of  these identities shortly below.\\

Finally, from (\ref{derivT}),  the divergence of  a vector density with  weight one  becomes after the twist,
\be
\partial_{A}K^{A}=e^{-2\lambda}\mD_{\mA}\mK^{\mA}=e^{-2\lambda}(\mpartial_{\mA}\mK^{\mA}+f_{\mA}\mK^{\mA})\,.
\label{Noether}
\ee
Thus, after the twist, the potentially anomalous terms    in the  supersymmetric variations of the $\cN=1$ or the $\cN=2$ $D=10$ SDFT Lagrangian~(\ref{var1SDFT}), (\ref{var2SDFT}) assume the following  generic form,
\be
\left<\delta_{\varepsilon}\mcL^{\scriptscriptstyle{\rm Twisted}}_{\scriptscriptstyle{\rm SDFT}}\right>_{\rm{anomalous}}=f_{\mA}\mK^{\mA}+
\left[~\mbullet\mD_{\mA}\mD^{\mA}\mbullet~
+~\mD_{A}\mbullet\mD^{\mA}\mbullet~\right]_{[2,1]}
\,.
\label{vartSDFT}
\ee
In order to ensure  the supersymmetry to be unbroken  after the twist, we need to show that these terms vanish up to  the twistability conditions. Fortunately, as discussed in section~\ref{SECSUSYextensions} and demonstrated in section~\ref{secSDFTtwist} later, these anomalous terms can be all  sufficiently  obtained  just   from the leading order supersymmetry.  \\

%%%%%%%%%%%%%%%%%%%%%%%%%%%%%%%%%%%%%%%%%%%%%%%%%%%%%%%%%%%%
%%%%%%%%%%%%%%%%%%%%%%%%%%%%%%%%%%%%%%%%%%%%%%%%%%%%%%%%%%%%
\subsection{Twistability conditions: closure of the diffeomorphisms\label{SECtwistability}}
Acting on the dotted twisted  fields,    \textit{twisted  diffeomorphism}  is generated by the  \textit{U-twisted generalized Lie derivative}, 
\be
\mcL_{\mX}\mT_{\mA_{1}\cdots \mA_{n}}:=\mX^{\mB}\mD_{\mB}\mT_{\mA_{1}\cdots \mA_{n}}+\omega\mD_{\mB}\mX^{\mB}\mT_{\mA_{1}\cdots \mA_{n}}+\sum_{i=1}^{n}(\mD_{\mA_{i}}\mX_{\mB}-\mD_{\mB}\mX_{\mA_{i}})
\mT_{\mA_{1}\cdots\mA_{i-1}}{}^{\mB}{}_{\mA_{i+1}\cdots  \mA_{n}}\,.
\label{brtcL}
\ee
In an identical manner to the twisting ansatz~(\ref{pullback}), this expression is    related to  the untwisted generalized Lie derivative~(\ref{tcL}) by
\be
\mcL_{\mX}\mT_{\mA_{1}\cdots\mA_{n}}=e^{2\omega\lambda}(U^{-1})_{\mA_{1}}{}^{A_{1}}\cdots(U^{-1})_{\mA_{n}}{}^{A_{n}}\hcL_{X}T_{A_{1}\cdots A_{n}}\,.
\label{mLhL}
\ee
The commutator of the U-twisted generalized Lie derivatives, without employing any section condition,  reads readily from (\ref{Cbracketclosure}),  \textit{c.f.~}Grana and Marques~\cite{Grana:2012rr},%\footnote{The overall sign in  \eqref{agB2br} is oppose to \cite{Grana:2012rr}.}
\be
\ba{ll}
\left([\mcL_{\mX},\mcL_{\mY}]-\mcL_{[\mX,\mY]_{\mrmC}}\right)\mT_{\mA_{1}\cdots \mA_{n}}=&\half(\mX^{\mN}\mD^{\mM}\mY_{\mN}
-\mY^{\mN}\mD^{\mM}\mX_{\mN})\mD_{\mM}
\mT_{\mA_{1}\cdots \mA_{n}}\\
{}&+\half\omega(\mX^{\mN}\mD_{\mM}\mD^{\mM}\mY_{\mN}
-\mY^{\mN}\mD_{\mM}\mD^{\mM}\mX_{\mN})
\mT_{\mA_{1}\cdots \mA_{n}}\\
{}&+\sum_{i=1}^{n}(\mD_{\mM}\mY_{\mA_{i}}\mD^{\mM}\mX_{\mB}
-\mD_{\mM}\mX_{\mA_{i}}\mD^{\mM}\mY_{\mB})
\mT_{\mA_{1}\cdots\mA_{i-1}}{}^{\mB}{}_{\mA_{i+1}\cdots  \mA_{n}}\,,
\ea
\label{agB2br}
\ee
where $[\mX,\mY]_{\mrmC}$ denotes the U-twisted C-bracket,
\be
[\mX,\mY]^{\mA}_{\mrmC}:= \mX^{\mB}\mD_{\mB}\mY^{\mA}-\mY^{\mB}\mD_{\mB}\mX^{\mA}
+\half\mY^{\mB}\mD^{\mA}\mX_{\mB}-\half \mX^{\mB}\mD^{\mA}\mY_{\mB}\,.
\ee
Clearly, if the condition of (\ref{secconD}) were imposed, the right hand side of (\ref{agB2br}) would vanish. Yet, we are after  other way of ensuring the closure. To this end, we dismantle   the U-derivative and display its connection explicitly: the U-twisted generalized Lie derivative (\ref{brtcL}) and the U-twisted C-bracket (\ref{agB2br}) can be  rewritten,  in terms of $f_{\mA}$~(\ref{f1}) and $f_{\mA\mB\mC}$~(\ref{f3}),  as
\be
\ba{ll}
\mcL_{\mX}\mT_{\mA_{1}\cdots \mA_{n}}=&\mX^{\mB}\mpartial_{\mB}\mT_{\mA_{1}\cdots \mA_{n}}+\omega\left(\mpartial_{\mB}\mX^{\mB}+
f_{\mB}\mX^{\mB}\right)
\mT_{\mA_{1}\cdots \mA_{n}}\\
{}&+\sum_{i=1}^{n}\left(\mpartial_{\mA_{i}}\mX_{\mB}-\mpartial_{\mB}\mX_{\mA_{i}}
+f_{\mA_{i}\mB\mC}\mX^{\mC}
\right)
\mT_{\mA_{1}\cdots\mA_{i-1}}{}^{\mB}{}_{\mA_{i+1}\cdots  \mA_{n}}\,,
\ea
\label{brtcL2}
\ee
and
\be
[\mX,\mY]^{\mA}_{\rm{C}}= \mX^{\mB}\mpartial_{\mB}\mY^{\mA}
-\mY^{\mB}\mpartial_{\mB}\mX^{\mA}
+\half\mY^{\mB}\mpartial^{\mA}\mX_{\mB}-\half \mX^{\mB}\mpartial^{\mA}\mY_{\mB}-f^{\mA}{}_{\mB\mC}\mX^{\mB}\mY^{\mC}\,.
\label{agB2br2}
\ee
Similarly,  the right hand side of the equality  in (\ref{agB2br}) reads
\be
\ba{l}
\left([\mcL_{\mX},\mcL_{\mY}]-\mcL_{[\mX,\mY]_{\rm{C}}}\right)\mT_{\mA_{1}\cdots \mA_{n}}\\
=\left(\half\mX^{\mN}\mpartial^{\mM}\mY_{\mN}
-\half\mY^{\mN}\mpartial^{\mM}\mX_{\mN}+\Omega^{\mM}{}_{\mN\mG}\mX^{\mN}\mY^{\mG}\right)\mpartial_{\mM}
\mT_{\mA_{1}\cdots \mA_{n}}\\
\ba{ll}
\quad~+\half\omega\Big[&\!\!\!\mX^{\mN}\mpartial_{\mM}\mpartial^{\mM}\mY_{\mN}
-\mY^{\mN}\mpartial_{\mM}\mpartial^{\mM}\mX_{\mN}
+2\mX^{\mN}\Omega^{\mM}{}_{\mN\mG}\mpartial_{\mM}\mY^{\mG}-2\mY^{\mN}\Omega^{\mM}{}_{\mN\mG}\mpartial_{\mM}\mX^{\mG}
\\
{}&\!\!\!\!\!+2\mX^{\mN}\mY^{\mG}\mpartial_{\mM}\Omega^{\mM}{}_{\mN\mG}
+f_{\mM}\left(\mX^{\mN}\mD^{\mM}\mY_{\mN}-\mY^{\mN}\mD^{\mM}\mX_{\mN}\right)\Big]
\mT_{\mA_{1}\cdots \mA_{n}}\ea\\
\ba{ll}
\quad~+\sum_{i=1}^{n}\Big[&\!\!\!\mpartial_{\mM}\mY_{\mA_{i}}\mpartial^{\mM}\mX_{\mB}
-\mpartial_{\mM}\mX_{\mA_{i}}\mpartial^{\mM}\mY_{\mB}
+3\Omega_{\mM[\mA_{i}\mB}\mX^{\mN}\mpartial^{\mM}
\mY_{\mN]}
-3\Omega_{\mM[\mA_{i}\mB}\mY^{\mN}\mpartial^{\mM}
\mX_{\mN]}\\
{}&\!\!\!\!\!\!\!\!\!\!
-\half\Omega_{\mM\mA_{i}\mB}
\left(\mX^{\mN}\mpartial^{\mM}\mY_{\mN}
-\mY^{\mN}\mpartial^{\mM}\mX_{\mN}\right)
-3\Omega_{\mM[\mB\mN}\Omega^{\mM}{}_{\mG]\mA_{i}}\mX^{\mN}\mY^{\mG}~
\Big]
\mT_{\mA_{1}\cdots\mA_{i-1}}{}^{\mB}{}_{\mA_{i+1}\cdots  \mA_{n}}\,,
\ea
\ea
\label{agB2br2}
\ee
which further becomes, using (\ref{partialOmega}) and (\ref{ffOO3}),
\be
\ba{l}
\left([\mcL_{\mX},\mcL_{\mY}]-\mcL_{[\mX,\mY]_{\rm{C}}}\right)\mT_{\mA_{1}\cdots \mA_{n}}\\
=\left(\half\mX^{\mN}\mpartial^{\mM}\mY_{\mN}
-\half\mY^{\mN}\mpartial^{\mM}\mX_{\mN}+\Omega^{\mM}{}_{\mN\mG}\mX^{\mN}\mY^{\mG}\right)\mpartial_{\mM}
\mT_{\mA_{1}\cdots \mA_{n}}\\
\ba{ll}
\quad~+\half\omega\Big[&\!\!\!\mX^{\mN}\mpartial_{\mM}\mpartial^{\mM}\mY_{\mN}
-\mY^{\mN}\mpartial_{\mM}\mpartial^{\mM}\mX_{\mN}
+2\mX^{\mN}\Omega^{\mM}{}_{\mN\mG}\mpartial_{\mM}\mY^{\mG}-2\mY^{\mN}\Omega^{\mM}{}_{\mN\mG}\mpartial_{\mM}\mX^{\mG}
\\
{}&\!\!\!\!\!+2\mX^{\mN}\mY^{\mG}\left(
\mpartial^{\mM}f_{\mM\mN\mG}+f^{\mM}f_{\mM\mN\mG}
+2\mpartial_{[\mN}f_{\mG]}
\right)
+f_{\mM}\left(\mX^{\mN}\mpartial^{\mM}\mY_{\mN}
-\mY^{\mN}\mpartial^{\mM}\mX_{\mN}\right)\Big]
\mT_{\mA_{1}\cdots \mA_{n}}\ea\\
\ba{ll}
\quad~+\sum_{i=1}^{n}\Big[&\!\!\!\mpartial_{\mM}\mY_{\mA_{i}}\mpartial^{\mM}\mX_{\mB}
-\mpartial_{\mM}\mX_{\mA_{i}}\mpartial^{\mM}\mY_{\mB}
-\half\Omega_{\mM\mA_{i}\mB}
\left(\mX^{\mN}\mpartial^{\mM}\mY_{\mN}
-\mY^{\mN}\mpartial^{\mM}\mX_{\mN}\right)\\
{}& +3\Omega_{\mM[\mA_{i}\mB}\mX^{\mN}\mpartial^{\mM}
\mY_{\mN]}
-3\Omega_{\mM[\mA_{i}\mB}\mY^{\mN}\mpartial^{\mM}
\mX_{\mN]}\\
{}&
+\mX^{\mN}\mY^{\mG}
\left(\mpartial_{\mA_{i}}f_{\mB\mN\mG}
-3f_{\mM[\mB\mN}f^{\mM}{}_{\mG]\mA_{i}}-3\mpartial_{[\mB}f_{\mN\mG]\mA_{i}}
\right)
\Big]
\mT_{\mA_{1}\cdots\mA_{i-1}}{}^{\mB}{}_{\mA_{i+1}\cdots  \mA_{n}}\,.\ea
\ea
\label{agB2br3}
\ee
Now we can easily  read off a set of conditions, or \textit{the twistability conditions},  which let  each  term in the right hand of the above equality vanish.  The twistability conditions which ensure the closure of the U-twisted generalized Lie derivative
\be
{}[\mcL_{\mX},\mcL_{\mY}]\equiv\mcL_{[\mX,\mY]_{\rm{C}}}\,,
\label{twclosure}
\ee
are as follows, \textit{c.f.~}\cite{Grana:2012rr,Berman:2013cli}.\footnote{Strictly speaking, our twistability conditions, especially (\ref{tsc2}), do not completely agree with the previous works.  Yet, with the ansatz~(\ref{split}) assumed, they agree.}
\begin{enumerate}
\item The section condition for all the dotted twisted fields,
\be
\mpartial_{\mM}\mpartial^{\mM}\equiv0\,.
\label{tsc1}
\ee
\item The orthogonality between  the connection and the derivatives  of  the dotted twisted fields,
\be
\Omega^{\mM}{}_{\mF\mG}\mpartial_{\mM}\equiv0\,.
\label{tsc2}
\ee
\item The Jacobi identity for $f_{\mA\mB\mC}=f_{[\mA\mB\mC]}$,
\be
f_{[\mA\mB}{}^{\mE}f_{\mC]\mD\mE}\equiv0\,.
\label{tsc3}
\ee
\item The constancy of the structure constant, $f_{\mA\mB\mC\,}$,
\be
\mpartial_{\mE}f_{\mA\mB\mC}\equiv0\,.
\label{tsc4}
\ee
\item The triviality  of $f_{\mA\,}$,
\be
f_{\mA}=\Omega^{\mC}{}_{\mC\mA}-2\mpartial_{\mA}\lambda=\partial_{C}U^{C}{}_{\mA}-2\mpartial_{\mA}\lambda\equiv0\,.
\label{tsc5}
\ee
\end{enumerate}
We stress that these five constraints, (\ref{tsc1}) -- (\ref{tsc5}), are the  natural requirement for the closure~(\ref{twclosure}) directly read off from  (\ref{agB2br3}).\footnote{Clearly  the  five constraints, (\ref{tsc1}) -- (\ref{tsc5}), are sufficient for the closure. It remains as an open question  whether they are also necessary.}  In principle, we should solve these constraints.  While we are currently lacking the most general form of  the solutions,  a class of solutions are well known which involve  dimensional reductions.  If we assume the $U$ matrix to be in a block diagonal form,
\be
U=\left(\ba{ll}1&0\\0&u\ea\right)\,,
\label{split}
\ee
the dotted $\ODD$ indices naturally split  into  effectively untwisted external indices and  truly twisted  internal indices. Letting all the twisted (or dotted) fields depend on the external coordinates while  allowing the twisting data, $u$, $\lambda$, to have only  the internal dependency,  the first condition~(\ref{tsc1}) is nothing but  the ordinary section condition for the twisted fields living in the dimensionally reduced, external doubled-yet-gauged  spacetime, and  the second condition~(\ref{tsc2}) is clearly satisfied. The remaining conditions (\ref{tsc3}), (\ref{tsc4}), (\ref{tsc5}) are then the genuine consistency conditions for the internal twisting data, $u$ and $\lambda$.  This `solution' then inevitably  implies the dimensional reduction of the section, from $D=10$ to a lower value. Namely, the twistability conditions 
consist of the ordinary section condition for the external spacetime and a set of  consistency conditions for the twisting data, $U$ and  $\lambda$, of  the orthogonal  internal ``manifold''.  It is interesting to explore other type of solution, if any,  generalizing the ansatz~(\ref{split}). Anyhow, all the forthcoming  analyses  require strictly the five constraints, (\ref{tsc1}) -- (\ref{tsc5}) only,  and do not necessarily  demand the ansatz~(\ref{split}).

It is worth while to note   from (\ref{partialOmega}), (\ref{ffOO3}),  that the twistability conditions imply
\be
\ba{l}
\mpartial^{\mC}\Omega_{\mC\mA\mB}\equiv0\,,\\
\Omega_{\mE[\mA\mB}\Omega^{\mE}{}_{\mC]\mD}\equiv0\,,\\

\mpartial_{\mA}\mpartial^{\mA}\lambda\equiv\half\mpartial_{\mA}\Omega_{\mB}{}^{\mB\mA}=
-\quarter\Omega_{\mA}{}^{\mA\mC}\Omega^{\mB}{}_{\mB\mC}-\quarter\Omega_{\mC\mB\mA}\Omega^{\mB\mC\mA}\,.
\ea
\label{follow}
\ee
Further, from  the integrability of the last condition~(\ref{tsc5}), we get
\be
\partial_{A}\left(U_{B}{}^{\mE}\partial_{C}U^{C}{}_{\mE}\right)\equiv
\partial_{B}\left(U_{A}{}^{\mE}\partial_{C}U^{C}{}_{\mE}\right)\,.
\ee
The U-twisted  generalized Lie derivative~(\ref{brtcL2})  reduces,  upon the twistability conditions, to
\be
\hcL_{\mX}\mT_{\mA_{1}\cdots \mA_{n}}\equiv\mX^{\mB}\mpartial_{\mB}\mT_{\mA_{1}\cdots \mA_{n}}+\omega\mpartial_{\mB}\mX^{\mB}
\mT_{\mA_{1}\cdots \mA_{n}}+\sum_{i=1}^{n}\left(2\mpartial_{[\mA_{i}}\mX_{\mB]}
+f_{\mA_{i}\mB\mC}\mX^{\mC}
\right)
\mT_{\mA_{1}\cdots\mA_{i-1}}{}^{\mB}{}_{\mA_{i+1}\cdots  \mA_{n}}\,,
\label{brtcL3}
\ee
which clearly has no `internal' coordinate dependency\footnote{With the internal/external splitting~(\ref{split}), the $\mpartial_{\mA}$ derivatives of   the  dotted fields are independent of the internal coordinates.}
and  decomposes into the external diffeomorphism and  internal gauge symmetry~\cite{Grana:2012rr,Berman:2013cli} (see also \cite{Hohm:2011ex}).\\

%%%%%%%%%%%%%%%%%%%%%%%%%%%%%%%%%%%%%%%%%%%%%%
%%%%%%%%%%%%%%%%%%%%%%%%%%%%%%%%%%%%%%%%%%%%%%
\subsection{Twisted semi-covariant formalism\label{SECTwistSemi}}      
The twisting of the semi-covariant formalism is straightforward.   The \textit{U-twisted master semi-covariant derivative} is 
\be
\mcD_{\mA}=\mna_{\mA}+\mPhi_{\mA}+\mbrPhi_{\mA}\,,
\label{UtwistMaster}
\ee
of which the twisted semi-covariant derivative and the twisted spin connections are given by
\be
\ba{lll}
\mna_{\mA}=\mD_{\mA}+\mGamma_{\mA}\,,\quad&\quad
\mPhi_{\mA pq}=\mV^{\mB}{}_{p}\mna_{\mA}\mV_{\mB q}\,,\quad&\quad
\mbrPhi_{\mA\brp\brq}=\mbrV^{\mB}{}_{\brp}\mna_{\mA}\mbrV_{\mB \brq}\,,
\ea
\ee
and the  twisted torsionless   connection reads
\be
\ba{ll}
\mGamma_{\mC\mA\mB}=&
2(\mP\mD_{\mC}\mP\mbrP)_{[\mA\mB]}
+2({{\mbrP}_{[\mA}{}^{\mD}{\mbrP}_{\mB]}{}^{\mE}}-{\mP_{[\mA}{}^{\mD}\mP_{\mB]}{}^{\mE}})\mD_{\mD}\mP_{\mE\mC}\\
{}&-\textstyle{\frac{4}{D-1}}(\mbrP_{\mC[\mA}\mbrP_{\mB]}{}^{\mD}+\mP_{\mC[\mA}\mP_{\mB]}{}^{\mD})
\left(\mD_{\mD}\md+(\mP\mD^{\mE}\mP\mbrP)_{[\mE\mD]}\right)\,,
\ea
\label{Gammat}
\ee
satisfying, in a completely parallel manner to the untwisted cases, (\ref{acompP}) -- (\ref{akernel}), 
\be
\ba{lll}
\mna_{\mA}\mP_{\mB\mC}=0\,,\quad&\quad\mna_{\mA}\mbrP_{\mB\mC}=0\,,\quad&\quad
\mGamma^{\mB}{}_{\mB\mA}=-2\mD_{\mA}\md\,,\\
\mGamma_{\mA(\mB\mC)}=0\,,&\quad\mGamma_{[\mA\mB\mC]}=0\,,\quad&\quad
(\mcP+\mbrcP)_{\mA\mB\mC}{}^{\mD\mE\mF}\mGamma_{\mD\mE\mF}=0\,,
\ea
\label{twtorsionless1} 
\ee
and, as  for the torsionless condition~(\ref{untwtorsionless}),
\be
\ba{ll}
\hcL_{X}(\mD)=\hcL_{X}(\mna)\,,\quad&\quad
[X,Y]_{\rmC}(\mD)=[X,Y]_{\rmC}(\mna)\,.
\ea
\label{twtorsionless2}
\ee  
Further, from (\ref{FABpq}), (\ref{defcG}),  in terms of
\be
\ba{ll}
\msF_{\mA\mB pq}&=\mna_{\mA}\mPhi_{\mB pq}-\mna_{\mB}\mPhi_{\mA pq}+\mPhi_{\mA p}{}^{r}\mPhi_{\mB rq}-\mPhi_{\mB p}{}^{r}\mPhi_{\mA rq}\\
{}&=\mcD_{\mA}\mPhi_{\mB pq}-\mcD_{\mB}\mPhi_{\mA pq}-\mPhi_{\mA p}{}^{r}\mPhi_{\mB rq}+\mPhi_{\mB p}{}^{r}\mPhi_{\mA rq}\,,\\
\msbrF_{\mA\mB \brp\brq}
&=\mna_{\mA}\mbrPhi_{\mB \brp\brq}-\mna_{\mB}\mbrPhi_{\mA \brp\brq}+\mbrPhi_{\mA \brp}{}^{\brr}\mbrPhi_{\mB \brr\brq}-\mbrPhi_{\mB \brp}{}^{\brr}\mbrPhi_{\mA \brr\brq}\\
{}&=\mcD_{\mA}\mbrPhi_{\mB \brp\brq}-\mcD_{\mB}\mbrPhi_{\mA \brp\brq}-\mbrPhi_{\mA \brp}{}^{\brr}\mbrPhi_{\mB \brr\brq}+\mbrPhi_{\mB \brp}{}^{\brr}\mbrPhi_{\mA \brr\brq}\,,
\ea
\ee
we have   \textit{the twisted spinorial semi-covariant four-index curvature}, 
\be
\mcG_{\mA\mB\mC\mD}:=\half\left[
(\msF+\msbrF )_{\mA\mB\mC\mD}
+(\msF+\msbrF )_{\mC\mD\mA\mB}+
(\mPhi+\mbrPhi)^{\mE}{}_{\mA\mB}(\mPhi+\mbrPhi)_{\mE\mC\mD}\right]\,.
\ee
Now, from (\ref{relationcGS}), it is useful to note
\be
\mcG_{\mA\mB\mC\mD}=\mS_{\mA\mB\mC\mD}+
\half(\mV_{\mA}{}^{p}\mD_{\mE}\mV_{\mB p}
+\mbrV_{\mA}{}^{\brp}\mD_{\mE}\mbrV_{\mB\brp})(\mV_{\mC}{}^{q}\mD^{\mE}\mV_{\mD q}
+\mbrV_{\mC}{}^{\brq}\mD^{\mE}\mbrV_{\mD\brq})\,,
\label{relationcGStwist}
\ee
and thus, upon the twistability conditions, 
\be
\mcG_{\mA\mB\mC\mD}\equiv\mS_{\mA\mB\mC\mD}+\half\Omega_{\mE\mA\mB}\Omega^{\mE}{}_{\mC\mD}\,.
\label{relationcGStwistequiv}
\ee
In the above, for sure,  we set
\be
\ba{l}
\mS_{\mA\mB\mC\mD}=\half\left(\mR_{\mA\mB\mC\mD}+\mR_{\mC\mD\mA\mB}-\mGamma^{\mE}{}_{\mA\mB}\mGamma_{\mE\mC\mD}\right)\,,\\
\mR_{\mA\mB\mC\mD}=\mpartial_{\mA}\mGamma_{\mB\mC\mD}-\mpartial_{\mB}\mGamma_{\mA\mC\mD}+\mGamma_{\mA\mC}{}^{\mE}\mGamma_{\mB\mE\mD}-\mGamma_{\mB\mC}{}^{\mE}\mGamma_{\mA\mE\mD}\,.
\ea
\ee
Thus,  in contrast to the untwisted case~(\ref{cGSsim}), $\cG_{\mA\mB\mC\mD}$ differs from $\mS_{\mA\mB\mC\mD}$ after the twist. In the twisted SDFT to be constructed below, we shall disregard the latter and employ the former only. The former  will be shown to be semi-covariant, while the latter is not.

Starting from the strict equality  of (\ref{relationcGStwist}) and using  (\ref{Svar0}), one can easily show nevertheless that the infinitesimal  transformation of $\mcG_{\mA\mB\mC\mD}$ induced by the variations of  its constituting all the twisted fields  coincides with that of $\mS_{\mA\mB\mC\mD}$,   up to the twistability conditions,
\be
\delta\mcG_{\mA\mB\mC\mD}\equiv\mna_{[\mA}\delta\mGamma_{\mB]\mC\mD}+\mna_{[\mC}\delta\mGamma_{\mD]\mA\mB}\equiv
\delta\mS_{\mA\mB\mC\mD}\,.
\label{varcGS}
\ee
This should be a naturally  expected result,  if we  focus on the variation of the   equivalence relation (\ref{relationcGStwistequiv}) rather than  the strict equality~(\ref{relationcGStwist}). Since $\Omega_{\mA\mB\mC}$ is not a field variable but  rather a  fixed data for a given  internal manifold, it is not taken to transform but must be inert under  any  `symmetry',\footnote{However,   $\,\hcL_{\mX}U_{A}{}^{\mB}=\mX^{\mC}\mD_{\mC}U_{A}{}^{\mB}+(\mD^{\mB}\mX_{\mC}-\mD_{\mC}\mX^{\mB})U_{A}{}^{\mC}=
\mD^{\mB}X_{A}-D_{A}\mX^{\mB}\neq0\,$.}
\be
\ba{lll}
\delta U_{A}{}^{\mB}=0\,,\quad&\quad\delta\Omega_{\mC\mA\mB}=0\,,\quad&\quad\delta(\mD_{\mC})=0\,.
\ea
\label{varU}
\ee
These are also consistent with (\ref{pullback}) and (\ref{mLhL}) with the identification of  `$\,\delta_{\mX}=\mL_{\mX}\,$' for  covariant twisted  fields.

\begin{itemize}
\item \textit{Complete covariantizations after the twist.}

Here we  focus on the twisted diffeomorphism. 
We twist the relation~(\ref{anomalpartial}) in order to obtain  the difference between the actual transformation of the U-derivative of a twisted field and its twisted  generalized Lie derivative,  
\be
\ba{l}
{}(\delta_{\mX}-\mcL_{\mX})\mD_{\mC}\mT_{\mA_{1}\cdots\mA_{n}}\\
=\left[\mD_{\mC}\,,\,\mcL_{\mX}\right]\mT_{\mA_{1}\cdots\mA_{n}}\\
=\mD^{\mB}\mX_{\mC}\mD_{\mB}\mT_{\mA_{1}\cdots \mA_{n}}+
\omega_{{\scriptscriptstyle{T\,}}}\mD_{\mC}\mD_{\mB}\mX^{\mB}
\mT_{\mA_{1}\cdots \mA_{n}}+
\dis{\sum_{i=1}^{n}\,2\mD_{\mC}\mD_{[\mA_{i}}\mX_{\mB]}
\mT_{\mA_{1}\cdots \mA_{i-1}}{}^{\mB}{}_{\mA_{i+1}\cdots \mA_{n}}\,.}
\ea
\label{anomalpartialtwist}
\ee
Writing the first equality above, we have  implicitly assumed  (\ref{varU}). It follows  for the twisted  connection~(\ref{Gammat}),
\be
\ba{ll}
(\delta_{\mX}{-\hcL_{\mX}})\mGamma_{\mC\mA\mB}&= 2\left[(\mcP+\mbrcP)_{\mC\mA\mB}{}^{\mF\mD\mE}-\delta_{\mC}^{~\mF}\delta_{\mA}^{~\mD}
\delta_{\mB}^{~\mE}\right]\mD_{\mF}\mD_{[\mD}\mX_{\mE]}\\
{}&\quad+2(\mcPp-\mcPq)_{\mC\mA\mB}{}^{\mF\mD\mE}
\mD^{\mG}\mP_{\mF\mD}\mD_{\mG}\mX_{\mE}
+2\mP_{[\mA}{}^{\mD}\mbrP_{\mB]}{}^{\mE}\mD^{\mG}\mP_{\mD\mE}\mD_{\mG}\mX_{\mC}\\
{}&\quad+\textstyle{\frac{2}{D-1}}(\mP_{\mC[\mA}\mP_{\mB]}{}^{\mE}+
\mbrP_{\mC[\mA}\mbrP_{\mB]}{}^{\mE})(\mD_{\mG}\mD^{\mG}\mX_{\mE}-2\mD^{\mG}\md\mD_{\mG}\mX_{\mE})\,.
\ea
\label{delhcLmG}
\ee
To simplify this expression up to the twistability conditions, we use (\ref{partialOmega}) and an identity,
\be
\ba{l}
\left[
2(\mcPp+\mcPq)_{\mC\mA\mB}{}^{\mF\mD\mE}
+2\mP_{[\mA}{}^{\mD}\mbrP_{\mB]}{}^{\mF}\delta_{\mC}^{~\mE}
+\textstyle{\frac{2}{D-1}}(\mP_{\mC[\mA}\mP_{\mB]}{}^{\mF}+
\mbrP_{\mC[\mA}\mbrP_{\mB]}{}^{\mF})\mcJ^{\mD\mE}\right]\Omega_{\mG\mF\mD}\Omega^{\mG}{}_{\mE\mK}\\
=3\left(\mP_{\mC}{}^{\mF}\mbrP_{\mA}{}^{\mD}\mbrP_{\mB}{}^{\mE}
+\mbrP_{\mC}{}^{\mF}\mP_{\mA}{}^{\mD}\mP_{\mB}{}^{\mE}\right)\Omega_{\mG[\mD\mE}\Omega^{\mG}{}_{\mF]\mK}+\left[(\mcP+\mbrcP)_{\mC\mA\mB}{}^{\mF\mD\mE}-\delta_{\mC}^{~\mF}\delta_{\mA}^{~\mD}
\delta_{\mB}^{~\mE}\right]\Omega_{\mG\mD\mE}\Omega^{\mG}{}_{\mF\mK}\\
=3\left(\mP_{\mC}{}^{\mF}\mbrP_{\mA}{}^{\mD}\mbrP_{\mB}{}^{\mE}
+\mbrP_{\mC}{}^{\mF}\mP_{\mA}{}^{\mD}\mP_{\mB}{}^{\mE}\right)
\left(f_{[\mD\mE}{}^{\mG}f_{\mF]\mK\mG}+\mpartial_{[\mD}f_{\mE\mF]\mK}-\textstyle{\frac{1}{3}}\mpartial_{\mK}f_{\mD\mE\mF}\right)\\
\quad+\left[(\mcP+\mbrcP)_{\mC\mA\mB}{}^{\mF\mD\mE}-\delta_{\mC}^{~\mF}\delta_{\mA}^{~\mD}
\delta_{\mB}^{~\mE}\right]\Omega_{\mG\mD\mE}\Omega^{\mG}{}_{\mF\mK}\,,
\ea
\ee
which follows  from (\ref{ffOO3}).     Eq.(\ref{delhcLmG})  can be then rewritten as
\be
\ba{l}
(\delta_{\mX}{-\hcL_{\mX}})\mGamma_{\mC\mA\mB}\\
=\left[(\mcP+\mbrcP)_{\mC\mA\mB}{}^{\mF\mD\mE}-\delta_{\mC}^{~\mF}\delta_{\mA}^{~\mD}
\delta_{\mB}^{~\mE}\right]\left(2\mD_{\mF}\mD_{[\mD}\mX_{\mE]}+\Omega_{\mG\mD\mE}\Omega^{\mG}{}_{\mF\mK}\mX^{\mK}\right)\\
\quad+2(\mcPp-\mcPq)_{\mC\mA\mB}{}^{\mF\mD\mE}
\left[(\mpartial_{\mG}\mP_{\mF\mD}+
\Omega_{\mG\mF}{}^{\mK}\mP_{\mK\mD}+\Omega_{\mG\mD}{}^{\mK}\mP_{\mF\mK})\mpartial^{\mG}\mX_{\mE}
+\mpartial_{\mG}\mP_{\mF\mD}\Omega^{\mG}{}_{\mE\mK}\mX^{\mK}\right]\\
\quad+2\mP_{[\mA}{}^{\mD}\mbrP_{\mB]}{}^{\mE}\left[(\mpartial_{\mG}\mP_{\mD\mE}+\Omega_{\mG\mD}{}^{\mK}\mP_{\mK\mE}+\Omega_{\mG\mE}{}^{\mK}\mP_{\mD\mK})
\mpartial^{\mG}\mX_{\mC}+\mpartial_{\mG}\mP_{\mD\mE}\Omega^{\mG}{}_{\mC\mK}\mX^{\mK}\right]\\
\quad+3\left(\mP_{\mC}{}^{\mF}\mbrP_{\mA}{}^{\mD}\mbrP_{\mB}{}^{\mE}
+\mbrP_{\mC}{}^{\mF}\mP_{\mA}{}^{\mD}\mP_{\mB}{}^{\mE}\right)
\left(f_{[\mD\mE}{}^{\mG}f_{\mF]\mK\mG}+\mpartial_{[\mD}f_{\mE\mF]\mK}-\textstyle{\frac{1}{3}}\mpartial_{\mK}f_{\mD\mE\mF}\right)\mX^{\mK}\\
\quad+\textstyle{\frac{2}{D-1}}(\mP_{\mC[\mA}\mP_{\mB]}{}^{\mE}+
\mbrP_{\mC[\mA}\mbrP_{\mB]}{}^{\mE})\left[\mpartial_{\mG}\mpartial^{\mG}\mX_{\mE}
+2\Omega_{\mG\mE\mK}\mpartial^{\mG}\mX^{\mK}
+(f_{\mG}-2\mpartial_{\mG}\md)(\mpartial^{\mG}\mX_{\mE}
+\Omega^{\mG}{}_{\mE\mK}\mX^{\mK})\right]\\
\quad+\textstyle{\frac{2}{D-1}}(\mP_{\mC[\mA}\mP_{\mB]}{}^{\mE}+
\mbrP_{\mC[\mA}\mbrP_{\mB]}{}^{\mE})
(\mpartial^{\mF}f_{\mF\mE\mK}+2\mpartial_{[\mE}f_{\mK]}
+2f^{\mF}\Omega_{[\mE\mK]\mF})\mX^{\mK}\,.
\ea
\ee
Hence, upon the twisted section conditions~(\ref{tsc1} -- \ref{tsc5}), we have a rather simple seminal expression, 
\be
(\delta_{\mX}{-\hcL_{\mX}})\mGamma_{\mC\mA\mB}
\equiv\left[(\mcP+\mbrcP)_{\mC\mA\mB}{}^{\mF\mD\mE}-\delta_{\mC}^{~\mF}\delta_{\mA}^{~\mD}
\delta_{\mB}^{~\mE}\right]\left(2\mD_{\mF}\mD_{[\mD}\mX_{\mE]}+\Omega_{\mG\mD\mE}\Omega^{\mG}{}_{\mF\mK}\mX^{\mK}\right)\,.
\ee
From this, the  diffeomorphic anomaly of the twisted semi-covariant derivative follows easily, 
\be
\ba{l}
(\delta_{\mX}-\hcL_{\mX})(\mna_{\mC}\mT_{\mA_{1}\cdots \mA_{n}})\\
=\mD^{\mE}\mX_{\mC}\mD_{\mE}\mT_{\mA_{1}\cdots \mA_{n}}
+\omega_{{\scriptscriptstyle{T\,}}}
\left[\mD_{C}\mD_{\mE}\mX^{\mE}-(\delta_{\mX}-\hcL_{\mX})
\mGamma^{\mE}{}_{\mE\mC}\right]\mT_{\mA_{1}\cdots \mA_{n}}\\
\quad+
\dis{\sum_{i=1}^{n}\,\left[2\mD_{\mC}\mD_{[\mA_{i}}X_{\mE]}+(\delta_{\mX}-\hcL_{\mX})
\mGamma_{\mC\mA_{i}\mE}\right]
\mT_{\mA_{1}\cdots\mA_{i-1}}{}^{\mE}{}_{\mA_{i+1}\cdots \mA_{n}}}\\
\equiv\omega_{{\scriptscriptstyle{T\,}}}
f_{\mK}\mD^{\mK}\mX_{\mC}\mT_{\mA_{1}\cdots \mA_{n}}+
\dis{\sum_{i=1}^{n}(\cP{+\brcP})_{\mC\mA_{i}}{}^{\mB\mF\mD\mE}
\left(2\mD_{\mF}\mD_{[\mD}\mX_{\mE]}+\Omega_{\mG\mD\mE}\Omega^{\mG}{}_{\mF\mK}\mX^{\mK}\right)\mT_{\cdots \mA_{i-1} \mB\mA_{i+1}\cdots}}\,.
\ea
\ee
Hence,    upon all the twistability conditions, finally we obtain
\be
(\delta_{\mX}-\hcL_{\mX})(\mna_{\mC}\mT_{\mA_{1}\cdots \mA_{n}})\equiv
\dis{\sum_{i=1}^{n}(\cP{+\brcP})_{\mC\mA_{i}}{}^{\mB}
\mT_{\mA_{1}\cdots \mA_{i-1} \mB\mA_{i+1}\cdots \mA_{n}}}\,,
\label{diffeoAnotwist}
\ee
where we have introduced  shorthand notations,
\be
\ba{l}
\mcP_{\mA\mB\mC}=
\mcP_{\mA\mB\mC}{}^{\mF\mD\mE}
(2\mD_{\mF}\mD_{[\mD}\mX_{\mE]}
+\Omega_{\mG\mD\mE}\Omega^{\mG}{}_{\mF\mK}\mX^{\mK})\,,\\
\mbrcP_{\mA\mB\mC}=
\mbrcP_{\mA\mB\mC}{}^{\mF\mD\mE}
(2\mD_{\mF}\mD_{[\mD}\mX_{\mE]}
+\Omega_{\mG\mD\mE}\Omega^{\mG}{}_{\mF\mK}\mX^{\mK})\,.
\ea
\ee
From (\ref{asymP6}) and (\ref{ffOO3}), they satisfy up to the  twistability conditions,
\be
\ba{ll}
\mcP_{[\mA\mB\mC]}\equiv 0\,,\quad&\quad
\mbrcP_{[\mA \mB\mC]}\equiv0\,.
\ea
\label{3asymP3}
\ee
Eq.(\ref{diffeoAnotwist})   immediately implies for the  spin connections,  
\be
\ba{l}
(\delta_{\mX}-\hcL_{\mX})
\mPhi_{\mA pq}=(\delta_{\mX}-\hcL_{\mX})(\mV^{\mB}{}_{p}\mna_{\mA}\mV_{\mB q})\equiv\mcP_{\mA pq}\,,\\
(\delta_{\mX}-\hcL_{\mX})
\mbrPhi_{\mA \brp\brq}=(\delta_{\mX}-\hcL_{\mX})(\mbrV^{\mB}{}_{\brp}\mna_{\mA}\mbrV_{\mB \brq})\equiv
\mbrcP_{\mA \brp\brq}\,.
\ea
\label{spinanomaltwist}
\ee
Although the final expressions of (\ref{diffeoAnotwist}) and (\ref{spinanomaltwist})  differ  in detail from what one would naively expect  by    `twisting' the results of (\ref{diffeoanormalous}) and (\ref{spinanomal}),\footnote{Twisting (\ref{diffeoanormalous}) or (\ref{spinanomal}) is naive, because they are not exact formulas. They   are  valid only  up to  the original section condition.} what remains still true  and crucial is that, once again \textit{the anomalies are all controlled by the index-six projection operators. Namely, they are still \underline{semi-covariant}.}   Thus, the cancellation mechanism is identical before and after the twist, and all the previous  completely covariant derivatives, (\ref{covT}),  (\ref{covT2}),  (\ref{Dpm}) and (\ref{covDirac}) are   still completely covariant after the twist. We recall them   exhaustively after  performing the twist,
\be
\ba{ll}
\mP_{\mC}{}^{\mD}{\mbrP}_{\mA_{1}}{}^{\mB_{1}}\cdots{\mbrP}_{\mA_{n}}{}^{\mB_{n}}
\mna_{\mD}\mT_{\mB_{1}\cdots \mB_{n}}\,,~&~
{\mbrP}_{\mC}{}^{\mD}\mP_{\mA_{1}}{}^{\mB_{1}}\cdots \mP_{\mA_{n}}{}^{\mB_{n}}
\mna_{\mD}\mT_{\mB_{1}\cdots \mB_{n}}\,,\\
\mP^{\mA\mB}{\brP}_{\mC_{1}}{}^{\mD_{1}}\cdots{\mbrP}_{\mC_{n}}{}^{\mD_{n}}\mna_{\mA}\mT_{\mB\mD_{1}\cdots \mD_{n}}\,,~&~
\mbrP^{\mA\mB}{\mP}_{\mC_{1}}{}^{\mD_{1}}\cdots{\mP}_{\mC_{n}}{}^{\mD_{n}}\mna_{\mA}\mT_{\mB\mD_{1}\cdots \mD_{n}}\quad~~(\mbox{divergences})\,,\\
\mP^{AB}{\mbrP}_{\mC_{1}}{}^{\mD_{1}}\cdots{\mbrP}_{\mC_{n}}{}^{\mD_{n}}
\mna_{\mA}\mna_{\mB}\mT_{\mD_{1}\cdots \mD_{n}}\,,~&~
{\mbrP}^{\mA\mB}\mP_{\mC_{1}}{}^{\mD_{1}}\cdots \mP_{\mC_{n}}{}^{\mD_{n}}
\mna_{\mA}\mna_{\mB}\mT_{\mD_{1}\cdots \mD_{n}}\quad(\mbox{Laplacians})\,,
\ea
\label{mcovT}
\ee
\be
\ba{llllll}
\mcD_{p}T_{\brq_{1}\cdots \brq_{n}}\,,\quad&
\mcD_{\brp}T_{q_{1}\cdots q_{n}}\,,\quad&
\mcD_{p}T^{p}{}_{\brq_{1}\cdots\brq_{n}}\,,\quad&
\mcD_{\brp}T^{\brp}{}_{q_{1}\cdots q_{n}}\,,\quad&
\mcD_{p}\mcD^{p}T_{\brq_{1}\cdots \brq_{n}}\,,\quad&
\mcD_{\brp}\mcD^{\brp}T_{q_{1}\cdots q_{n}}\,,
\ea
\label{mcovT2}
\ee
\be
\ba{lll}
\mcD_{+}\cT\,,\quad&\quad\mcD_{-}\cT\,,\quad&\quad\mcF=\mcD_{+}\cC\,,
\ea
\label{mcovDpm}
\ee
and
\be
\ba{lllll}
\gamma^{p}\mcD_{p}\rho\,,~~&~~
\gamma^{p}\mcD_{p}\psi_{\brp}\,,~~&~~\mcD_{\brp}\rho\,,~~&~~\mcD_{\brp}\psi^{\brp}\,,
~~&~~\brpsi^{\mA}\gamma_{p}(\mcD_{\mA}\psi_{\brq}-\half\mcD_{\brq}\psi_{\mA})\,,\\
\brgamma^{\brp}\mcD_{\brp}\rhop\,,~~&~~
\brgamma^{\brp}\mcD_{\brp}\psi^{\prime}_{p}\,,~~&~~
\mcD_{p}\rhop\,,~~&~~\mcD_{p}\psip{}^{p}\,,~&~
\brpsi^{\prime \mA}\brgamma_{\brp}(\mcD_{\mA}\psi^{\prime}_{q}-\half\mcD_{q}\psi^{\prime}_{\mA})\,.
\ea
\label{mcovDirac}
\ee
Now we turn to the curvatures.  The relations, (\ref{3asymP3}), (\ref{spinanomaltwist}),   give sequently, 
\be
\ba{ll}
(\delta_{\mX}-\hcL_{\mX})\msF_{\mA\mB pq}\equiv
2\mcD_{[\mA}\mcP_{\mB] pq}-(\mcP+\mbrcP)^{\mE}{}_{\mA \mB}\mPhi_{\mE pq}\,,\\
(\delta_{\mX}-\hcL_{\mX})\msbrF_{\mA\mB \brp\brq}\equiv
2\mcD_{[\mA}\mbrcP_{\mB] \brp\brq}-(\mcP+\mbrcP)^{\mE}{}_{\mA \mB}\mbrPhi_{\mE \brp\brq}\,,\\
(\delta_{\mX}-\hcL_{\mX})(\msF+\msbrF)_{\mA\mB \mC\mD}\equiv
2\mcD_{[\mA}(\mcP+\mbrcP)_{\mB] \mC\mD}-(\mcP+\mbrcP)^{\mE}{}_{\mA \mB}(\mPhi+\mbrPhi)_{\mE\mC\mD}\,,
\ea
\ee
and thus,  another  crucial result follows
\be
(\delta_{\mX}-\hcL_{\mX})\mcG_{\mA\mB \mC\mD}\equiv
\mcD_{[\mA}(\mcP+\mbrcP)_{\mB] \mC\mD}
+\mcD_{[\mC}(\mcP+\mbrcP)_{\mD] \mA\mB}\,.
\label{anomalcGtwist}
\ee
This shows that,   in an identical manner  to  the untwisted case~(\ref{anomalcurv}),  $\mcG_{\mA\mB \mC\mD}$ is  still \textit{semi-covariant   after the twist}. The completely covariant index-two (``Ricci")  and index-zero (scalar) twisted  curvatures are as untwisted cases, 
\be
\ba{llll}
\mcG_{pr\brq}{}^{r}\,,\quad&\quad
\mcG_{p\brr\brq}{}^{\brr}\,,\quad&\quad
\mcG_{pq}{}^{pq}\,,\quad&\quad
\mcG_{\brp\brq}{}^{\brp\brq}\,.
\ea
\label{aRicciScalartwist}
\ee
Finally we look into  $\mS_{\mA\mB\mC\mD\,}$.  It is straightforward to check that it is \textit{not}  semi-covariant.  It produces  additional  anomalous  terms  which  are not governed by the index-six projectors,\footnote{Putting the three relations  (\ref{varcGS}), (\ref{varU}) and (\ref{anomalcGtwist})  together, we conjecture   an equivalence relation, 
\[\mcL_{\mX}(\Omega_{\mE\mA\mB}\Omega^{\mE}{}_{\mC\mD})\equiv 
-2\mD_{[\mA}\left(\Omega^{\mE}{}_{\mB]\mK}\Omega_{\mE\mC\mD}\mX^{\mK}\right)
-2\mD_{[\mC}\left(\Omega^{\mE}{}_{\mD]\mK}\Omega_{\mE\mA\mB}\mX^{\mK}\right)\,,\]
of which a direct proof  is desirable.}
\be
\ba{l}
(\delta_{\mX}-\hcL_{\mX})\mS_{\mA\mB\mC\mD}\\
=\mna_{[A}(\delta_{\mX}-\hcL_{\mX})\mGamma_{\mB]\mC\mD}
+2\mna_{[\mA}\mD_{\mB]}\mD_{[\mC}X_{\mD]}
+\mD_{\mE}\mX_{[\mA}\mD^{\mE}\mGamma_{\mB]\mC\mD}~
+\Big[\,(\mA,\mB)\,\leftrightarrow\,(\mC,\mD)\,\Big]\\
\equiv \mna_{[A}(\mcP+\mbrcP)_{\mB]\mC\mD}
-\mD_{[\mA}\left(\Omega^{\mE}{}_{\mB]\mK}
\Omega_{\mE\mC\mD}\mX^{\mK}\right)~+\Big[\,(\mA,\mB)\,\leftrightarrow\,(\mC,\mD)\,\Big]\,.
\ea
\ee
We conclude that $\mS_{\mA\mB\mC\mD}$ is of no use in the twisted double field theory.  We discard it and keep  $\cG_{\mA\mB\mC\mD}$ only.

\item \textit{Identities which still hold after the twist.}

Straightforward yet useful implications of the twistability conditions include
\be
\ba{ll}
\mD_{\mA}\mD^{\mA}-2\mD_{\mA}\md\mD^{\mA}\equiv f^{\mA}\mD_{\mA}\equiv0\,,\quad&\quad
\mGamma^{\mC}{}_{p\brq}\mpartial_{\mC}\equiv0\,.
\ea
\ee

From  (\ref{BianchiS}), (\ref{follow}) and (\ref{relationcGStwist}), the  Bianchi identity of $\cG_{\mA\mB\mC\mD}$ is valid  upon  the twistability conditions, 
\be
\mcG_{\mA[\mB\mC\mD]}\equiv 0\,.
\ee
Further from (\ref{vanishingS}), it is straightforward to show
\be
\ba{lll}
\mcG_{pq\brr\brs}\equiv
\textstyle{\frac{3}{2}}\Omega_{\mE\mA[\mB}\Omega^{\mE}{}_{\mC\mD]}\mV^{\mA}{}_{p}\mV^{\mB}{}_{q}\mbrV^{\mC}{}_{\brr}\mbrV^{\mD}{}_{\brs}\equiv0\,,\quad&\quad
\mcG_{p\brq r\brs}\equiv 0\,,\quad&\quad
\mcG_{pr\brq}{}^{r}\equiv\mcG_{p\brr\brq}{}^{\brr}\equiv\half\mcG_{p\brq}\,,
\ea
\label{vanishingG}
\ee
and notably,
\be
\mcG_{pq}{}^{pq}+\mcG_{\brp\brq}{}^{\brp\brq}\equiv\textstyle{\frac{1}{6}}
f_{\mA\mB\mC}f^{\mA\mB\mC}\,.
\label{GGfsquare}
\ee
That is to say, replacing $\mS_{\mA\mB\mC\mD}$ by $\mcG_{\mA\mB\mC\mD}$,  almost all the properties of the four-index curvature~(\ref{vanishingS}) still hold after the twist,  up to the twistability conditions. The only  exception is  (\ref{GGfsquare}) and this is  also  crucial.

The  relations between the completely covariant curvatures and the completely covariant derivatives~(\ref{vectorREP}), (\ref{GG1}), (\ref{GG2}) still hold after the twist, 
\be
\ba{ll}
\half[\mcD_{p},\mcD_{\brq}]T^{p}\equiv\mcG_{pr\brq}{}^{r}T^{p}\,,\quad&\quad
\half[\mcD_{p},\mcD_{\brq}]T^{\brq}\equiv-\mcG_{p\brr\brq}{}^{\brr}T^{\brq}\,,\\
{}[\gamma^{p}\mcD_{p},\mcD_{\brq}]\varepsilon\equiv\mcG_{pr\brq}{}^{r}\gamma^{p}\varepsilon\,,\quad&\quad
{}[\mcD_{p},\brgamma^{\brq}\mcD_{\brq}]\varepsilon^{\prime}\equiv-\mcG_{p\brr\brq}{}^{\brr}\gamma^{\brq}\varepsilonp\,,\\
(\gamma^{p}\mcD_{p})^{2}\varepsilon
+\mcD_{\brp}\mcD^{\brp}\varepsilon\equiv-\quarter\mcG_{pq}{}^{pq}\varepsilon\,,\quad&\quad
(\brgamma^{\brp}\mcD_{\brp})^{2}\varepsilonp
+\mcD_{p}\mcD^{p}\varepsilonp\equiv-\quarter\mcG_{\brp\brq}{}^{\brp\brq}\varepsilonp\,.
\label{GGtwist}
\ea
\ee
But, in contrast to  (\ref{nilpotent}), we  get after the twist,
\be
(\mcD_{\pm})^{2}\cT\equiv-\textstyle{\frac{1}{24}}
f_{\mA\mB\mC}f^{\mA\mB\mC}\cT\,.
\label{twistnilpotent}
\ee
This indicates that in addition to the twistability conditions,  in the presence of the R-R sector,  in order to ensure the nilpotency of the differential operators, $\mcD_{\pm}$, which should  define   the twisted R-R gauge symmetry or the `twisted R-R cohomology' consistently,  we should separately  impose
\be
f_{\mA\mB\mC}f^{\mA\mB\mC}\equiv 0\,.
\label{fsquare}
\ee
For a relevant  previous work we refer the readers to \cite{Geissbuhler:2013uka} where the R-R sector was treated as an $\Ott$ spinor which can be related to our treatment after the diagonal gauge fixing of the twofold local Lorentz symmetries~\cite{Jeon:2012kd}. We shall see shortly  that this extra condition is also required for the supersymmetric invariance of the  twisted maximal  SDFT.\newpage

\item \textit{``Effective connection" and internal coordinate independence.}

Writing explicitly,
\be
\ba{ll}
\mD_{\mA}=\mpartial_{\mA}+\Omega_{\mA}\,,\quad&\quad
\mna_{\mA}=\mpartial_{\mA}
+\Omega_{\mA}+\mGamma_{\mA}\,,
\ea
\ee
it may appear plausible   to  view $\Omega_{\mA}+\mGamma_{\mA}$ as the ``effective connection''.   Upon the twistability  conditions~(\ref{tsc1})  -- (\ref{tsc5}), this  ``effective connection'' reads explicitly,  \textit{c.f.~}\cite{Berman:2013cli},
\be
\ba{ll}
\Omega_{\mC\mA\mB}+\mGamma_{\mC\mA\mB}\equiv
&2(\mP\mpartial_{\mC}\mP\mbrP)_{[\mA\mB]}
+2({{\mbrP}_{[\mA}{}^{\mD}{\mbrP}_{\mB]}{}^{\mE}}-{\mP_{[\mA}{}^{\mD}\mP_{\mB]}{}^{\mE}})\mpartial_{\mD}\mP_{\mE\mC}\\
{}&-\textstyle{\frac{4}{D-1}}(\mbrP_{\mC[\mA}\mbrP_{\mB]}{}^{\mD}+\mP_{\mC[\mA}\mP_{\mB]}{}^{\mD})\left(\mpartial_{\mD}\md
+(\mP\mpartial^{\mE}\mP\mbrP)_{[\mE\mD]}\right)\\
{}&+(\mbrP_{\mC}{}^{\mD}\mP_{\mA}{}^{\mE}\mP_{\mB}{}^{\mF}+\mP_{\mC}{}^{\mD}\mbrP_{\mA}{}^{\mE}\mbrP_{\mB}{}^{\mF})f_{\mD\mE\mF}\\
{}&+(\mcP+\mbrcP)_{\mC\mA\mB}{}^{\mD\mE\mF}\Omega_{\mD\mE\mF}\,.
\ea
\label{Gammat2}
\ee
In particular, we have
\be
\ba{ll}
\mV^{\mA}{}_{p}\mV^{\mB}{}_{q}\mV^{\mC}{}_{r}(\Omega_{[\mA\mB\mC]}+\mGamma_{[\mA\mB\mC]})=\textstyle{\frac{1}{3}}
f_{pqr}\,,\quad&\quad 
\mbrV^{\mA}{}_{\brp}\mbrV^{\mB}{}_{\brq}\mbrV^{\mC}{}_{\brr}(\Omega_{[\mA\mB\mC]}+\mGamma_{[\mA\mB\mC]})=\textstyle{\frac{1}{3}}f_{\brp\brq\brr}\,.
\ea
\label{Gammat2pr}
\ee
We may view the  last two lines  of (\ref{Gammat2}) as  ``effective torsions''. They satisfy the desired properties~(\ref{torsion}). In this ``effective" point of view, the ``torsion" should be  defined from the  difference, $\hcL_{X}(\mna)-\hcL_{X}(\mpartial)$,  instead of the trivial one~(\ref{twtorsionless2}),  $\hcL_{X}(\mna)-\hcL_{X}(\mD){=0}$.  For further related discussion, we refer readers to  section 3.4.2 of \cite{Coimbra:2011nw}.

We also note that only the last line, \textit{i.e.~}the six-index projection of $\Omega_{\mD\mE\mF}$, may depend on the internal coordinates,  which could be problematic. However, as seen  in (\ref{Gammat2pr}),  it is easy to see that this potentially dangerous internal coordinate  dependency disappears   thoroughly       inside the completely covariant derivatives, listed in   (\ref{mcovT}), (\ref{mcovT2}), (\ref{mcovDpm}) and  (\ref{mcovDirac}).  This  cancellation further  implies  that the completely covariant curvatures~(\ref{aRicciScalartwist})  are also independent of the internal coordinates, since the completely covariant curvatures can be constructed from  the quadratic completely covariant differential operators,  see (\ref{GGtwist}) with (\ref{mcovDirac}).

The above expression of the effective connection~(\ref{Gammat2}) is also comparable to the  torsionful  connection proposed   by Berman and Lee in \cite{Berman:2013cli}.  With the intention of handling  the twisted generalized Lie derivative~\cite{Grana:2012rr,Berman:2013cli}, \textit{i.e.~}(\ref{brtcL3}),  they introduced torsions by clever guess work.   Their  torsionful  connection  differs from our effective connection~(\ref{Gammat2}). Yet,  it nevertheless satisfies (\ref{Gammat2pr}) and  the difference amounts to  certain  six-index projection terms. Accordingly,  their proposal is  practically  consistent  with our result.   A  novel contribution  of the present work is to derive the effective connection~(\ref{Gammat2}) straightforwardly by applying   the U-twisting ansatz~(\ref{pushback}) to the semi-covariant formalism, without any   ambiguity.

\end{itemize}

%%%%%%%%%%%%%%%%%%%%%%%%%%%%%%%%%%%%%%%%%%%%%%%%%%%
%%%%%%%%%%%%%%%%%%%%%%%%%%%%%%%%%%%%%%%%%%%%%%%%%%%
\section{Twisted supersymmetric double field theory\label{secSDFTtwist}}
Here we present explicitly   half-maximal (\textit{i.e.~}sixteen) and maximal (\textit{i.e.~}thirty two) supersymmetric gauged double field theories as the twists  of the previously constructed  $\cN=1$ and $\cN=2$,  $D=10$ supersymmetric double field theories. All the fields satisfy the twistability conditions, (\ref{tsc1}) --( \ref{tsc5}), and in the case of  the maximal supersymmetric twist, one extra condition, \textit{i.e.}~(\ref{fsquare}) must be also met in order to ensure both the  R-R gauge symmetry and the  $32$ supersymmetries unbroken.

\subsection{Half-maximal supersymmetric gauged double field theory} 
After replacing $S_{ABCD}$ by $\cG_{ABCD}$ and adding  section-condition-vanishing purely bosonic terms of (\ref{cGpluscG}),  
we twist the  $\cN=1$ $D=10$  SDFT which was constructed in   \cite{Jeon:2011sq} to the full order in fermions.   The twist leads to a  half-maximal supersymmetric gauged double field theory  of which the  Lagrangian is
\be
\mcL_{\scriptscriptstyle{\rm Twisted~SDFT}}^{\scriptscriptstyle{\rm Half-maximal}}
=e^{-2\md}\Big[\textstyle{\frac{1}{4}}\mcG_{pq}{}^{pq}
+i\half\brrho \gamma^{p}\mcD_{p}\rho
-i\brpsi^{\brp}\mcD_{\brp}\rho
-i\half\brpsi^{\brp}\gamma^{q}\mcD_{q}\psi_{\brp}
\Big]\,.
\label{halfmaximal}
\ee
Each   term  in the Lagrangian  is completely covariant with respect to    the twisted diffeomorphisms, (\ref{brtcL}) or  (\ref{brtcL3}), the  ${\Spint\times\oSpint}$ local Lorentz symmetries, and a subgroup of $\Ott$  which preserves the structure constant, $f_{\mA\mB\mC\,}$. Being completely covariant, each term is also independent of the internal coordinates.

The  leading order half-maximal (\textit{i.e.~}sixteen) twisted supersymmetry transformation rules are,  for the twisted bosons,
\be
\ba{lll}
\deltaS  \md=-i\half\brvarepsilon\rho\,,\quad&\quad
\deltaS \mV_{Ap}=-i\mbrV_{A}{}^{\brq}\bar{\varepsilon}\gamma_{p}\psi_{\brq}\,,\quad&\quad
\deltaS \mbrV_{A\brp}=+i\mV_{A}{}^{q}
\bar{\varepsilon}\gamma_{q}\psi_{\brp}\,,
\ea
\label{halfsusyB}
\ee
and for the `untwisted'  fermions,
\be
\ba{ll}
\deltaS  \rho= -\gamma^{p}\mcD_{p}\varepsilon\,,\quad&\quad
\deltaS  \psi_{\brp}=\mcD_{\brp}\varepsilon\,.
\ea
\label{halfsusyF}
\ee
The supersymmetry works, as the induced leading order variation of the Lagrangian vanishes, up to  total derivatives and the  twistability conditions, thanks to (\ref{GGtwist}), 
\be
\deltaS\mcL_{\scriptscriptstyle{\rm Twisted~SDFT}}^{\scriptscriptstyle{\rm Half-maximal}}\equiv
-ie^{-2\md}\brrho\left[(\gamma^{p}\mcD_{p})^{2}
+\mcD_{\brp}\mcD^{\brp}+\quarter\mcG_{pq}{}^{pq}\right]\varepsilon+ie^{-2\md}\brpsi^{\brp}\left[
\mcG_{\brp rq}{}^{r}\gamma^{q}+[\mcD_{\brp},\gamma^{q}\mcD_{q}]\right]\varepsilon\equiv0\,.
\ee
As discussed at the end of section~\ref{SECUderivative},  the leading order supersymmetric invariance is sufficient to guarantee the full order completion. Outsourcing from the full order untwisted  $\cN=1$ $D=10$ SDFT~\cite{Jeon:2011sq}, we only need to add the  quartic   fermions therein  to the twisted  Lagrangian~(\ref{halfmaximal})  and the cubic  fermions to the twisted  supersymmetry transformation rules for the fermions~(\ref{halfsusyF}).

As in the untwisted SDFT~\cite{Jeon:2011sq},  the conventional  Rarita-Schwinger  term is forbidden, and this is due to the hybrid nature of the  gravitino indices, $\psi_{\brp}^{\alpha}$: one $\oSpint$ vectorial and the other $\Spint$ spinorial. Simply they cannot be mixed.  Nonetheless, the $\cN=1$ $D=10$ SDFT reduces  consistently  to the minimal supergravity in ten-dimensions after the diagonal gauge fixing, $\Spint\times\oSpint\rightarrow\Spint_{D\,}$, see the appendix of \cite{Jeon:2011sq} for  details.

It is worth while to note from the $Z_{2}$ symmetry which  exchanges the two spin groups, $\Spint\leftrightarrow\oSpint$, there is a parallel formulation of the half-maximal SDFT, 
\be
\overline{{\mcL}_{\scriptscriptstyle{\rm Twisted~SDFT}}^{\scriptscriptstyle{\rm Half-maximal}}}
=e^{-2\md}\Big[-\textstyle{\frac{1}{4}}\mcG_{\brp\brq}{}^{\brp\brq}
-i\half\brrhop \brgamma^{\brp}\mcD_{\brp}\rhop
+i\brpsip{}^{p}\mcD_{p}\rhop
+i\half\brpsip{}^{p}\brgamma^{\brq}\mcD_{\brq}\psip_{p}\Big]\,.
\label{barredhalfmaximal}
\ee
The supersymmetry is realized by
\be
\ba{lllll}
\deltaS  \md=-i\half\brvarepsilonp\rhop\,,~&
\deltaS \mV_{\mA p}=+i\brvarepsilonp\bar{\gamma}_{\mA}\psi^{\prime}_{p}
\,,~&
\deltaS \mbrV_{\mA\brp}=-i\brvarepsilonp\bar{\gamma}_{\brp}\psi^{\prime}_{\mA}\,,~&
\deltaS  \rhop= -\bar{\gamma}^{\brp}\mcD_{\brp}\varepsilonp\,,~&
\deltaS  \psi^{\prime}_{p}=\mcD_{p}\varepsilonp\,.
\ea
\label{brhalfmaximalsusyF}
\ee

%%%%%%%%%%%%%%%%%%%%%%%%%%%%%%%%%%%%%%%%%%%%%%%%%%%

\subsection{Maximal supersymmetric gauged double field theory} 
The  twisting  of the  $\cN=2$ $D=10$  SDFT which was constructed  to the full order in fermions in   \cite{Jeon:2012hp}, leads to the following  maximal supersymmetric gauged double field theory  Lagrangian, 
\be
\ba{l}
\mcL_{\scriptscriptstyle{\rm Twisted~SDFT}}^{\scriptscriptstyle{\rm Maximal}}=e^{-2\md}\Big[~\textstyle{\frac{1}{8}}(\mcG_{pq}{}^{pq}-\mcG_{\brp\brq}{}^{\brp\brq})+\half\Tr(\mcF\bar{\mcF})-i\brrho\mcF\rhop+i\brpsi_{\brp}\gamma_{q}
\mcF\brgamma^{\brp}\psip{}^{q}\\
\quad\qquad\qquad\qquad+i\half\brrho \gamma^{p}\mcD_{p}\rho
-i\brpsi^{\brp}\mcD_{\brp}\rho
-i\half\brpsi^{\brp}\gamma^{q}\mcD_{q}\psi_{\brp}
-i\half\brrhop \brgamma^{\brp}\mcD_{\brp}\rhop
+i\brpsip{}^{p}\mcD_{p}\rhop
+i\half\brpsip{}^{p}\brgamma^{\brq}\mcD_{\brq}\psip_{p}~
\Big]\,.
\ea
\label{mtwistSDFT}
\ee
As in the half-maximal case~(\ref{halfmaximal}), each  term  in the Lagrangian  is independent of the internal coordinates, and is  completely covariant with respect to    the twisted diffeomorphisms, the  ${\Spint\times\oSpint}$ local Lorentz symmetries, the structure constant preserving  subgroup of $\Ott$, and further the  R-R gauge symmetry  provided the extra condition of (\ref{fsquare}), 
\be
\ba{lll}
\delta\cC=\mcD_{+}\Lambda&\quad\longrightarrow\quad&
\delta\mcF=(\mcD_{+})^{2}\Lambda\equiv -\textstyle{\frac{1}{24}}f_{\mA\mB\mC}f^{\mA\mB\mC}\Lambda\equiv0\,.
\ea
\ee
The leading order maximal (\textit{i.e.~}thirty two)  twisted  supersymmetry transformation rules are, for the bosons,
\be
\ba{lll}
\deltaS  \md=-i\half(\brvarepsilon\rho+\brvarepsilonp\rhop)\,,\quad&\quad
\deltaS \mV_{\mA p}=i\mbrV_{\mA}{}^{\brq}(\brvarepsilonp\bar{\gamma}_{\brq}\psi^{\prime}_{p}
-\bar{\varepsilon}\gamma_{p}\psi_{\brq})\,,\quad&\quad
\deltaS \mbrV_{\mA\brp}=i\mV_{\mA}{}^{q}(
\bar{\varepsilon}\gamma_{q}\psi_{\brp}-\brvarepsilonp\bar{\gamma}_{\brp}\psi^{\prime}_{q})\,,\\
\multicolumn{3}{c}{
\deltaS  \cC= i \half 
(\gamma^{p}\varepsilon \brpsi^{\prime}_{p}-\varepsilon\brrhop
-\psi_{\brp}\brvarepsilonp\bar{\gamma}^{\brp}+\rho\brvarepsilonp)
+\cC\deltaS \md 
-\half(\mbrV^{\mA}{}_{\brq\,}\deltaS  \mV_{\mA p})\gamma^{(11)}\gamma^{p}\cC \bar{\gamma}^{\brq}\,,}
\ea
\label{maximalsusyB}
\ee
and for the fermions,
\be
\ba{llll}
\deltaS  \rho= -\gamma^{p}\mcD_{p}\varepsilon\,,\quad&\quad
\deltaS  \rhop= -\bar{\gamma}^{\brp}\mcD_{\brp}\varepsilonp\,,\quad&\quad
\deltaS  \psi_{\brp}=\mcD_{\brp}\varepsilon+\mcF\brgamma_{\brp} \varepsilonp\,,\quad&\quad
\deltaS  \psi^{\prime}_{p}=\mcD_{p}\varepsilonp + \bar{\mcF}\gamma_{p}\varepsilon\,.
\ea
\label{maximalsusyF}
\ee
Ignoring  total derivatives and  up to the twistability conditions,  the  supersymmetric infinitesimal variation  of the Lagrangian is, from 
(\ref{vanishingG}), (\ref{GGfsquare}), (\ref{GGtwist}), (\ref{twistnilpotent}) and the appendices  of  \cite{Jeon:2012hp},  
\be
\ba{lll}
\deltaS\mcL_{\scriptscriptstyle{\rm Twisted~SDFT}}^{\scriptscriptstyle{\rm Maximal}}&\equiv& 
i\textstyle{\frac{1}{8}}e^{-2\md}(\brrho\varepsilon-\brrhop\varepsilonp)(\mcG_{pq}{}^{pq}+\mcG_{\brp\brq}{}^{\brp\brq})
-i\half e^{-2\md}(\brpsi^{\brq}\gamma^{p}\varepsilon
+\brpsip{}^{p}\brgamma^{\brq}\varepsilonp)(\mcG_{pr\brq}{}^{r}-\mcG_{p\brr\brq}{}^{\brr})\\
{}&{}&+i\half e^{-2\md}\Tr\left[(
\rhop\brvarepsilon+\psip_{p}\brvarepsilon\gamma^{p}+
\varepsilonp\brrho+\brgamma^{\brp}\varepsilonp\brpsi_{\brp})(\cD_{+})^{2}\cC\right]\\
{}&{}&+i\textstyle{\frac{1}{8}} e^{-2d}(
\brvarepsilon\gamma_{p}\psi_{\brq}
-\brvarepsilonp\brgamma_{\brq}\psip_{p})
\Tr\left(\gamma^{p}\mcF_{-}\brgamma^{\brq}\overline{\mcF_{-}}\,\right)\\
{}&\equiv&
i\textstyle{\frac{1}{48}}e^{-2\md}
\left(\brrho\varepsilon-\brrhop\varepsilonp+
\brvarepsilon\cC\rhop+\brvarepsilon\gamma^{p}\cC\psip_{p}+
\brrho\cC\varepsilonp+\brpsi_{\brp}\cC\brgamma^{\brp}\varepsilonp
\right)\times f_{\mA\mB\mC}f^{\mA\mB\mC}\\
{}&{}&+i\textstyle{\frac{1}{8}} e^{-2d}(
\brvarepsilon\gamma_{p}\psi_{\brq}
-\brvarepsilonp\brgamma_{\brq}\psip_{p})
\Tr\left(\gamma^{p}\mcF_{-}\brgamma^{\brq}\overline{\mcF_{-}}
\,\right)\,.
\ea
\ee
Here $\mcF_{-}$ denotes the (leading order) self-dual part of the R-R field strength, \textit{c.f.~}(\ref{SD1}), 
\be
\mcF_{-}:=(1-\gamma^{\eleven})\mcF\,.
\ee
Thus, requiring the extra condition~(\ref{fsquare}) which we recall here,
\be
f_{\mA\mB\mC}f^{\mA\mB\mC}\equiv0\,,
\ee 
the action is supersymmetric invariant modulo the self-duality,\footnote{For consistency , the supersymmetric  variation  of the self-duality  relation is, even in the full order supersymmetric completion, precisely  closed by  the gravitino  equations of motion~\cite{Jeon:2012hp}.}  up to surface integrals.   
%%%
%%\be
%%\deltaS\mcL_{\scriptscriptstyle{\rm Twisted~SDFT}}^{\scriptscriptstyle{\rm Maximal}}\equiv
%%i\textstyle{\frac{1}{8}} e^{-2d}(\brvarepsilon\gamma_{p}\psi_{\brq}
%%-\brvarepsilonp\brgamma_{\brq}\psip_{p})\Tr\left(\gamma^{p}\mcF_{-}\brgamma^{\brq}\overline{\mcF_{-}}
%%\,\right)\,.
%%\ee
%%%
Once again,  the leading order supersymmetric invariance  guarantees  the full order completion.

\subsection{Explicit comparison with the untwisted case}   
To compare with the untwisted DFT and  to identify the newly added terms after the U-twist, we dismantle the U-derivatives, $\mD_{\mA}$, explicitly and obtain up to the twistability conditions, 
\be
\ba{ll}
+\mcG_{pq}{}^{pq}\equiv&
{\textstyle{\frac{1}{16}}}\mcH^{\mA\mB}\mpartial_{\mA}\mcH_{\mC\mD}\mpartial_{\mB}\mcH^{\mC\mD}
+\quarter\mcH^{\mA\mB}\mpartial^{\mC}\mcH_{\mA\mD}\mpartial^{\mD}\mcH_{\mB\mC}
-\half\mpartial_{\mA}\mpartial_{\mB}\mcH^{\mA\mB}\\
{}&-2\mcH^{\mA\mB}\mpartial_{\mA}\md\mpartial_{\mB}\md
+2\mcH^{\mA\mB}\mpartial_{\mA}\mpartial_{\mB}\md
+2\mpartial_{\mA}\mcH^{\mA\mB}\mpartial_{\mB}\md\\
{}&+\textstyle{\frac{1}{8}}f_{\mA\mB\mC}f^{\mA\mB}{}_{\mD}\mcH^{\mC\mD}
-{\textstyle{\frac{1}{24}}}f_{\mA\mB\mC}f_{\mD\mE\mF}\mcH^{\mA\mD}\mcH^{\mB\mE}\mcH^{\mC\mF}
-\quarter f_{\mA\mB\mC}\mcH^{\mB\mD}\mcH^{\mC\mE}\mpartial_{\mD}\mcH_{\mE}{}^{\mA}\\
{}& +\textstyle{\frac{1}{12}}f_{\mA\mB\mC}f^{\mA\mB\mC}\,,\\

-\mcG_{\brp\brq}{}^{\brp\brq}\equiv&{\textstyle{\frac{1}{16}}}\mcH^{\mA\mB}\mpartial_{\mA}\mcH_{\mC\mD}\mpartial_{\mB}\mcH^{\mC\mD}
+\quarter\mcH^{\mA\mB}\mpartial^{\mC}\mcH_{\mA\mD}\mpartial^{\mD}\mcH_{\mB\mC}
-\half\mpartial_{\mA}\mpartial_{\mB}\mcH^{\mA\mB}\\
{}&-2\mcH^{\mA\mB}\mpartial_{\mA}\md\mpartial_{\mB}\md
+2\mcH^{\mA\mB}\mpartial_{\mA}\mpartial_{\mB}\md
+2\mpartial_{\mA}\mcH^{\mA\mB}\mpartial_{\mB}\md\\
{}&+\textstyle{\frac{1}{8}}f_{\mA\mB\mC}f^{\mA\mB}{}_{\mD}\mcH^{\mC\mD}
-{\textstyle{\frac{1}{24}}}f_{\mA\mB\mC}f_{\mD\mE\mF}\mcH^{\mA\mD}\mcH^{\mB\mE}\mcH^{\mC\mF}
-\quarter f_{\mA\mB\mC}\mcH^{\mB\mD}\mcH^{\mC\mE}\mpartial_{\mD}\mcH_{\mE}{}^{\mA}\\
{}& -\textstyle{\frac{1}{12}}f_{\mA\mB\mC}f^{\mA\mB\mC}\,.
\ea
\label{mcGtwo}
\ee
It follows  that  the sum, $\mcG_{pq}{}^{pq}+\mcG_{\brp\brq}{}^{\brp\brq}=\textstyle{\frac{1}{6}}f_{\mA\mB\mC}f^{\mA\mB\mC}$,  indeed  gives  (\ref{GGfsquare}),  and the difference reads
\be
\ba{l}
\mcG_{pq}{}^{pq}-\mcG_{\brp\brq}{}^{\brp\brq}\\
{}\equiv
{\textstyle{\frac{1}{8}}}\mcH^{\mA\mB}\mpartial_{\mA}\mcH_{\mC\mD}\mpartial_{\mB}\mcH^{\mC\mD}
+\half\mcH^{\mA\mB}\mpartial^{\mC}\mcH_{\mA\mD}\mpartial^{\mD}\mcH_{\mB\mC}
-\mpartial_{\mA}\mpartial_{\mB}\mcH^{\mA\mB}\\
{}\quad~~-4\mcH^{\mA\mB}\mpartial_{\mA}\md\mpartial_{\mB}\md
+4\mcH^{\mA\mB}\mpartial_{\mA}\mpartial_{\mB}\md
+4\mpartial_{\mA}\mcH^{\mA\mB}\mpartial_{\mB}\md\\
{}\quad~~+\textstyle{\frac{1}{4}}f_{\mA\mB\mC}f^{\mA\mB}{}_{\mD}\mcH^{\mC\mD}
-{\textstyle{\frac{1}{12}}}f_{\mA\mB\mC}f_{\mD\mE\mF}\mcH^{\mA\mD}\mcH^{\mB\mE}\mcH^{\mC\mF}
-\half f_{\mA\mB\mC}\mcH^{\mB\mD}\mcH^{\mC\mE}\mpartial_{\mD}\mcH_{\mE}{}^{\mA}\,.
\ea
\label{mcGdiff}
\ee
In the above, \textit{i.e.~}(\ref{mcGtwo}) and (\ref{mcGdiff}),  the first two lines on the right hand sides essentially  correspond to  the original untwisted DFT Lagrangian~\cite{Hohm:2010pp}~\textit{i.e.~}(\ref{SgH}) written in terms of the generalized metric. The third line then matches with the literature~\cite{Hohm:2011ex,Geissbuhler:2011mx,Aldazabal:2011nj,Grana:2012rr,Berman:2013cli}. The last lines  in (\ref{mcGtwo}) correspond to the DFT cosmological constant~\cite{Jeon:2011cn} which is  apparently  the special feature  of the  \textit{half-maximal} supersymmetric DFT~\cite{Geissbuhler:2011mx,Aldazabal:2011nj,Grana:2012rr,Berman:2013cli}.  Depending on the choice of $+\mcG_{pq}{}^{pq}$ or $-\mcG_{\brp\brq}{}^{\brp\brq}$ we may freely fix the sign of it.\\

It is further worth while to note
\be
\mcF=\mcD_{+}\cC=\left.\mcD_{+}\cC\right|_{\mD}\equiv\left.\mcD_{+}\cC\right|_{\mpartial}+
\textstyle{\frac{1}{12}}f_{pqr}\gamma^{pqr}\cC-\quarter f_{p\brq\brr}\gamma^{p}\cC\brgamma^{\brq\brr}
-\textstyle{\frac{1}{12}}f_{\brp\brq\brr}\gamma^{\eleven}\cC\brgamma^{\brp\brq\brr}+\quarter f_{pq\brr}\gamma^{\eleven}\gamma^{pq}\cC\brgamma^{\brr}\,,
\label{twcF}
\ee
\be
\ba{l}
 \gamma^{p}\mcD_{p}\rho=\left.\gamma^{p}\mcD_{p}\rho\right|_{\mD}\equiv\left.\gamma^{p}\mcD_{p}\rho\right|_{\mpartial}+\textstyle{\frac{1}{12}}f_{pqr}\gamma^{pqr}\rho\,,\\
 \mcD_{\brp}\rho=\left.\mcD_{\brp}\rho\right|_{\mD}\equiv\left.\mcD_{\brp}\rho\right|_{\mpartial}+\textstyle{\frac{1}{4}}f_{\brp qr}\gamma^{qr}\rho\,,\\
 \gamma^{q}\mcD_{q}\psi_{\brp}=\left.\gamma^{q}\mcD_{q}\psi_{\brp}\right|_{\mD}\equiv\left.\gamma^{q}\mcD_{q}\psi_{\brp}\right|_{\mpartial}+\textstyle{\frac{1}{12}}f_{qrs}\gamma^{qrs}\psi_{\brp}+f_{r\brp\brq}\gamma^{r}\psi^{\brq}\,,
 \ea
 \label{twFK}
 \ee
 and
 \be
 \ba{l}
 \brgamma^{\brp}\mcD_{\brp}\rhop=\left.\brgamma^{\brp}\mcD_{\brp}\rhop\right|_{\mD}\equiv\left.\brgamma^{\brp}\mcD_{\brp}\rhop\right|_{\mpartial}+\textstyle{\frac{1}{12}}f_{\brp\brq\brr}\brgamma^{\brp\brq\brr}\rhop\,,\\
 \mcD_{p}\rhop=\left.\mcD_{p}\rhop\right|_{\mD}\equiv\left.\mcD_{p}\rhop\right|_{\mpartial}+\textstyle{\frac{1}{4}}f_{p\brq\brr}\brgamma^{\brq\brr}\rhop\,,\\
 \brgamma^{\brq}\mcD_{\brq}\psip_{p}=\left.\brgamma^{\brq}\mcD_{\brq}\psip_{p}\right|_{\mD}\equiv
 \left.\brgamma^{\brq}\mcD_{\brq}\psip_{p}\right|_{\mpartial}+\textstyle{\frac{1}{12}}f_{\brq\brr\brs}\brgamma^{\brq\brr\brs}\psip_{p}
 +f_{\brr pq}\brgamma^{\brr}\psip{}^{q}\,.
\ea
\label{twFKp}
\ee
As expected from  the consistency of the ``effective connection", (\ref{twFK}) and (\ref{twFKp}) agree with Berman and Lee~\cite{Berman:2013cli}, while (\ref{twcF}) is a new result we report in this work.\\

%%%%%%%%%%%%%%%%%%%%%%%%%%%%%%%%%%%%%%%%%%%%%%%%%%%
\section{Discussion\label{secDiscussion}}
In this paper, we  have  successfully twisted  the semi-covariant formulations of the $\cN=2$ and the $\cN=1$, $D=10$ SDFT constructed in \cite{Jeon:2011sq,Jeon:2012hp}, and systematically   derived  the gauged maximal and half-maximal  supersymmetric double field theories, (\ref{halfmaximal}) (\ref{barredhalfmaximal}), (\ref{mtwistSDFT}), along with their  supersymmetry transformation rules, (\ref{halfsusyB}), (\ref{halfsusyF}), (\ref{maximalsusyB}), (\ref{barredhalfmaximal}),  (\ref{maximalsusyF}).    Our derivation is systematic in the sense that, we  only applied the twisting ansatz~(\ref{pushback}) to the untwisted SDFT of \cite{Jeon:2011sq,Jeon:2012hp}, and then without any ambiguity the gauged   supersymmetric double field theories were straightforwardly derived. Further, just   like the untwisted SDFT yet now subject to   the twistability conditions, (\ref{tsc1})  -- (\ref{tsc5}) and also   (\ref{fsquare}) for the maximal supersymmetric twist,   each term in the constructed Lagrangian is completely covariant. Namely, the NS-NS curvature term,  the fermionic kinetic terms and the R-R kinetic term  are  all completely covariant,   with respect to the twisted diffeomorphisms, the $\Spint\times\oSpint$ local Lorentz symmetries, the R-R gauge symmetry for the maximal case, and  a subgroup of $\Ott$  which preserves the structure constant.   The twofold  Lorentz symmetries are `local' with respect to the dimensionally reduced external spacetime.   The  twisted and hence   gauged  SDFTs are   completely fixed  by  requiring the  supersymmetry to be  unbroken,  in the precisely same manner as the untwisted SDFTs.

The nilpotency of the twisted R-R cohomology differential operators~(\ref{twistnilpotent}), (\ref{fsquare}), implies the Bianchi identity for the twisted R-R flux,
\be
\mcD_{+}\mcF=(\mcD_{+})^{2}\cC\equiv0\,.
\ee
As demonstrated   in the section~4.3 of \cite{Jeon:2012kd}, one may take the diagonal gauge fixing of the  local Lorentz symmetry, expand the R-R potential in terms of the conventional $p$-form fields coupled to gamma matrices in a `democratic' manner~\cite{Bergshoeff:2001pv}, and compute  the R-R field strengths  explicitly. The above Bianchi identity is then naturally expected to  produce the   `tensor hierarchy'~\cite{Bergshoeff:2009ph,FernandezMelgarejo:2012ne,Fernandez-Melgarejo:2013xua}. 

It is worth while to note that, while the twist breaks the $\Ott$ T-duality to its subgroup which preserves the structure constant, $f_{\mA\mB\mC\,}$,  the  $\Spint\times\oSpint$ local Lorentz symmetries are still all unbroken after the twist and the dimensional reduction. 

When the twisting data, $U_{A}{}^{\mA}, \lambda$,\,  do not satisfy the original section condition, the corresponding background cannot be  identified as a solution to  the untwisted `$D=10$'  supersymmetric double field theories.  This might well  motivate one to wonder about  the existence of unknown genuinely ten-dimensional  ``generalized double field theory" with ``relaxed'' section conditions.  However,  the twistability conditions seem to admit only lower dimensional sections,  as the  non-trivial solutions.   In those  lower dimensions,    the standard section condition must be    obeyed,  see (\ref{tsc1}), and   its  doubled  coordinates are still to be \textit{gauged}. We regard the twist not as an indication of the existence of  any   unknown  $D=10$ ``generalized  DFT''  but as  a lower dimensional  deformation of the  known rigid untwisted  $D=10$ theories,  \textit{i.e.~}\cite{Jeon:2011sq,Jeon:2012hp}. A well known such example is the  massive supersymmetric deformations  of the super Yang-Mills   quantum mechanics~\cite{Berenstein:2002jq,Kim:2006wg}.  The deformations do not  necessarily mean that the  parental super Yang-Mills field theories can be likely  deformed.

In this work,  the R-R sector is taken as $\Ott$ singlet and assumes  the    $\Spint\times\oSpint$ local Lorentz bi-spinorial representation~\cite{Hassan:1999bv,Hassan:1999mm,Hassan:2000kr,
Berkovits:2001ue,Coimbra:2011nw,Coimbra:2012yy,
Jeon:2012kd,Jeon:2012hp}.\footnote{Alternative approach puts the R-R sector into  the $\Ott$ spinorial representation~\cite{Fukuma:1999jt,Hassan:1999mm,Hohm:2011zr}, to which   the bi-spinorial  approach  actually reduces  after taking the  diagonal gauge fixing, $\Spint\times\oSpint\rightarrow\Spint_{D}$~\cite{Jeon:2012kd}. Yet, it is not clear how to  twist the $\Ott$ spinor and then   couple to the $\Spint$ or $\oSpint$ fermions,  \textit{c.f.~}\cite{Geissbuhler:2011mx,Geissbuhler:2013uka}.  } This made   the twisting of the R-R sector rather trivial. Essentially,   the R-R potential, $\cC^{\alpha}{}_{\bralpha}$ , is \textit{not}  twisted,  like other fermions. Only the R-R field strength, $\mcF=\mcD_{+}\cC$, is influenced by the twist  through  the  twisted nilpotent differential operator.   We expect that this feature should change when  the U-duality group is twisted in  $\cM$-theory setup, but this goes beyond the scope of the present work. \\

%%%%%%%%%%%%%%%%%%%%%%%%%%%%%%%%%%%%%%%%%%%%%%%%%%%

\section*{Acknowledgements}
We wish to thank David Geissb{\"u}hler, Kanghoon Lee and  Diego Marqu\'es for helpful discussions. 
This work was  supported by the Fundaci\'on S\'eneca - Talento Investigador Program, and also by  the National Research Foundation of Korea (NRF)   with the Grants,  2012R1A2A2A02046739, 2013R1A1A1A05005747,  2015K1A3A1A21000302.

\newpage

%%%\appendix

%%%%%%%%%%%%%%%%%%%%%%%%%%%%%%%%%%%%%%%%%%%%%%%%%%%
%%%%%%%%%%%%%%%%%%%%%%%%%%%%%%%%%%%%%%%%%%%%%%%%%%%

\providecommand{\href}[2]{#2}\begingroup\raggedright\endgroup

%\bibliography{references}

\begin{thebibliography}{10}

\bibitem{Siegel:1993th}
W.~Siegel, ``{Superspace duality in low-energy superstrings},''
  \href{http://dx.doi.org/10.1103/PhysRevD.48.2826}{{\em Phys.Rev.} {\bf D48}
  (1993)  2826--2837},
\href{http://arxiv.org/abs/hep-th/9305073}{{\tt arXiv:hep-th/9305073
  [hep-th]}}.
%%CITATION = HEP-TH/9305073;%%.

\bibitem{Hull:2009mi}
C.~Hull and B.~Zwiebach, ``{Double Field Theory},''
  \href{http://dx.doi.org/10.1088/1126-6708/2009/09/099}{{\em JHEP} {\bf 0909}
  (2009)  099},
\href{http://arxiv.org/abs/0904.4664}{{\tt arXiv:0904.4664 [hep-th]}}.
%%CITATION = ARXIV:0904.4664;%%.

\bibitem{Hohm:2010jy}
O.~Hohm, C.~Hull, and B.~Zwiebach, ``{Background independent action for double
  field theory},'' \href{http://dx.doi.org/10.1007/JHEP07(2010)016}{{\em JHEP}
  {\bf 1007} (2010)  016},
\href{http://arxiv.org/abs/1003.5027}{{\tt arXiv:1003.5027 [hep-th]}}.
%%CITATION = ARXIV:1003.5027;%%.

\bibitem{Hohm:2010pp}
O.~Hohm, C.~Hull, and B.~Zwiebach, ``{Generalized metric formulation of double
  field theory},'' \href{http://dx.doi.org/10.1007/JHEP08(2010)008}{{\em JHEP}
  {\bf 1008} (2010)  008},
\href{http://arxiv.org/abs/1006.4823}{{\tt arXiv:1006.4823 [hep-th]}}.
%%CITATION = ARXIV:1006.4823;%%.

\bibitem{Duff:1989tf}
M.~Duff, ``{Duality Rotations in String Theory},''
\href{http://dx.doi.org/10.1016/0550-3213(90)90520-N}{{\em Nucl.Phys.} {\bf
  B335} (1990)  610}.
%%CITATION = NUPHA,B335,610;%%.

\bibitem{Tseytlin:1990nb}
A.~A. Tseytlin, ``{Duality Symmetric Formulation of String World Sheet
  Dynamics},''
\href{http://dx.doi.org/10.1016/0370-2693(90)91454-J}{{\em Phys.Lett.} {\bf
  B242} (1990)  163--174}.
%%CITATION = PHLTA,B242,163;%%.

\bibitem{Tseytlin:1990va}
A.~A. Tseytlin, ``{Duality symmetric closed string theory and interacting
  chiral scalars},''
\href{http://dx.doi.org/10.1016/0550-3213(91)90266-Z}{{\em Nucl.Phys.} {\bf
  B350} (1991)  395--440}.
%%CITATION = NUPHA,B350,395;%%.

\bibitem{Park:2013mpa}
J.-H. Park, ``{Comments on double field theory and diffeomorphisms},''
  \href{http://dx.doi.org/10.1007/JHEP06(2013)098}{{\em JHEP} {\bf 1306} (2013)
   098},
\href{http://arxiv.org/abs/1304.5946}{{\tt arXiv:1304.5946 [hep-th]}}.
%%CITATION = ARXIV:1304.5946;%%.

\bibitem{Hohm:2012gk}
O.~Hohm and B.~Zwiebach, ``{Large Gauge Transformations in Double Field
  Theory},'' \href{http://dx.doi.org/10.1007/JHEP02(2013)075}{{\em JHEP} {\bf
  1302} (2013)  075},
\href{http://arxiv.org/abs/1207.4198}{{\tt arXiv:1207.4198 [hep-th]}}.
%%CITATION = ARXIV:1207.4198;%%.

\bibitem{Lee:2013hma}
K.~Lee and J.-H. Park, ``{Covariant action for a string in "doubled yet gauged"
  spacetime},'' \href{http://dx.doi.org/10.1016/j.nuclphysb.2014.01.003}{{\em
  Nucl.Phys.} {\bf B880} (2014)  134--154},
\href{http://arxiv.org/abs/1307.8377}{{\tt arXiv:1307.8377 [hep-th]}}.
%%CITATION = ARXIV:1307.8377;%%.

\bibitem{Hohm:2013bwa}
O.~Hohm, D.~L\"ust, and B.~Zwiebach, ``{The Spacetime of Double Field Theory:
  Review, Remarks, and Outlook},''
  \href{http://dx.doi.org/10.1002/prop.201300024}{{\em Fortsch.Phys.} {\bf 61}
  (2013)  926--966},
\href{http://arxiv.org/abs/1309.2977}{{\tt arXiv:1309.2977 [hep-th]}}.
%%CITATION = ARXIV:1309.2977;%%.

\bibitem{Berman:2014jba}
D.~S. Berman, M.~Cederwall, and M.~J. Perry, ``{Global aspects of double
  geometry},'' \href{http://dx.doi.org/10.1007/JHEP09(2014)066}{{\em JHEP} {\bf
  1409} (2014)  066},
\href{http://arxiv.org/abs/1401.1311}{{\tt arXiv:1401.1311 [hep-th]}}.
%%CITATION = ARXIV:1401.1311;%%.

\bibitem{Hull:2014mxa}
C.~M. Hull, ``{Finite Gauge Transformations and Geometry in Double Field
  Theory},'' \href{http://dx.doi.org/10.1007/JHEP04(2015)109}{{\em JHEP} {\bf
  1504} (2015)  109},
\href{http://arxiv.org/abs/1406.7794}{{\tt arXiv:1406.7794 [hep-th]}}.
%%CITATION = ARXIV:1406.7794;%%.

\bibitem{Naseer:2015tia}
U.~Naseer, ``{A note on large gauge transformations in double field theory},''
\href{http://arxiv.org/abs/1504.05913}{{\tt arXiv:1504.05913 [hep-th]}}.
%%CITATION = ARXIV:1504.05913;%%.

\bibitem{Hull:2006qs}
C.~Hull, ``{Global aspects of T-duality, gauged sigma models and T-folds},''
  \href{http://dx.doi.org/10.1088/1126-6708/2007/10/057}{{\em JHEP} {\bf 0710}
  (2007)  057},
\href{http://arxiv.org/abs/hep-th/0604178}{{\tt arXiv:hep-th/0604178
  [hep-th]}}.
%%CITATION = HEP-TH/0604178;%%.

\bibitem{Hull:2006va}
C.~M. Hull, ``{Doubled Geometry and T-Folds},''
  \href{http://dx.doi.org/10.1088/1126-6708/2007/07/080}{{\em JHEP} {\bf 0707}
  (2007)  080},
\href{http://arxiv.org/abs/hep-th/0605149}{{\tt arXiv:hep-th/0605149
  [hep-th]}}.
%%CITATION = HEP-TH/0605149;%%.

\bibitem{Berman:2013eva}
D.~S. Berman and D.~C. Thompson, ``{Duality Symmetric String and M-Theory},''
  \href{http://dx.doi.org/10.1016/j.physrep.2014.11.007}{{\em Phys.Rept.} {\bf
  566} (2014)  1--60},
\href{http://arxiv.org/abs/1306.2643}{{\tt arXiv:1306.2643 [hep-th]}}.
%%CITATION = ARXIV:1306.2643;%%.

\bibitem{deWit:2005ub}
B.~de~Wit, H.~Samtleben, and M.~Trigiante, ``{Magnetic charges in local field
  theory},'' \href{http://dx.doi.org/10.1088/1126-6708/2005/09/016}{{\em JHEP}
  {\bf 0509} (2005)  016},
\href{http://arxiv.org/abs/hep-th/0507289}{{\tt arXiv:hep-th/0507289
  [hep-th]}}.
%%CITATION = HEP-TH/0507289;%%.

\bibitem{Scherk:1979zr}
J.~Scherk and J.~H. Schwarz, ``{How to Get Masses from Extra Dimensions},''
\href{http://dx.doi.org/10.1016/0550-3213(79)90592-3}{{\em Nucl.Phys.} {\bf
  B153} (1979)  61--88}.
%%CITATION = NUPHA,B153,61;%%.

\bibitem{Kaloper:1999yr}
N.~Kaloper and R.~C. Myers, ``{The Odd story of massive supergravity},''
  \href{http://dx.doi.org/10.1088/1126-6708/1999/05/010}{{\em JHEP} {\bf 9905}
  (1999)  010},
\href{http://arxiv.org/abs/hep-th/9901045}{{\tt arXiv:hep-th/9901045
  [hep-th]}}.
%%CITATION = HEP-TH/9901045;%%.

\bibitem{Geissbuhler:2011mx}
D.~Geissbuhler, ``{Double Field Theory and N=4 Gauged Supergravity},''
  \href{http://dx.doi.org/10.1007/JHEP11(2011)116}{{\em JHEP} {\bf 1111} (2011)
   116},
\href{http://arxiv.org/abs/1109.4280}{{\tt arXiv:1109.4280 [hep-th]}}.
%%CITATION = ARXIV:1109.4280;%%.

\bibitem{Aldazabal:2011nj}
G.~Aldazabal, W.~Baron, D.~Marques, and C.~Nunez, ``{The effective action of
  Double Field Theory},'' \href{http://dx.doi.org/10.1007/JHEP11(2011)052,
  10.1007/JHEP11(2011)109}{{\em JHEP} {\bf 1111} (2011)  052},
\href{http://arxiv.org/abs/1109.0290}{{\tt arXiv:1109.0290 [hep-th]}}.
%%CITATION = ARXIV:1109.0290;%%.

\bibitem{Schon:2006kz}
J.~Schon and M.~Weidner, ``{Gauged N=4 supergravities},''
  \href{http://dx.doi.org/10.1088/1126-6708/2006/05/034}{{\em JHEP} {\bf 0605}
  (2006)  034},
\href{http://arxiv.org/abs/hep-th/0602024}{{\tt arXiv:hep-th/0602024
  [hep-th]}}.
%%CITATION = HEP-TH/0602024;%%.

\bibitem{Dibitetto:2012rk}
G.~Dibitetto, J.~Fernandez-Melgarejo, D.~Marques, and D.~Roest, ``{Duality
  orbits of non-geometric fluxes},''
  \href{http://dx.doi.org/10.1002/prop.201200078}{{\em Fortsch.Phys.} {\bf 60}
  (2012)  1123--1149},
\href{http://arxiv.org/abs/1203.6562}{{\tt arXiv:1203.6562 [hep-th]}}.
%%CITATION = ARXIV:1203.6562;%%.

\bibitem{Coimbra:2011nw}
A.~Coimbra, C.~Strickland-Constable, and D.~Waldram, ``{Supergravity as
  Generalised Geometry I: Type II Theories},''
  \href{http://dx.doi.org/10.1007/JHEP11(2011)091}{{\em JHEP} {\bf 1111} (2011)
   091},
\href{http://arxiv.org/abs/1107.1733}{{\tt arXiv:1107.1733 [hep-th]}}.
%%CITATION = ARXIV:1107.1733;%%.

\bibitem{Coimbra:2012yy}
A.~Coimbra, C.~Strickland-Constable, and D.~Waldram, ``{Generalised Geometry
  and type II Supergravity},''
  \href{http://dx.doi.org/10.1002/prop.201100096}{{\em Fortsch.Phys.} {\bf 60}
  (2012)  982--986},
\href{http://arxiv.org/abs/1202.3170}{{\tt arXiv:1202.3170 [hep-th]}}.
%%CITATION = ARXIV:1202.3170;%%.

\bibitem{Grana:2012rr}
M.~Grana and D.~Marques, ``{Gauged Double Field Theory},''
  \href{http://dx.doi.org/10.1007/JHEP04(2012)020}{{\em JHEP} {\bf 1204} (2012)
   020},
\href{http://arxiv.org/abs/1201.2924}{{\tt arXiv:1201.2924 [hep-th]}}.
%%CITATION = ARXIV:1201.2924;%%.

\bibitem{Geissbuhler:2013uka}
D.~Geissbuhler, D.~Marques, C.~Nunez, and V.~Penas, ``{Exploring Double Field
  Theory},'' \href{http://dx.doi.org/10.1007/JHEP06(2013)101}{{\em JHEP} {\bf
  1306} (2013)  101},
\href{http://arxiv.org/abs/1304.1472}{{\tt arXiv:1304.1472 [hep-th]}}.
%%CITATION = ARXIV:1304.1472;%%.

\bibitem{Berman:2013cli}
D.~S. Berman and K.~Lee, ``{Supersymmetry for Gauged Double Field Theory and
  Generalised Scherk-Schwarz Reductions},''
  \href{http://dx.doi.org/10.1016/j.nuclphysb.2014.02.015}{{\em Nucl.Phys.}
  {\bf B881} (2014)  369--390},
\href{http://arxiv.org/abs/1305.2747}{{\tt arXiv:1305.2747 [hep-th]}}.
%%CITATION = ARXIV:1305.2747;%%.

\bibitem{Berman:2013uda}
D.~S. Berman, C.~D. Blair, E.~Malek, and M.~J. Perry, ``{The $O_{D,D}$ geometry
  of string theory},'' \href{http://dx.doi.org/10.1142/S0217751X14500808}{{\em
  Int.J.Mod.Phys.} {\bf A29} (2014) no.~15, 1450080},
\href{http://arxiv.org/abs/1303.6727}{{\tt arXiv:1303.6727 [hep-th]}}.
%%CITATION = ARXIV:1303.6727;%%.

\bibitem{Aldazabal:2013sca}
G.~Aldazabal, D.~Marques, and C.~Nunez, ``{Double Field Theory: A Pedagogical
  Review},'' \href{http://dx.doi.org/10.1088/0264-9381/30/16/163001}{{\em
  Class.Quant.Grav.} {\bf 30} (2013)  163001},
\href{http://arxiv.org/abs/1305.1907}{{\tt arXiv:1305.1907 [hep-th]}}.
%%CITATION = ARXIV:1305.1907;%%.

\bibitem{Hohm:2010xe}
O.~Hohm and S.~K. Kwak, ``{Frame-like Geometry of Double Field Theory},''
  \href{http://dx.doi.org/10.1088/1751-8113/44/8/085404}{{\em J.Phys.} {\bf
  A44} (2011)  085404},
\href{http://arxiv.org/abs/1011.4101}{{\tt arXiv:1011.4101 [hep-th]}}.
%%CITATION = ARXIV:1011.4101;%%.

\bibitem{Hohm:2011nu}
O.~Hohm and S.~K. Kwak, ``{N=1 Supersymmetric Double Field Theory},''
  \href{http://dx.doi.org/10.1007/JHEP03(2012)080}{{\em JHEP} {\bf 1203} (2012)
   080},
\href{http://arxiv.org/abs/1111.7293}{{\tt arXiv:1111.7293 [hep-th]}}.
%%CITATION = ARXIV:1111.7293;%%.

\bibitem{Jeon:2010rw}
I.~Jeon, K.~Lee, and J.-H. Park, ``{Differential geometry with a projection:
  Application to double field theory},''
  \href{http://dx.doi.org/10.1007/JHEP04(2011)014}{{\em JHEP} {\bf 1104} (2011)
   014},
\href{http://arxiv.org/abs/1011.1324}{{\tt arXiv:1011.1324 [hep-th]}}.
%%CITATION = ARXIV:1011.1324;%%.

\bibitem{Jeon:2011cn}
I.~Jeon, K.~Lee, and J.-H. Park, ``{Stringy differential geometry, beyond
  Riemann},'' \href{http://dx.doi.org/10.1103/PhysRevD.84.044022}{{\em
  Phys.Rev.} {\bf D84} (2011)  044022},
\href{http://arxiv.org/abs/1105.6294}{{\tt arXiv:1105.6294 [hep-th]}}.
%%CITATION = ARXIV:1105.6294;%%.

\bibitem{Jeon:2011sq}
I.~Jeon, K.~Lee, and J.-H. Park, ``{Supersymmetric Double Field Theory: Stringy
  Reformulation of Supergravity},''
  \href{http://dx.doi.org/10.1103/PhysRevD.86.089903,
  10.1103/PhysRevD.85.081501, 10.1103/PhysRevD.85.089908}{{\em Phys.Rev.} {\bf
  D85} (2012)  081501},
\href{http://arxiv.org/abs/1112.0069}{{\tt arXiv:1112.0069 [hep-th]}}.
%%CITATION = ARXIV:1112.0069;%%.

\bibitem{Jeon:2012hp}
I.~Jeon, K.~Lee, J.-H. Park, and Y.~Suh, ``{Stringy Unification of Type IIA and
  IIB Supergravities under N=2 D=10 Supersymmetric Double Field Theory},''
  \href{http://dx.doi.org/10.1016/j.physletb.2013.05.016}{{\em Phys.Lett.} {\bf
  B723} (2013)  245--250},
\href{http://arxiv.org/abs/1210.5078}{{\tt arXiv:1210.5078 [hep-th]}}.
%%CITATION = ARXIV:1210.5078;%%.

\bibitem{Jeon:2012kd}
I.~Jeon, K.~Lee, and J.-H. Park, ``{Ramond-Ramond Cohomology and O(D,D)
  T-duality},'' \href{http://dx.doi.org/10.1007/JHEP09(2012)079}{{\em JHEP}
  {\bf 1209} (2012)  079},
\href{http://arxiv.org/abs/1206.3478}{{\tt arXiv:1206.3478 [hep-th]}}.
%%CITATION = ARXIV:1206.3478;%%.

\bibitem{Gualtieri:2003dx}
M.~Gualtieri, ``{Generalized complex geometry},''
\href{http://arxiv.org/abs/math/0401221}{{\tt arXiv:math/0401221 [math-dg]}}.
%%CITATION = MATH/0401221;%%.

\bibitem{Grana:2008yw}
M.~Grana, R.~Minasian, M.~Petrini, and D.~Waldram, ``{T-duality, Generalized
  Geometry and Non-Geometric Backgrounds},''
  \href{http://dx.doi.org/10.1088/1126-6708/2009/04/075}{{\em JHEP} {\bf 0904}
  (2009)  075},
\href{http://arxiv.org/abs/0807.4527}{{\tt arXiv:0807.4527 [hep-th]}}.
%%CITATION = ARXIV:0807.4527;%%.

\bibitem{Hull:2009zb}
C.~Hull and B.~Zwiebach, ``{The Gauge algebra of double field theory and
  Courant brackets},''
  \href{http://dx.doi.org/10.1088/1126-6708/2009/09/090}{{\em JHEP} {\bf 0909}
  (2009)  090},
\href{http://arxiv.org/abs/0908.1792}{{\tt arXiv:0908.1792 [hep-th]}}.
%%CITATION = ARXIV:0908.1792;%%.

\bibitem{Jeon:2011vx}
I.~Jeon, K.~Lee, and J.-H. Park, ``{Incorporation of fermions into double field
  theory},'' \href{http://dx.doi.org/10.1007/JHEP11(2011)025}{{\em JHEP} {\bf
  1111} (2011)  025},
\href{http://arxiv.org/abs/1109.2035}{{\tt arXiv:1109.2035 [hep-th]}}.
%%CITATION = ARXIV:1109.2035;%%.

\bibitem{Andriot:2013xca}
D.~Andriot and A.~Betz, ``{$\beta$-supergravity: a ten-dimensional theory with
  non-geometric fluxes, and its geometric framework},''
  \href{http://dx.doi.org/10.1007/JHEP12(2013)083}{{\em JHEP} {\bf 1312} (2013)
   083},
\href{http://arxiv.org/abs/1306.4381}{{\tt arXiv:1306.4381 [hep-th]}}.
%%CITATION = ARXIV:1306.4381;%%.

\bibitem{Andriot:2014qla}
D.~Andriot and A.~Betz, ``{Supersymmetry with non-geometric fluxes, or a
  $\beta$-twist in Generalized Geometry and Dirac operator},''
  \href{http://dx.doi.org/10.1007/JHEP04(2015)006}{{\em JHEP} {\bf 1504} (2015)
   006},
\href{http://arxiv.org/abs/1411.6640}{{\tt arXiv:1411.6640 [hep-th]}}.
%%CITATION = ARXIV:1411.6640;%%.

\bibitem{Hohm:2011si}
O.~Hohm and B.~Zwiebach, ``{On the Riemann Tensor in Double Field Theory},''
  \href{http://dx.doi.org/10.1007/JHEP05(2012)126}{{\em JHEP} {\bf 1205} (2012)
   126},
\href{http://arxiv.org/abs/1112.5296}{{\tt arXiv:1112.5296 [hep-th]}}.
%%CITATION = ARXIV:1112.5296;%%.

\bibitem{Jeon:2011kp}
I.~Jeon, K.~Lee, and J.-H. Park, ``{Double field formulation of Yang-Mills
  theory},'' \href{http://dx.doi.org/10.1016/j.physletb.2011.05.051}{{\em
  Phys.Lett.} {\bf B701} (2011)  260--264},
\href{http://arxiv.org/abs/1102.0419}{{\tt arXiv:1102.0419 [hep-th]}}.
%%CITATION = ARXIV:1102.0419;%%.

\bibitem{Peeters:2006kp}
K.~Peeters, ``{A Field-theory motivated approach to symbolic computer
  algebra},'' \href{http://dx.doi.org/10.1016/j.cpc.2007.01.003}{{\em
  Comput.Phys.Commun.} {\bf 176} (2007)  550--558},
\href{http://arxiv.org/abs/cs/0608005}{{\tt arXiv:cs/0608005 [cs.SC]}}.
%%CITATION = CS/0608005;%%.

\bibitem{Peeters:2007wn}
K.~Peeters, ``{Introducing Cadabra: A Symbolic computer algebra system for
  field theory problems},''
\href{http://arxiv.org/abs/hep-th/0701238}{{\tt arXiv:hep-th/0701238
  [HEP-TH]}}.
%%CITATION = HEP-TH/0701238;%%.


\bibitem{13044294}
M.  Garcia-Fernandez, ``{Torsion-free generalized connections and Heterotic Supergravity},''
\href{http://arxiv.org/abs/1304.4294}{{\tt arXiv:1304.4294  [math.DG]}}.





\bibitem{Hohm:2011ex}
O.~Hohm and S.~K. Kwak, ``{Double Field Theory Formulation of Heterotic
  Strings},'' \href{http://dx.doi.org/10.1007/JHEP06(2011)096}{{\em JHEP} {\bf
  1106} (2011)  096},
\href{http://arxiv.org/abs/1103.2136}{{\tt arXiv:1103.2136 [hep-th]}}.
%%CITATION = ARXIV:1103.2136;%%.

\bibitem{Bergshoeff:2001pv}
E.~Bergshoeff, R.~Kallosh, T.~Ortin, D.~Roest, and A.~Van~Proeyen, ``{New
  formulations of D = 10 supersymmetry and D8 - O8 domain walls},''
  \href{http://dx.doi.org/10.1088/0264-9381/18/17/303}{{\em Class.Quant.Grav.}
  {\bf 18} (2001)  3359--3382},
\href{http://arxiv.org/abs/hep-th/0103233}{{\tt arXiv:hep-th/0103233
  [hep-th]}}.
%%CITATION = HEP-TH/0103233;%%.

\bibitem{Bergshoeff:2009ph}
E.~A. Bergshoeff, J.~Hartong, O.~Hohm, M.~Huebscher, and T.~Ortin, ``{Gauge
  Theories, Duality Relations and the Tensor Hierarchy},''
  \href{http://dx.doi.org/10.1088/1126-6708/2009/04/123}{{\em JHEP} {\bf 0904}
  (2009)  123},
\href{http://arxiv.org/abs/0901.2054}{{\tt arXiv:0901.2054 [hep-th]}}.
%%CITATION = ARXIV:0901.2054;%%.

\bibitem{FernandezMelgarejo:2012ne}
J.~Fernandez-Melgarejo, T.~Ortin, and E.~Torrente-Lujan, ``{Maximal Nine
  Dimensional Supergravity, General gaugings and the Embedding Tensor},''
  \href{http://dx.doi.org/10.1002/prop.201200039}{{\em Fortsch.Phys.} {\bf 60}
  (2012)  1012--1018},
\href{http://arxiv.org/abs/1209.3774}{{\tt arXiv:1209.3774 [hep-th]}}.
%%CITATION = ARXIV:1209.3774;%%.

\bibitem{Fernandez-Melgarejo:2013xua}
J.~J. Fernandez-Melgarejo, ``{Gaugings and other aspects in supergravity},''
\href{http://arxiv.org/abs/1311.7145}{{\tt arXiv:1311.7145 [hep-th]}}.
%%CITATION = ARXIV:1311.7145;%%.

\bibitem{Berenstein:2002jq}
D.~E. Berenstein, J.~M. Maldacena, and H.~S. Nastase, ``{Strings in flat space
  and pp waves from N=4 superYang-Mills},''
  \href{http://dx.doi.org/10.1088/1126-6708/2002/04/013}{{\em JHEP} {\bf 0204}
  (2002)  013},
\href{http://arxiv.org/abs/hep-th/0202021}{{\tt arXiv:hep-th/0202021
  [hep-th]}}.
%%CITATION = HEP-TH/0202021;%%.

\bibitem{Kim:2006wg}
N.~Kim and J.-H. Park, ``{Massive super Yang-Mills quantum mechanics:
  Classification and the relation to supermembrane},''
  \href{http://dx.doi.org/10.1016/j.nuclphysb.2006.10.005}{{\em Nucl.Phys.}
  {\bf B759} (2006)  249--282},
\href{http://arxiv.org/abs/hep-th/0607005}{{\tt arXiv:hep-th/0607005
  [hep-th]}}.
%%CITATION = HEP-TH/0607005;%%.

\bibitem{Hassan:1999bv}
S.~Hassan, ``{T duality, space-time spinors and RR fields in curved
  backgrounds},'' \href{http://dx.doi.org/10.1016/S0550-3213(99)00684-7}{{\em
  Nucl.Phys.} {\bf B568} (2000)  145--161},
\href{http://arxiv.org/abs/hep-th/9907152}{{\tt arXiv:hep-th/9907152
  [hep-th]}}.
%%CITATION = HEP-TH/9907152;%%.

\bibitem{Hassan:1999mm}
S.~Hassan, ``{SO(d,d) transformations of Ramond-Ramond fields and space-time
  spinors},'' \href{http://dx.doi.org/10.1016/S0550-3213(00)00337-0}{{\em
  Nucl.Phys.} {\bf B583} (2000)  431--453},
\href{http://arxiv.org/abs/hep-th/9912236}{{\tt arXiv:hep-th/9912236
  [hep-th]}}.
%%CITATION = HEP-TH/9912236;%%.

\bibitem{Hassan:2000kr}
S.~Hassan, ``{Supersymmetry and the systematics of T duality rotations in type
  II superstring theories},''
  \href{http://dx.doi.org/10.1016/S0920-5632(01)01539-0}{{\em
  Nucl.Phys.Proc.Suppl.} {\bf 102} (2001)  77--82},
\href{http://arxiv.org/abs/hep-th/0103149}{{\tt arXiv:hep-th/0103149
  [hep-th]}}.
%%CITATION = HEP-TH/0103149;%%.

\bibitem{Berkovits:2001ue}
N.~Berkovits and P.~S. Howe, ``{Ten-dimensional supergravity constraints from
  the pure spinor formalism for the superstring},''
  \href{http://dx.doi.org/10.1016/S0550-3213(02)00352-8}{{\em Nucl.Phys.} {\bf
  B635} (2002)  75--105},
\href{http://arxiv.org/abs/hep-th/0112160}{{\tt arXiv:hep-th/0112160
  [hep-th]}}.
%%CITATION = HEP-TH/0112160;%%.

\bibitem{Fukuma:1999jt}
M.~Fukuma, T.~Oota, and H.~Tanaka, ``{Comments on T dualities of Ramond-Ramond
  potentials on tori},'' \href{http://dx.doi.org/10.1143/PTP.103.425}{{\em
  Prog.Theor.Phys.} {\bf 103} (2000)  425--446},
\href{http://arxiv.org/abs/hep-th/9907132}{{\tt arXiv:hep-th/9907132
  [hep-th]}}.
%%CITATION = HEP-TH/9907132;%%.

\bibitem{Hohm:2011zr}
O.~Hohm, S.~K. Kwak, and B.~Zwiebach, ``{Unification of Type II Strings and
  T-duality},'' \href{http://dx.doi.org/10.1103/PhysRevLett.107.171603}{{\em
  Phys.Rev.Lett.} {\bf 107} (2011)  171603},
\href{http://arxiv.org/abs/1106.5452}{{\tt arXiv:1106.5452 [hep-th]}}.
%%CITATION = ARXIV:1106.5452;%%.

\end{thebibliography}
%\bibliographystyle{utphys}

%%%%%%%%%%%%%%%%%%%%%%%%%%%%%%%%%%%%%%%%%%%%%%%%%%%%%%%%%%%%%%%
%%%%%%%%%%%%%%%%%%%%%%%%%%%%%%%%%%%%%%%%%%%%%%%%%%%%%%%%%%%%%%%

\end{document}